\documentclass{aastex}
\usepackage{spr-astr-addons}
\usepackage{url}
\graphicspath{{fig/}{fig/inclusive_leptons/}}
\usepackage{ifdraft}
\usepackage{xspace}
\usepackage{color}
\usepackage{ifthen}
\definecolor{rossoCP3}{cmyk}{0,.88,.77,.40}
\definecolor{darkBlue}{rgb}{0, 0, 0.8}
\usepackage{hyperref}
\hypersetup{
    bookmarksopen=true,     
    bookmarksopenlevel=1,
    unicode=false,          
    pdftoolbar=true,        
    pdfmenubar=true,        
    colorlinks=true,        
    linkcolor=darkBlue,     
    citecolor=darkBlue,     
    filecolor=darkBlue,     
    urlcolor=darkBlue       
}

\newcommand{\ee}{\ensuremath{e^+e^-}\xspace}
\newcommand{\pX}[1]{\mbox{{\sl p}\hspace{0.1em}-#1}\xspace}
\newcommand{\piX}[1]{\mbox{$\pi$-#1}\xspace}
\newcommand{\hX}[1]{\mbox{{\sl h}\hspace{0.1em}-#1}\xspace}
\newcommand{\pp}{\pX{p}}
\newcommand{\pPb}{\pX{Pb}}
\newcommand{\pA}{\pX{A}}
\newcommand{\pO}{\pX{O}}
\newcommand{\XX}[2]{\mbox{#1\hspace{0.1em}-#2}\xspace}
\newcommand{\XeXe}{\XX{Xe}{Xe}}
\newcommand{\AuAu}{\XX{Au}{Au}}
\newcommand{\PbPb}{\XX{Pb}{Pb}}
\newcommand{\OO}{\XX{O}{O}}
\newcommand{\xmax}{\ensuremath{X_\text{max}}\xspace}
\newcommand{\nmu}{\ensuremath{N_\mu}\xspace}
\newcommand{\nmult}{N_\text{mult}}
\newcommand{\lnn}{\ln\!\nmu}
\newcommand{\lna}{\ln\!A}
\newcommand{\avg}[1]{\langle #1 \rangle}
\newcommand{\mlna}{\ensuremath{\avg{\lna}}\xspace}

\newcommand{\fg}[1]{Fig.~\ref{fig:#1}\xspace}
\newcommand{\tb}[1]{Table~\ref{tab:#1}\xspace}
\newcommand{\eq}[1]{Eq.~\ref{eq:#1}\xspace}
\newcommand{\sect}[1]{Section~\ref{sec:#1}\xspace}
\newcommand{\dd}{\text{d}}
\newcommand{\pt}{p_\text{T}}
\newcommand{\pbar}{\ensuremath{\bar{p}}\xspace}
\newcommand{\nbar}{\ensuremath{\bar{n}}\xspace}

\newcommand{\si}[1]{\ensuremath{\,\mathrm{#1}}\xspace}
\newcommand{\gevc}{\si{GeV}\,\ensuremath{c^{-1}}}
\newcommand{\tev}{\si{TeV}}

\newcommand{\gev}{\si{GeV}}
\newcommand{\pev}{\si{PeV}}
\newcommand{\eev}{\si{EeV}}

\newcommand{\fe}{\text{Fe}}

\newcommand{\sqrtsnn}{\ensuremath{\sqrt{s_\text{NN}}}\xspace}

\newcommand{\piz}{\ensuremath{\pi^0}\xspace}
\newcommand{\rhoz}{\ensuremath{\rho^0}\xspace}
\newcommand{\kzero}{K\ensuremath{^0}\xspace}
\newcommand{\vzero}{\ensuremath{V^0}\xspace}
\newcommand{\kshort}{K\ensuremath{^0_S}\xspace}
\newcommand{\rhozero}{\ensuremath{\rho^{0}}\xspace}

\newcommand{\beq}{\begin{equation}}
\newcommand{\eeq}{\end{equation}}

\newcommand{\NASixtyOne}{NA61\slash SHINE\xspace}

\newcommand{\sibyll}[1]{\mbox{\textsc{Sibyll}}#1\xspace}
\newcommand{\dpmjet}[1]{\mbox{\textsc{DPMJet\ifthenelse{\equal{#1}{}}{}{-#1}}}\xspace}
\newcommand{\qgsjet}[1]{\mbox{\textsc{QGSJet#1}}\xspace}
\newcommand{\gheisha}{\textsc{Gheisha}\xspace}
\newcommand{\epos}{\textsc{EPOS}\xspace}
\newcommand{\eposlhc}{\textsc{EPOS-LHC}\xspace}
\newcommand{\pythia}[1]{\mbox{\textsc{Pythia}\ifthenelse{ \equal{#1}{} }{}{\,#1}}\xspace}
\newcommand{\mceq}{{\sc MCEq}\xspace}

\newcommand{\aires}{\textsc{Aires}\xspace}
\newcommand{\corsika}{\textsc{Corsika}\xspace}
\newcommand{\conex}{\textsc{Conex}\xspace}
\newcommand{\mocca}{\textsc{Mocca}\xspace}
\newcommand{\seneca}{\textsc{Seneca}\xspace}
\newcommand{\cosmos}{\textsc{Cosmos}\xspace}
\newcommand{\proposal}{\textsc{Proposal}\xspace}
\newcommand{\emca}{\textsc{EmCa}\xspace}

\newcommand{\xsinel}{\sigma_\text{inel}}

\begin{document}


\title{The Muon Puzzle in cosmic-ray induced air showers and its connection to the Large Hadron Collider}

\shorttitle{Muon Puzzle in air showers and the LHC}
\shortauthors{Albrecht \textit{et al.}}

\author{Johannes Albrecht\altaffilmark{1}}
\author{Lorenzo Cazon\altaffilmark{{2}}}
\author{Hans Dembinski\altaffilmark{1,$\star$}}
\author{Anatoli Fedynitch\altaffilmark{3}}
\author{Karl-Heinz Kampert\altaffilmark{4}}
\author{Tanguy Pierog\altaffilmark{5}}
\author{Wolfgang Rhode\altaffilmark{1}}
\author{Dennis Soldin\altaffilmark{6}}
\author{Bernhard Spaan\altaffilmark{1}}
\author{Ralf Ulrich\altaffilmark{5}}
\author{Michael Unger\altaffilmark{5}}
\email{hans.dembinski@tu-dortmund.de}

\altaffiltext{1}{Faculty of Physics, TU Dortmund University, Germany}
\altaffiltext{2}{LIP, Lisbon, Portgual}
\altaffiltext{3}{Institute for Cosmic Ray Research, The University of Tokyo,
5-1-5 Kashiwa-no-ha, Kashiwa, Chiba 277-8582, Japan}
\altaffiltext{4}{Department of Physics, Wuppertal University, Germany}
\altaffiltext{5}{Institute for Astroparticle Physics, Karlsruhe Institute of Technology (KIT), Karlsruhe, Germany}
\altaffiltext{6}{Bartol Research Institute and Dept. of Physics and Astronomy, University of Delaware, Newark, DE 19716, USA}
\altaffiltext{$\star$}{Corresponding author: \url{hans.dembinski@tu-dortmund.de}}

\begin{abstract}
High-energy cosmic rays are observed indirectly by detecting the extensive air showers initiated in Earth's atmosphere. The interpretation of these observations relies on accurate models of air shower physics, which is a challenge and an opportunity to test QCD under extreme conditions. Air showers are hadronic cascades, which give rise to a muon component through hadron decays. The muon number is a key observable to infer the mass composition of cosmic rays. Air shower simulations with state-of-the-art QCD models show a significant muon deficit with respect to measurements; this is called the Muon Puzzle. By eliminating other possibilities, we conclude that the most plausible cause for the muon discrepancy is a deviation in the composition of secondary particles produced in high-energy hadronic interactions from current model predictions. The muon discrepancy starts at the TeV scale, which suggests that this deviation is observable at the Large Hadron Collider. An enhancement of strangeness production has been observed at the LHC in high-density events, which can potentially explain the puzzle, but the impact of the effect on forward produced hadrons needs further study, in particular with future data from oxygen beam collisions.
\end{abstract}


\section{Introduction}

\begin{figure*}
\centering
\includegraphics[width=0.9\textwidth]{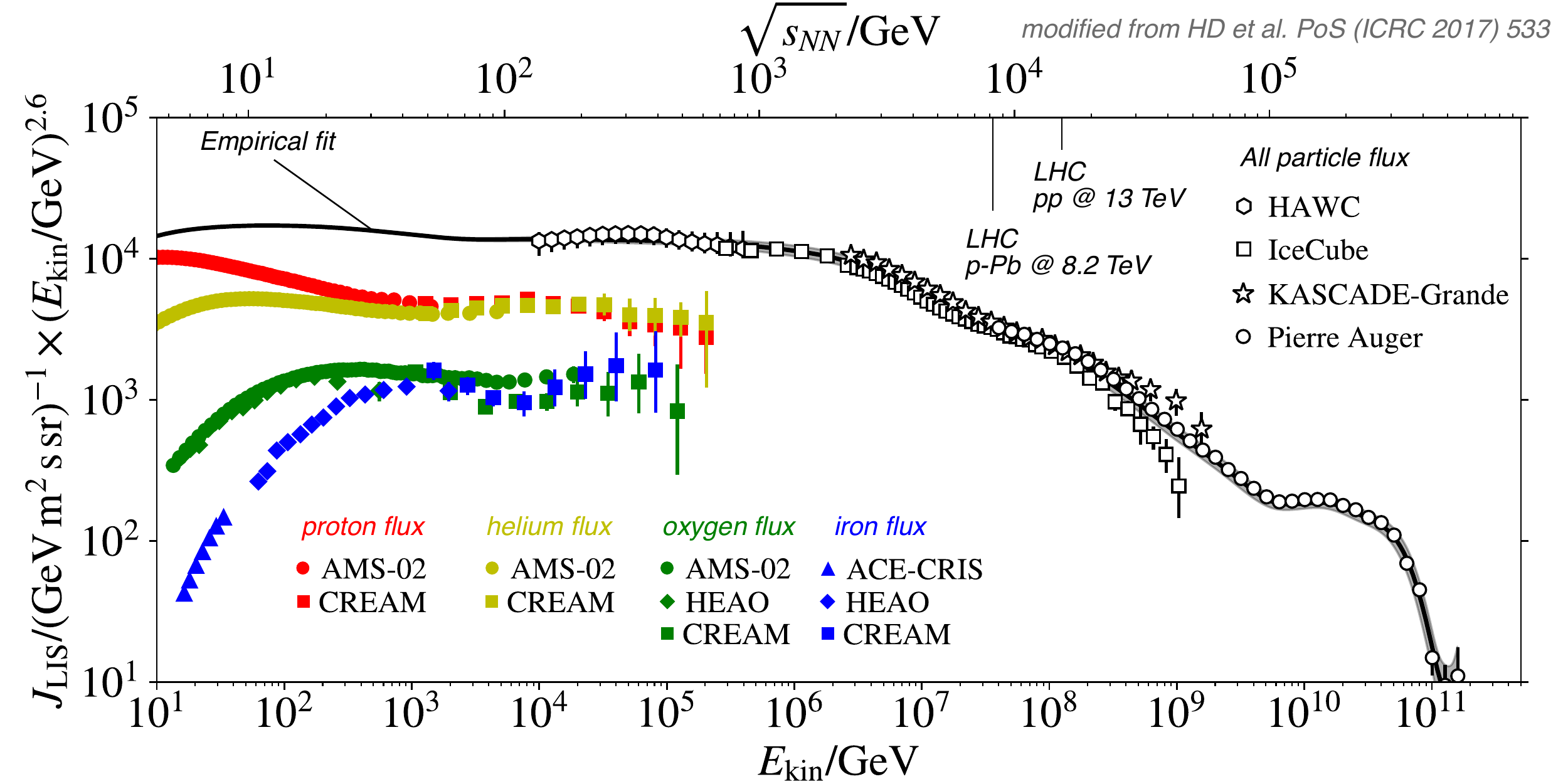}
\caption{Selected measurements of the cosmic ray flux as a function of kinetic energy, see \cite{Dembinski:2017zsh,Schroder:2019rxp} for the full data set and references. The steeply falling fluxes are scaled by $E_\text{kin}^{2.6}$. Open symbols show the all-particle cosmic-ray flux measured by air shower experiments. Coloured solid symbols show the fluxes of individual elements measured by balloon- and satellite-borne experiments. The line is an empirical fit to these data. The upper axis shows the equivalent centre-of-mass energy of a nucleon-nucleon collision. Determining the elemental composition of the cosmic-ray flux above $10^6$\,GeV motivates this project, which requires precise measurements of forward hadron production at the LHC.}
\label{fig:cr_flux}
\end{figure*}

\begin{figure*}
\centering
\includegraphics[width=0.48\textwidth]{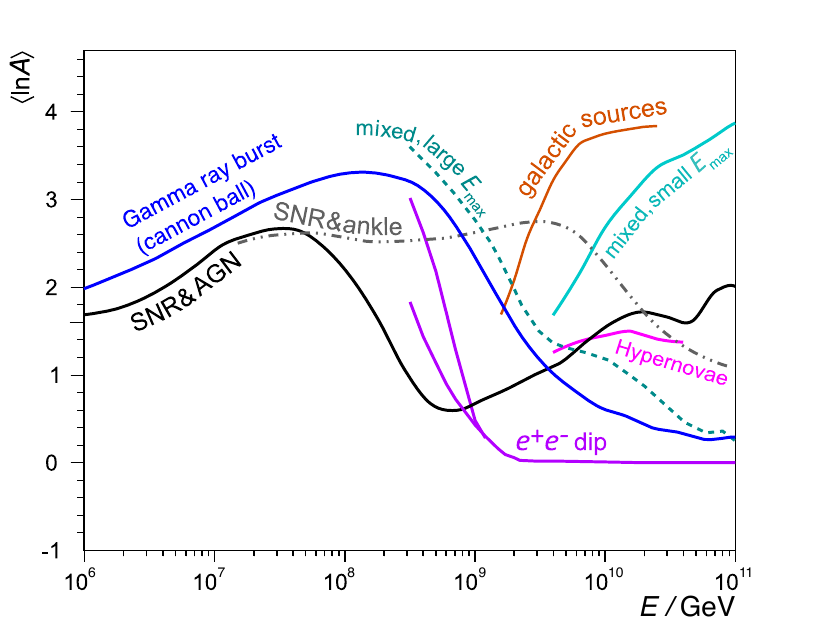}
\includegraphics[width=0.48\textwidth]{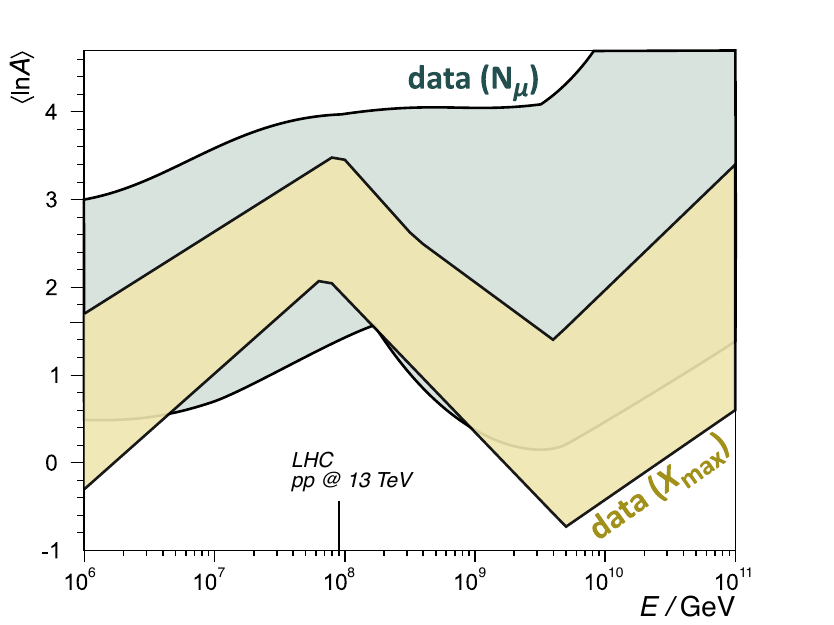}
\caption{\emph{Left}: Predictions of the mean-logarithmic mass \mlna of cosmic rays as a function of their energy from several theories. \emph{Right}: Two bands that cover the ranges of measurements, grouped by the mass-sensitive variable used (\xmax or \nmu, explained in the text). The vertical line indicates the equivalent energy of \pp interaction at $13\,$TeV at the LHC. The width of the data bands is dominantly caused by theoretical uncertainties of forward hadron production. These uncertainties prevent the exclusion of theories on the origins of cosmic rays. Data and model lines were taken from \cite{Kampert:2012mx}. }
\label{fig:lna}
\end{figure*}

\begin{figure*}
\centering
\includegraphics[width=\textwidth]{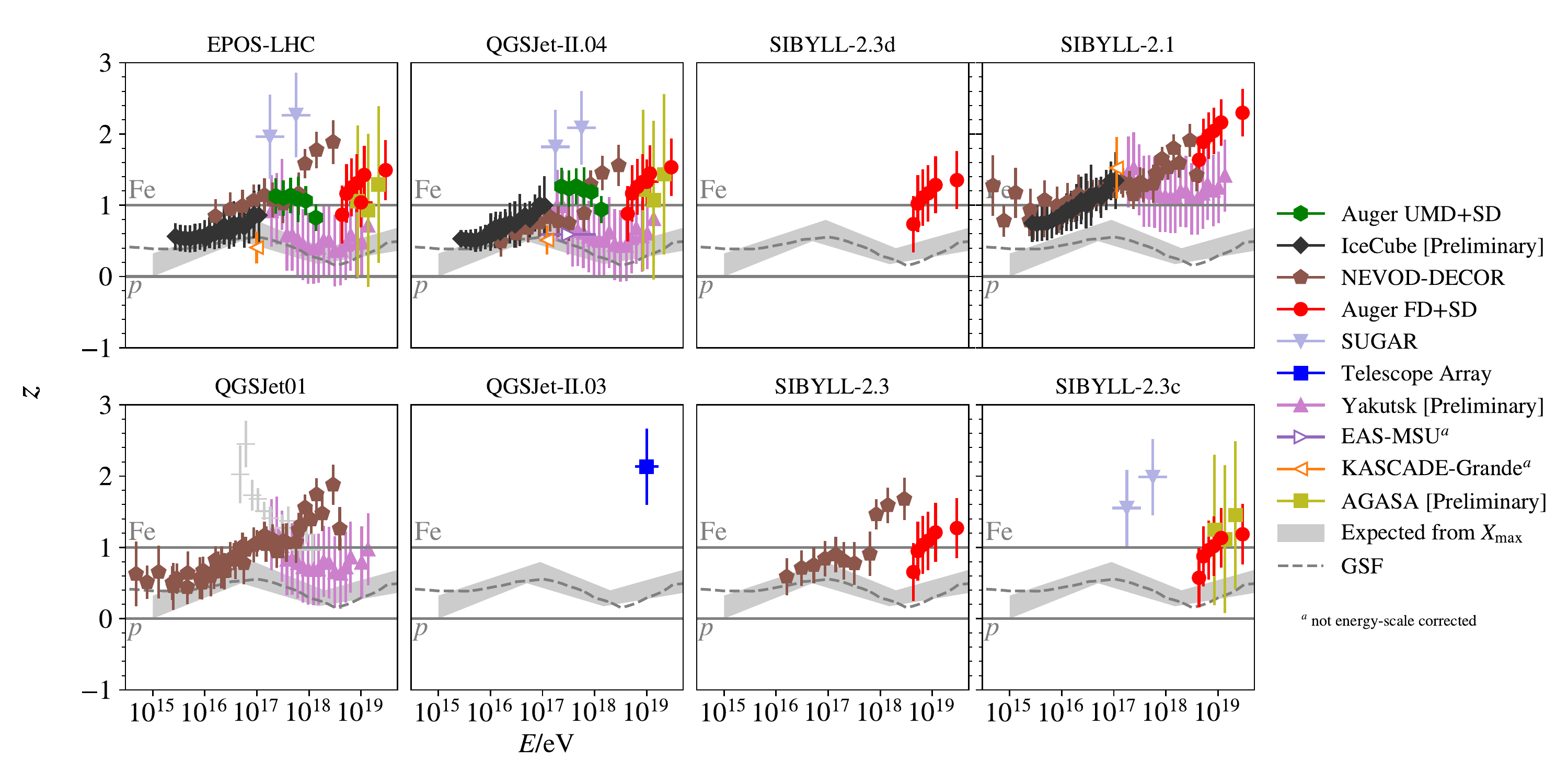}
\caption{Compilation of muon measurements converted to the abstract $z$-scale and after cross-calibrating the energy scales of the experiments as described in the text (image from \cite{Soldin:2021wyv}). Shown for comparison are predicted $z_\text{mass}$-values based on air shower simulations and \xmax-measurements (grey band). The prediction from the GSF model \citep{Dembinski:2017zsh} for $z_\text{mass}$ is also shown (dashed line).}
\label{fig:z_rescaled}
\end{figure*}

Cosmic rays are fully-ionised nuclei with relativistic kinetic energies that arrive at Earth. The elements range from proton to iron, with a negligible fraction of heavier nuclei. Cosmic rays originate from unknown sources outside of our solar system and are messengers of the high-energy universe. The cosmic-ray energy flux weighted with $E^{2.6}$ to compress the enormous scale is shown in \fg{cr_flux}. It spans over 11 orders in energy and over more than 30 orders in flux intensity, which can only be covered by multiple experiments using different measurement techniques. The particle with the highest energy ever reported was a cosmic ray \citep{Bird:1994mp} with $(320\pm 90)\eev = (3.2 \pm 0.9) \cdot 10^{11}\gev$ ($1\eev = 10^9\gev$). Cosmic rays with energies below $1\pev = 10^6\gev$ are commonly assumed to be produced by shock acceleration in super-nova remnants \citep{Blasi:2010gr,Caprioli_2012}. The origins of cosmic rays with higher energies are unclear and many mechanisms have been proposed, see e.g.\ \cite{Anchordoqui:2018qom%
} for a recent review. Ultra-high energy cosmic rays with energies exceeding \eev are of unknown extra-galactic origin.

Classic astronomy with cosmic rays has not yet been established. Although cosmic rays are likely produced in well-identifiable point-sources, these sources do not appear point-like in the sky. The incoming flux of cosmic rays is very isotropic \citep{Aab:2014ila}. The reason is that cosmic rays are charged and scattered by inhomogeneous galactic and extra-galactic fields on their way to Earth. Their movement through space resembles a diffusive flow and their arrival directions at Earth are largely random. The average angle of deflection decreases with energy, however, and evidence of anisotropies has been found above the EeV scale \citep{Aab:2017tyv,Aab:2016ban,Aab:2018chp,Abbasi:2014lda}.

Up to particle energies of about 100\tev, cosmic rays are observed directly by space-based experiments, like AMS-02 \citep{AMS02}, and high-altitude balloons, like CREAM \citep{2011ApJ...728..122Y}. At higher energies the flux is too low for direct observation and ground-based experiments with huge apertures (up to 3000\si{km^2}) like the Pierre Auger Observatory \citep{ThePierreAuger:2015rma} and Telescope Array \citep{AbuZayyad:2012kk,Tokuno:2012mi} are used. Ground-based experiments observe cosmic rays indirectly through the particle showers (extensive air showers) produced in Earth's atmosphere. How air showers arise from cosmic rays and how observable air shower features are linked to the properties of the cosmic ray, its direction, energy $E$, and nuclear mass $A$ is described in Sections \ref{sec:cascade-equations} and \ref{sec:heitler}.

In regard to determining the dominant sources of cosmic rays, an important complementary approach to anisotropy studies is to measure the energy-dependent elemental (or mass) composition of cosmic rays. The fluxes of individual elements can be directly measured with suitable satellite- and balloon-borne experiments, but this is not equally possible with indirect air shower observations. The mass has to be inferred from air shower features in the latter case, which change depending on the mass and are subject to stochastic randomisations because of intrinsic fluctuations in the shower. These fluctuations overwhelm the small average shower differences between neighbouring elements. The composition above the PeV scale is therefore often summarised by a single number, the mean-logarithmic mass \mlna. In \fg{lna} left-hand side, predictions for \mlna are shown for several proposed source classes (lines) \citep[and references therein]{Kampert:2012mx}. Precise measurements of \mlna can rule out many of these competing theories. In particular, whether the cosmic rays with the highest energies are light or heavy is of crucial importance for the design of the next generation of cosmic ray and cosmic neutrino observatories, see e.g.\ \cite{Aloisio:2009sj, AlvesBatista:2019tlv}.

Two main features of an air shower are used to estimate the mass, its depth of shower maximum \xmax, and the number of muons \nmu produced in the shower. The two bands in \fg{lna} right-hand side represent an envelope of the measurements carried out by various air shower experiments \citep{Kampert:2012mx}. The composition estimates derived from measurements of \nmu have particularly large systematic uncertainties and hardly constrain the models. This is very unsatisfactory, since \nmu discriminates better between light and heavy primaries shower-by-shower at the EeV scale \citep{Muller:2018zoc} than \xmax, and it is useful to collect large statistics especially above $10^{19.5}$~eV, where observations of \xmax with fluorescence telescopes run out of statistics as these telescopes can only be operated in dark nights and therefore have a duty cycle of about 15\,\%, while muons can be observed with a duty cycle of 100\,\%.

A closer inspection reveals that most of the uncertainty is not experimental. Hybrid experiments in particular are able to make precise air shower measurements. An overview on these experiments is given in \sect{air-shower-detection}. The experimental uncertainty on the measured value of \nmu is around 10\,\%, which is a factor 2.5 to 4 (depending on the energy) smaller than the width of the band shown in \fg{lna}. Instead, most of the uncertainty is of theoretical nature and originates from the air shower simulations that are used to infer \mlna from \xmax and \nmu. These simulations are essential for the interpretation of air shower measurements, since there is no astrophysically identifiable source in the sky with a known mass composition that could be used for calibration.

The uncertainty does not originate from the particle transport in the atmosphere itself, which is comparably well understood. A variety of independently developed air shower simulation programs is in use and have been compared against each other. An overview is given in \sect{shower-codes}. These codes show only small variations at the level of 5\,\% in regard to \nmu. The uncertainty originates from the evolution of the hadronic cascade that drives the air shower evolution and the muon production at the end of that cascade. The cascade is dominated by hadronic collisions with small momentum transfer, which cannot be calculated with perturbative quantum chromodynamics (pQCD). Effective theories and phenomenology are used to predict the rates of these interactions and the spectra of secondary particles produced in them. These take the form of software codes which are called hadronic interaction models or generators. A variety of generators is in use which follow different ideas and which vary in their predictions. An overview is given in \sect{generators}.

LHC data were used to improve the latest versions of these generators, which would reduce the width of the bands in \fg{lna} if all measurements based on older versions were updated accordingly. However, a detailed comparison of measurements with predictions from the latest generators then revealed that the generators consistently predict a lower muon production in air showers than is observed. This was established with a nearly model-independent measurement for the first time by the Pierre Auger Observatory \citep{Aab:2014pza,Aab:2016hkv}. The measurement was recently updated and now includes also a measurement of the intrinsic shower-to-shower fluctuations of \nmu \citep{Aab:2021zfr}, which has been measured for the first time in any air shower experiment. While there are also theoretical uncertainties in the simulation of \xmax that need to be further reduced, there are no such blatant discrepancies between the simulated and observed values of \xmax. No large discrepancies are observed between the predicted and measured \nmu-fluctuations, which disfavours some exotic explanations of the muon deficit.

The Auger publication from 2015 triggered several follow-up measurements and the re-analysis of existing air shower data. In particular, the IceCube Neutrino Observatory was used to perform a nearly model-independent study of \nmu at lower energies \citep{Dembinski:2017zkb}. The wealth of newly published data over a wide range in shower energies made a review necessary and the Working group on Hadronic Interactions and Shower Physics (WHISP) was founded by members of eight experimental collaborations \citep{Dembinski:2019uta}. To facilitate the comparison of very diverse muon measurements, the group defined an abstract $z$-scale. The $z$-values are proportional to the logarithm of \nmu and can be computed for each pair of experiment and generator. In a comprehensive meta-analysis, simulations with six hadronic interaction models were tested against the combined data (three from the previous generation and three state-of-the-art). The results are shown in \fg{z_rescaled}. The data points exceed the expectation for showers above 10\pev. More precisely, a positive linear slope with $8\si{\sigma}$ significance was found. This significance is higher than that of any individual measurement, where the significance for a muon deficit does not exceed $3\si{\sigma}$. More details on the WHISP meta-analysis are given in \sect{meta-analysis}.

This muon discrepancy is called the \emph{Muon Puzzle}, because the authors of the generators have been unable to resolve the discrepancy by parameter tuning. The required changes to existing models would violate either data constraints from accelerators or the consistency between air shower simulations and the other air shower features. This suggests that a physical effect is missing in the generators that governs soft hadronic interactions at high energies. This point is explored in \sect{impact} and \ref{sec:solutions}. Also remarkable is the early onset of the muon discrepancy, which starts smoothly above the knee and increases with the logarithm of the shower energy. The onset of the discrepancy is seen in showers where the first interaction of the cosmic ray has a centre-of-mass (cms) energy of about 8\tev in the nucleon-nucleon system. This suggests that the source of the Muon Puzzle can be studied at the LHC.

Given the wealth of existing LHC data described in \sect{lhc-measurements}, this poses the question why the source of the discrepancy was not yet observed, which is the other aspect that turns the muon discrepancy into a puzzle. One likely answer is that we have not yet looked in the right place. Since the study of soft hadronic interactions was not driving their design, most LHC experiments focus their instrumentation on the mid-rapidity region where new heavy particles such as the Higgs are best observed, while the air shower development is strongly dominated by particles produced in the forward region. The relevant phase-space starts at pseudo-rapidity\footnote{In this experimental review we have chosen to work with pseudo-rapidity instead of rapidity because full particle identification is not provided by the majority of experiments in the very forward region of phase space.} $\eta \gtrsim 2$. LHCb is the notable exception in that it is fully instrumented in the pseudo-rapidity range $2 < \eta < 5$. Important input is further provided by the CMS experiment with its CASTOR forward calorimeter covering $-6.6 < \eta < -5.2$ and from the TOTEM and LHCf experiments. An overview of the acceptances of LHC experiments is given in \sect{lhc-experiments}.

It is well-known that the number of muons \nmu produced at the end of the hadronic cascade is very sensitive to the energy fraction carried away by photons, which are primarily produced in \piz decays, as explained in Sections \ref{sec:heitler} and \ref{sec:impact}. A reduction of the \piz-fraction has a compounding effect in the shower cascade, so that only a comparably small change is required to obtain a notable change in the muon number. Why the \piz-fraction cannot be easily changed without introducing new physics is explained in \sect{solutions}, together with extensions of the standard soft QCD picture in high-multiplicity interactions which achieve just that. Such extensions seem inevitable based on basic QCD arguments and are required to match recent LHC data.

We may have already seen a glimpse of the solution to the puzzle. The ALICE experiment has observed a universal enhancement of strangeness production \citep{ALICE:2016fzo,Vasileiou:2020rov} in high-multiplicity events at mid-rapidity, an effect that was previously only observed in heavy-ion collisions \citep{Koch:1986ud}. An increase in strangeness leads to a corresponding relative decrease of the pion yield, including the \piz yield. Preliminary studies suggest that this effect could potentially solve the Muon Puzzle \citep{Baur:2019cpv,Anchordoqui:2019laz}, if it is present also in the forward region that drives the air shower development.

One of the most pressing questions is therefore whether there is also a universal enhancement of strange particle production in the forward region $\eta > 2$. The production cross-sections for strange hadrons need to be measured as a function of charged particle multiplicity in several collisions systems (\pp, \pPb, and \PbPb). The LHCb experiment is in a unique position to perform such measurements with its high-precision particle tracking with particle identification in the forward region. Such studies can be done by analysing already recorded LHC data. More generally speaking, precise measurements of the energy fraction carried by forward photons are needed. The CASTOR component of CMS and the LHCf experiment are ideal for such measurements.

Similarly important would be the study of soft hadronic interactions in future \pO collisions at the LHC, which directly mimic interactions of the cosmic ray with air, which have been proposed in \cite{Citron:2018lsq} for the upcoming Run 3 of the LHC. It is clear that soft hadron production at the TeV scale is full of unexpected phenomena and the scaling of hadron production from a \pp system to a \pO system is not well understood. Current generators show a large variation in the charged particle multiplicity of $\pm 25\,\%$ in the \pO system despite good agreement in the \pp system, which could be reduced five-fold to 5\,\% or better with LHC data. While it may be possible to interpolate to \pO from reference measurements in \pp and \pPb, this has to be confirmed with an explicit high-precision measurement. The case for \pO collisions at the LHC is discussed in \sect{oxygen}. Finally, there are also opportunities to further improve our knowledge on hadronic interactions below the TeV scale with fixed-target experiments that are described in \sect{fixed-target}.

A solution for the Muon Puzzle from the LHC would have a large impact on the field of astroparticle physics. The sizes of the \nmu-bands in \fg{lna} could shrink by factors 2.5 to 4 (depending on the energy). The ambiguity of the cosmic-ray mass composition at the EeV scale would be resolved and the LHC measurements on hadron production would also further decrease the uncertainty of \xmax-predictions. The changes in air shower predictions would have an immediate impact on the interpretation of existing air shower data and trigger the re-evaluation of existing data. The improvements to the generators and the increased accuracy of the estimated cosmic-ray mass composition would in turn make predictions of atmospheric lepton fluxes more precise, the main background for neutrino observatories like IceCube. Air shower measurements connected to the Muon Puzzle are summarised in \sect{connected-results}.

\section{Cosmic-ray detection via air showers: Muon Puzzle and connected challenges}
\label{sec:astro-part}

Cosmic rays with EeV energy produce extensive air showers (EAS) in Earth's atmosphere with lateral extensions that reach tens of kilometres. The angular spread of the produced particles with respect to the direction of the cosmic ray is initially very small due to relativistic boost, but increases as the average energy per particle drops and eventually reaches a few degrees. These particles define the \emph{shower axis}. The axis is characterised by two angles, the zenith angle $\theta$ and the azimuthal angle $\phi$. The zenith angle measures the inclination of the axis, $0^\circ$ corresponds to a vertical down-going shower. At any given point in time, the shower particles are concentrated in a thin disc with a thickness of a few tens of meters perpendicular to the shower axis that moves with the speed of light. The fastest particles (dominantly muons) form the \emph{shower front}. The position where the shower axis intersects the ground is called the \emph{shower core}.

Air showers above PeV energies are highly regular and can be characterised by a few parameters, a feature sometimes referred to as \emph{shower universality} \citep{Patterson:1983qj,Giller:2005qz,Lipari:2008td,Yushkov:2010un}. Statistical fluctuations in the first few interactions have an important effect on the development of the rest of the shower, but the later shower stages show no substructure since fluctuations average out due to the large number of particles interactions. This makes it feasible to build cosmic-ray observatories with huge apertures that cover thousands of square-kilometres by building sparse arrays of particle detectors at the ground. In principle, a shower can be fully characterised with measurements at only three spatially separated points. The angles $\theta$ and $\phi$ of the shower axis can be accurately triangulated from the arrival times, while the measured local densities determine the shower core and the amplitude of its lateral profile of particle density.

After the first interaction, the shower develops two partially coupled parallel cascades: the hadronic and the electromagnetic cascade. The hadronic cascade is formed by interactions of long-lived hadrons (mostly pions, but also nucleons and strange hadrons), while the electromagnetic cascade is fed by photons which are mostly generated by decays of neutral pions. The electromagnetic cascade develops independently via electron pair-production and bremsstrahlung. There is only a small feedback from the electromagnetic into the hadronic cascade from rare photo-nuclear interactions.

The air shower produces long-lived hadrons, electrons, photons, muons, and neutrinos. The average energy of particles in the shower continuously decreases, since every interaction spreads the energy of the parent onto several children. Most of the muons are produced at the end of the hadronic cascade when the Lorentz factor gets so small that decay becomes more probable than another interaction. A small fraction of muons is produced electromagnetically by direct pair-production. Neutrinos are completely decoupled once produced. A small fraction of muons decays to feed the electromagnetic component and neutrino component. Muons trace the development of the hadronic cascade.

Charged particles produced by the shower ionise air and thereby loose energy. The ionised air molecules eventually fall back to their ground state. Some of that released energy is converted into isotropically emitted fluorescence light; the process has been measured precisely in the laboratory \citep{Ave:2012ifa}. This light is used in dark nights to observe air showers at several kilometres distance with sensitive telescopes. The advantage of fluorescence light is that the shower can be seen from the side, which allows a single telescope to observe a large volume of air. Air showers further generate Cherenkov light flashes in a cone around the shower axis. The aperture achieved with Cherenkov light is smaller, so this technique is mostly used for the detection of sub-EeV air showers.

The air shower detection with particle detectors at the ground and with telescopes detecting fluorescence and Cherenkov light has become the established standard in the field. To measure the muon number, particle detectors are required and the telescopes provide important additional information. It was the combination of telescopes with particle detectors in a large scale experiment, the Pierre Auger Observatory, that allowed the field to unambiguously establish the muon production as the source of several discrepancies that had been observed in the last twenty years. A third more recently developed technique is the detection of air showers via radiowave emissions \citep{Falcke:2005tc,Buitink:2014eqa,Kostunin:2015taa}, which could become an alternative to the detection via fluorescence and Cherenkov light with a 100\,\% duty cycle.

In the following subsections, we give an overview on how air showers are modelled and detected, and highlight key results in regard to the Muon Puzzle.

\subsection{Calculating air showers with cascade equations}
\label{sec:cascade-equations}

The transport, production and decay of particles in the atmosphere can be characterised by coupled differential equations. A generic introduction into this subject is given by \cite{Gaisser:2016cr}. These equations are conveniently expressed as a function of slant depth $X$. A slant depth interval is the product of the geometric length interval $\dd s$ travelled and the air density, $\dd X = \rho_\text{air} \, \dd s$. Let $n$ be the density of particles of type $k$ in an infinitesimal energy interval $\dd E$, then the change of that density over a travel distance over a slant depth interval is given by
\begin{align}\label{eq:cascade-equations}
\frac{\dd n}{\dd X} =
 &-\left( \frac{1}{\lambda_{\text{int},k}(E)} + \frac{1}{\lambda_{\text{dec},k}(E, X)} \right) n(E,X) \nonumber\\
 &-\frac{\dd}{\dd E}\big(\mu_k(E) \, n(E,X)\big) \nonumber\\
 &+\sum_\ell{\int_E^\infty \dd E_\ell~\frac{c_{\ell \to k}(E_\ell,E)}{\lambda_{\text{int}, \ell}(E_\ell)}} n_\ell(E_\ell,X) \nonumber\\
 &+\sum_\ell{\int_E^\infty \dd E_\ell~\frac{d_{\ell \to k}(E_\ell,E)}{\lambda_{\text{dec}, \ell}(E_\ell, X)}} n_\ell(E_\ell,X),
\end{align}
with $n \equiv n_k, E \equiv E_k$. This equation describes the longitudinal shower development. The lateral development is integrated out, but can in principle be included as well.
The first two terms describe losses; particle loss from interaction and decay with the lengths $\lambda_{\text{int},k}$ and $\lambda_{\text{dec},k}$, respectively, and energy loss from ionisation where the specifics are captured by the function $\mu_k(E)$. The last two terms describe gains from interactions and decays of other particle species $\ell$, which depends on the interaction and decay lengths and the transfer probabilities $c_{\ell \to k}$ and $d_{\ell \to k}$. The integrals go formally to infinity, but are cut off by the upper energy limit in $n_\ell(E_\ell, X)$.

\begin{figure}
\centering
\includegraphics[width=\columnwidth]{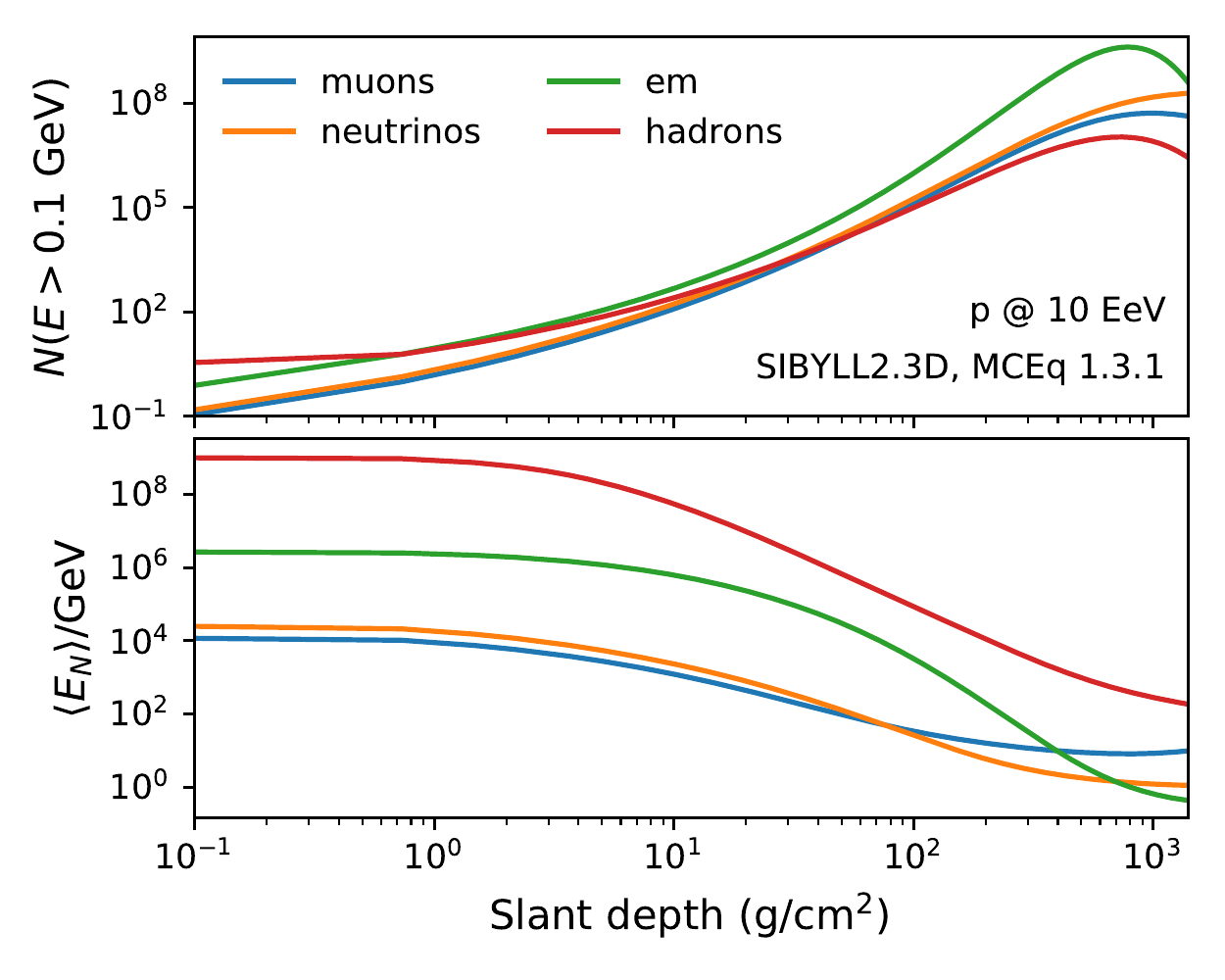}
\caption{Average particle count (top) and mean energy (bottom) in an air shower initiated by a 10 EeV proton at a zenith angle $\theta = 45^\circ$. In the legend, \emph{em} corresponds to electrons and photons. The maximum of this component is the observable shower maximum with slant depth \xmax.}
\label{fig:particle_cascade_demo}
\end{figure}

Cascade equations are primarily used to compute atmospheric lepton fluxes, where the initial densities $n_k(E, 0)$ are given by the energy spectra of cosmic nuclei, but they are also used to calculate particle densities in an average air shower. The air shower solution is obtained with the initial condition $n_k(E, 0) = \delta(E-E_0)$. The solution for a 10\eev proton calculated with the program \mceq \citep{Fedynitch:2015zma} is shown in \fg{particle_cascade_demo}. The particle densities initially increase exponentially as the energy of the initial cosmic ray is distributed to more and more secondary particles. The cascading process stops when most particles drop below a critical energy, where decay (in case of hadrons) or absorption (in case of electrons) becomes dominant over re-interaction. At this point the exponential growth is replaced by exponential loss. The loss rate for electrons is much higher than for muons due to the much smaller mass and correspondingly larger pair-production cross section. Muons start off with energies of tens of GeV and are minimum ionising with an energy loss of only a few \si{MeV/(g\, cm^{-2})}.

The interaction length $\lambda_{\text{int},k}$ is the ratio of the average mass of an air nucleus $m_\text{air}$ and the inelastic cross-section, $\lambda_{\text{int},k} = m_\text{air} / \sigma_{\text{inel},k}(E)$. The decay length $\lambda_{\text{dec},k}$ depends on the gamma factor $\gamma = E/m$ of the particle, the decay length in its rest frame $c \tau$ and the local air density, $\lambda_\text{dec} = E/m \, c \, \tau \, \rho_\text{air}(X)$. These losses compete; we have $\lambda_{\text{int},k} \ll \lambda_{\text{dec},k}$ at high energy, but the situation reverses eventually at low energies.

Hadrons, electrons, and photons are produced dominantly by interactions,
\begin{equation}
c_{\ell \to k}(E_\ell,E_k) = \frac{\dd N_{\ell + air \to k + X}}{\dd E} (E_\ell, E_k).
\end{equation}
In hadronic processes, the couplings $c_{\ell \to h}$ correspond to inclusive differential production cross-sections or decay energy distributions \citep{Fedynitch:2018cbl}. Theoretical uncertainties in regard to hadron production enter here. For electromagnetic processes, the couplings correspond to pair production, annihilation, bremsstrahlung, Moeller, Bhabha and Compton scattering \citep{Bergmann:2006yz}, which are well understood and can be calculated from first principles. Muons and neutrinos are produced dominantly by pion decays. The transfer is given by the decay energy distributions,
\begin{equation}
d_{\ell \to k}(E_\ell,E_k) = \frac{\dd N_{\ell \to k + X}}{\dd E} (E_\ell, E_k),
\end{equation}
which are also well known for the decays of long-lived particles. Short-lived particles are not explicitly tracked, they implicitly contribute to these transfers or they are neglected if their production cross-sections are small.

Cascade equations are ideal to study the influence of changes in hadron production on average air shower variables. The equations offer exact solutions if all relevant processes are implemented, and solving them takes only a few seconds with modern programs. Since the mapping between inputs and outputs is analytical, one can also perform error propagation of inputs to outputs. In order to reproduce shower-to-shower fluctuations of quantities like \xmax and \nmu, another approach is required where the cascade equations are not solved numerically, but with Monte-Carlo methods. Some codes use hybrid calculations, where the initial stages of the shower are solved with the Monte-Carlo method, then cascade equations are used when the particle count is large, and finally Monte-Carlo methods are used again in the final stages. An overview of the available codes is given in \sect{shower-codes}. In the next subsection, we discuss a useful toy model of an air shower, which has simple analytical solutions.

\subsection{Heitler model of air shower development}
\label{sec:heitler}

\begin{figure}
\centering
\includegraphics[width=0.45\textwidth,trim=20 -30 0 0,clip]{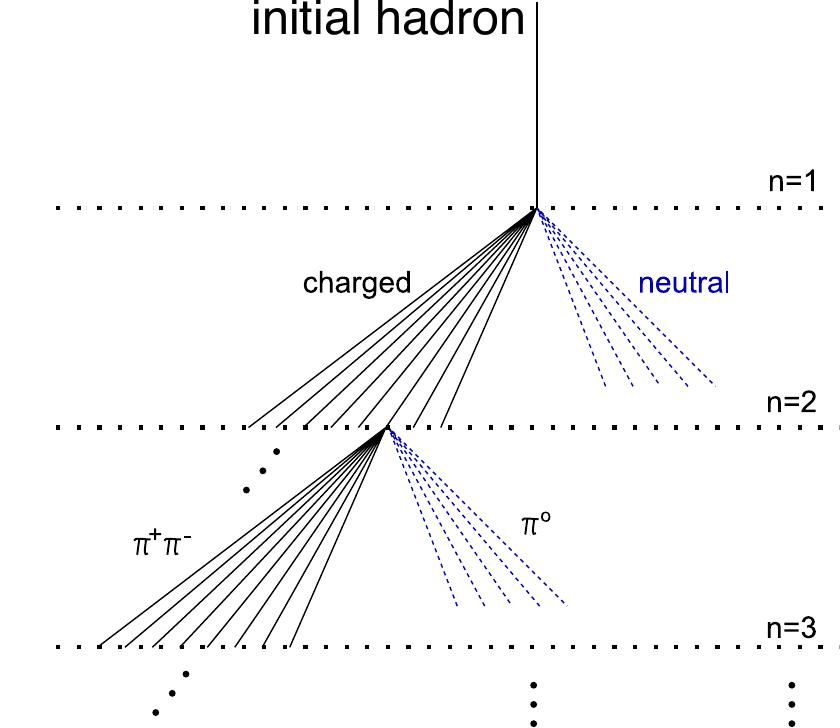}
\caption{First three steps of a hadronic cascade in the simplified Heitler-Matthews model (image from \cite{Ulrich:2010rg}). Solid (dashed) lines represent charged (neutral) pions. Only one of the splittings is shown after the second step.}
\label{fig:shower_physics}
\end{figure}

Air showers can be computed accurately using cascade equations, which are solved either numerically or by Monte Carlo methods. To gain basic insights, simplified air shower models of the Heitler type have been very successful. They are not viable alternatives to solving cascade equations, but reveal essential relationships. In some cases, the derived formulas produce accurate predictions if the coefficients are taken from full air shower simulations. We briefly outline how the basic Heitler model is constructed and list key results that are relevant for the Muon Puzzle. For more details, we refer to the accessible treatments by \cite{Matthews:2005sd} and \cite{Ulrich:2010rg}.

\cite{heitler1984quantum} introduced a simplified model of an electromagnetic cascade consisting only of photons and electrons. Starting from a primary particle of energy $E$, the particle count doubles after travelling a fixed amount of slant depth. The energy of the parent is distributed equally among the children. The doubling stops when the energy per particle falls below a critical energy, which is reached when absorption becomes equally likely to another splitting. At this point, all particles are absorbed in the medium.

The concept was generalised to a cosmic-ray induced air shower by \cite{Matthews:2005sd}. The shower is approximated by a pure pion shower, as sketched in \fg{shower_physics}. Charged and neutral pions are produced. The total number of pions produced is $\nmult$. Neutral pions decay immediately into photon pairs, which are further treated with the aforementioned electromagnetic cascade. We call the energy fraction $\alpha$ that is retained in charged pions. In the basic Heitler-Matthews model, $\alpha$ is exactly $2/3$, but it is instructive to keep $\alpha$ as a parameter in the equations that can take on other values. In a real shower also other long-lived hadrons are produced, which increases $\alpha$. Charged pions travel though a fixed amount of slant depth and then produce a fixed number $\nmult$ of new pions. The energy of the parent is distributed equally among the children. The hadronic cascade stops when the energy per pion reaches the critical energy $\xi_h$, which is reached when the decay length of a charged pion becomes equal to the interaction length. At this point, all charged pions decay into muons. Cosmic-ray nuclei can be included with the superposition model, which treats a shower initiated by a nucleus with $A$ nucleons as $A$ independent showers with energy $E/A$.

The Heitler-Matthews model connects basic air shower quantities, the primary energy $E$ and mass $A$ of the cosmic ray, the muon number \nmu, and the depth \xmax of the maximum of the electromagnetic cascade, with features of hadronic interactions, the inelastic cross-section $\xsinel$, the hadron multiplicity $\nmult$, and the energy fraction $\alpha$ retained in long-lived hadrons.
\begin{itemize}
\item The number of steps $k$ in the hadronic cascade increases logarithmically with the energy $E$ as
\beq
k = \frac{\ln (E / \xi_h)}{\ln \nmult}
\eeq
which gives with $\xi_h \approx 10\gev$ and $\nmult \approx 50$ values between 3 at 1\pev and 5 at 1\eev.
\item The muon number scales sub-linearly with the cosmic-ray energy,
\beq
\label{eq:nmu_e_a}
\nmu(E, A) = A^{(1-\beta)} \left(\frac{E}{\xi_h} \right)^\beta \text{ with } \beta = \frac{\ln (\alpha \, \nmult)}{\ln \nmult}.
\eeq
The fact that $\beta \approx 0.9$ is close to but less than 1 is the consequence of the energy transfer from the hadronic to the electromagnetic cascade without an equivalent feedback.
\item There is a linear relationship between the mean-logarithmic mass \mlna and the mean-logarithmic muon number
\beq
\avg\lnn(E, A) = \avg\lnn(E, 1) + (1-\beta) \, \mlna,
\eeq
where $\avg\lnn(E, 1)$ is the mean-logarithmic muon number for proton showers. This follows from \eq{nmu_e_a}, see \cite{Dembinski:2017kpa} for a more detailed discussion. In practice, $\beta$ is taken from air shower simulations. Iron showers have about 40\,\% more muons than proton showers at the EeV scale.
\item Likewise, there is a linear relationship between the shower depth \xmax and \mlna
\beq
\avg\xmax(E, A) = \avg\xmax(E, 1) - D_p \, \mlna,
\eeq
where $D_p = \dd \avg{\xmax}(E,1) / \dd \ln E$ is the so-called elongation rate for proton showers, which taken from air shower simulations in practice. Proton showers develop deeper by about $100\si{g\,cm^{-2}}$ on average than iron showers at the EeV scale. A more detailed discussion is given in \cite{Kampert:2012mx,Abreu:2013env}.
\end{itemize}

Further conclusions can be drawn about the dependence of muon production in air showers and microscopic features of hadronic interactions which are listed below without derivation. They are confirmed overall by detailed simulations as described in \sect{impact}, with minor modifications.
\begin{itemize}
\item Both \nmu and \xmax depend weakly on the hadron multiplicity $\nmult$.
\item The muon number \nmu is independent of the inelastic cross-section $\xsinel$ for pion interactions, while \xmax is very sensitive to it.
\item The muon number is very sensitive to $\alpha$, while \xmax is (nearly) independent of $\alpha$. The relative increase in \nmu from \eq{nmu_e_a} for a small change $\Delta \alpha$ is to first order
\beq
\frac{\Delta \nmu}\nmu \approx \frac{\ln(E / \xi_h)}{\ln \nmult} \, \frac{\Delta \alpha}\alpha = k \, \frac{\Delta \alpha}\alpha,
\eeq
where $k$ is the number of steps of the hadronic cascade. For an \eev air shower, $k \approx 5$, which implies that a $10\,\%$ change in $\alpha$ introduces a $50\,\%$ change in the muon number. This implies that we need to measure $\alpha$ very precisely over a wide energy range and understand its extrapolation toward higher energies.
\end{itemize}

The Heitler-Matthews model is useful to build an intuition about air showers, but it is important to keep the approximations and simplifications in mind to not overinterpret the results. We summarise them here.
\begin{itemize}
\item All secondaries receive the same energy fraction. In reality, the energy depends strongly on the pseudorapidity of the particles. Particles produced at forward pseudorapidity in the cms-system of a hadron-air collision carry the largest energies in the lab frame, and therefore quantities like $\nmult$ and $\alpha$ should be understood as averages of the subset of forward produced particles.
\item Hadronic interactions produce other long-lived particles in addition to pions. Also important are kaons, protons, and neutrons. The relative fractions of these other hadrons could be the key for solving the Muon Puzzle, since effects which enhance strangeness and baryon production keep more energy in the hadronic cascade and increase $\alpha$.
\item The hadronic interaction length and the hadron multiplicity $\nmult$ are not constant but weakly energy dependent. The impact of this was further studied by \cite{Montanus:2014lya}.
\item The atmosphere does not have constant density which has an impact on the critical energy $\xi_h$, which depends on the zenith angle of the shower. Vertical air showers develop in denser atmosphere and have a lower critical energy than inclined air showers. This effect is best described by full simulations with a realistic atmosphere, but it can be ignored if only showers with a fixed zenith angle are considered. For an isothermal atmosphere, $\xi_h$ can be calculated analytically, see \cite{Kampert:2012mx}.
\item Since each random process is replaced by its average process, the model describes an average air shower. Extensions of the basic model are needed to describe intrinsic shower fluctuations.
\end{itemize}

Several authors have refined the Heitler-Matthews model to make its predictions more accurate or performed additional calculations based on the model. \cite{Grimm:2017svd} has investigated the leading particle effect that \cite{Matthews:2005sd} discusses only briefly, which is important in real air showers. \cite{Montanus:2014lya} has studied the effect of the weak energy dependence of certain hadronic interaction features that are set constant in the original treatment. \cite{Kampert:2012mx} present formulas for the first two moments of the cosmic-ray mass composition from measurements of depth \xmax of the shower maximum, and discuss the energy-dependence of $\nmult$ and the hadronic interaction length. \cite{Dembinski:2017kpa} presents a discussion of the first two moments of the muon number \nmu, and the impact of additional fluctuations in the measurement on these estimates.

\subsection{Air shower simulation codes}
\label{sec:shower-codes}

\begin{table*}[tb!]
\caption{Comparison of dedicated air shower simulation codes that use full Monte-Carlo simulation (MC) and/or numerical solutions of cascade equations (CE) to compute hadronic and electromagnetic cascades up to ultra-high energies. Model legend. B:Bertini, D: DPMJET, EGS4: \cite{Nelson:1994zb}, E: EPOS, F: FLUKA, G: GHEISHA, HS: Hillas-Splitting, J: JAM, JQ: JQMD, LPM: LPM-effect (see text), Q: QGSJet, S: Sibyll, Tsai: \cite{Tsai:1973py}, U: UrQMD. Details and references for the hadronic models are given in the text and in \sect{generators}. A star ($\star$) indicates that development has been discontinued.}
\label{tab:shower-codes}
\centering
\setlength{\tabcolsep}{3pt}
\begin{tabular}{lcccccccc}
\hline
        &           &        &          &          & \multicolumn{2}{c}{Hadronic model} \\
Program & Reference & Method & Language & EM model & Low $E$ & High $E$ & \\
\hline
\aires & \cite{Sciutto:1999jh} & MC & Fortran & custom+LPM & HS & S, Q, E & \\
\conex & \cite{Bergmann:2006yz} & MC+CE & Fortran & EGS4+LPM & U & S, Q, E & \\
\textsc{Cosmos} & \cite{Kasahara:2007jwa} & MC & Fortran & Tsai+LPM & J, JQ, B, D & S, Q, E, D & \\
\corsika & \cite{Heck:1998vt} & MC+CE & Fortran & EGS4+LPM & G, F, U & S, Q, E, D & \\
\corsika 8 & \cite{Engel:2018akg} & MC+CE & C++ & \proposal & U & S, Q & \\
\mceq & \cite{Fedynitch:2015zma} & CE & Python & \emca (Tsai) & D & S, Q, E, D & \\
\mocca ($\star$) & \cite{Hillas:1997tf} & MC & Pascal & custom & HS & HS & \\
\textsc{Seneca} ($\star$) & \cite{Drescher:2002cr} & MC+CE & Fortran & EGS4 & G & Q \\
\hline
\end{tabular}
\end{table*}

Accurate predictions of observable air shower features requires solving the cascade equations from \sect{cascade-equations} numerically or via Monte-Carlo methods. The full complexity with particle collisions, decays, hadronic and electromagnetic physics, elastic scattering, continuous energy losses as well as deflection in magnetic fields can only be handled by computer programs. Monte-Carlo methods are required when the full four-dimensional structure of the shower needs to be known and when shower-to-shower fluctuations are of interest.
%
%
The field profits from a set of well-maintained and widely used air shower simulation programs that offered robust predictions over long periods of time, providing the foundation to compare data from past and present experiments to model predictions, as discussed in \sect{meta-analysis}.

Special event generators are developed to describe hadron interactions up to the highest energies, but these models loose validity at low (\gev{} level) energies. Simulation codes therefore switch from a high-energy to a low-energy model when the lab energies fall below about 100\gev. Available choices for the low-energy model are \textsc{UrQMD} \citep{Lang:2016jpe}, \textsc{Fluka} \citep{Bohlen:2014buj}, and the now disfavoured \gheisha \citep{Fesefeldt:1985yw}. These generators are mostly limited to lab energies below \tev or less.

Both low-energy and high-energy generators have an impact on the resulting air shower cascades. Their impact approximately factorises in high-energy showers, but this is not true for showers below \tev energy, where a strong interplay between the high-energy and low-energy model is observed \citep{Parsons:2019hol,Pastor-Gutierrez:2021lxs}.

The low-energy models have a large impact on the lateral density profile of muons at the ground level \citep{Swordy:2002df,Drescher:2002vp,Drescher:2003gh}, but minor impact on the total muon number. A discrepancy in the low-energy models cannot explain the observed increase in the muon discrepancy with the shower energy that is described in \sect{meta-analysis}. The number of steps in the hadronic cascade calculated with the high-energy model increases with the shower energy, but remains constant for the low energy model. The low-energy model can therefore only contribute a constant to the discrepancy. More details about this point can be found in \sect{heitler} and \sect{impact}. Accordingly, this review focusses on the high-energy models, which are discussed in \sect{generators}, where also the references are given. However, there are still opportunities to further improve the low-energy models with experimental data, which are discussed in \sect{fixed-target}.

There are only minor variations in the outputs of air shower codes, if the same hadronic interaction models are used. The outputs of several programs have been compared \citep{Knapp:2002vs,Roh:2013ev,Bergmann:2006yz,Ortiz:2004gb} and generally agree within 5\,\% in the predicted muon number \nmu. It is therefore unlikely that the Muon Puzzle originates entirely from an inaccuracy in the air shower codes, but further reductions of these uncertainties are desirable. Apart from the Muon Puzzle several other deviations between muon measurements and simulations have been observed, which are discussed in \sect{connected-results}.

There is an ongoing effort to reduce the uncertainties in the lepton transport codes, especially in regard to simulation for underground detectors. Leptons which propagate through kilometres of dense matter undergo hundreds of interactions so that even small errors have a large compounding effect \citep{Chirkin:2004hz}. \proposal is a recent lepton propagation code with the goal to compute atmospheric lepton fluxes to 1\,\% accuracy \citep{Koehne:2013gpa,Dunsch:2018nsc}. The calculation includes radiative corrections \citep{Sandrock:2017yeq} to the average energy loss, higher-order corrections to bremsstrahlung and pair-production cross-sections in $Z\alpha$, and corrections due to the finite size of the nucleus \citep{Sandrock:2018ivj}. These predictions can be tested to the percent level with large samples of neutrino-induced muon tracks collected by large underground detectors \citep{Soedingrekso:2020yzf}. \proposal can be used standalone and as the lepton propagator in the new air shower program \corsika~8 \citep{Engel:2018akg}. \emca \citep{Meighen-Berger:2019cxt} is another new code for the high-accuracy simulation of electromagnetic cascades. The \emca cross-sections are also used in \mceq \citep{Fedynitch:2015zma}.

An overview of widely used air shower programs is given in \tb{shower-codes} and they are briefly introduced below.
\begin{itemize}

\item
\aires \citep{Sciutto:1999jh} is the successor of \mocca \citep{1977ICRC....8..460H,Hillas:1997tf}. It was extended in particular to interface with different hadronic interaction models. \aires is an air shower program with a feature set similar to \corsika. It supports a wide range of high-energy hadronic interaction models, but does not allow the user to change the low-energy model. Low-energy hadronic interactions are implemented with the Hillas-Splitting-Algorithm \citep{Hillas:1981} tuned to reproduce \gheisha. The electromagnetic cascades include the Landau-Pomeranchuk-Midgal (LPM) effect \citep{Landau:1953um,Migdal:1956tc} and are simulated with custom code. Radio emission is generated with the \textsc{ZHAireS} extension \citep{AlvarezMuniz:2011bs}. \aires and \corsika have been used to simulate air showers for the Pierre Auger Observatory and were compared by \cite{Knapp:2002vs}. Predictions for the electromagnetic energy deposit and the produced muon number agree within 2-3\,\%. \aires is faster than \corsika by a factor of 3 to 4, probably in part due to its simplified treatment of low-energy hadronic interactions.

\item
\conex offers fast simulation of air showers by solving the cascade equations numerically \citep{Bergmann:2006yz}. The computing time is independent of energy and less than a minute per event. \conex does not simulate the shower in four dimensions, only its longitudinal development along the shower axis. This limits the application since air shower experiments typically need full simulations, but \conex is ideal for studying the effect of hadronic interactions on longitudinal shower features, like the profile of electrons and photons and the depth \xmax of shower maximum. \conex was developed in parallel with \corsika and supports the most relevant subset of interaction models implemented in \corsika. It is further able to perform hybrid simulations where a full Monte-Carlo simulation of the first interactions is performed and the output is used to feed cascade equations. This allows one to simulate shower-to-shower fluctuations.

\item
\corsika is an air shower program \citep{Heck:1998vt} originally developed for the detector design and the physics interpretation of the KASCADE experiment \citep{Antoni:2003gd}. It has since been used by most air shower experiments in the last 30 years to simulate showers from cosmic rays and \pev gamma rays, owing its widespread use to a significant amount of continuously invested resources to make it the most complete, well documented, and comprehensive tool for air shower simulations. It was designed from the beginning as a tool open to the community for adaptation and improvement. \corsika supports the widest range of low-energy and high-energy hadronic interaction models. Electromagnetic cascades are simulated with the EGS4 code \citep{Nelson:1994zb} that was extended with the LPM effect. 
Particle decays are taken from \pythia{6} \citep{Sjostrand:2006za}.
Cherenkov photon emission is simulated with the Bernlöhr package \citep{Bernloehr:2008xd} and radio emissions are generated with \textsc{CoREAS} \citep{Huege:2006kd}.

\item
\corsika~8 is a modern reinvention of \corsika rewritten from scratch in C++ to allow for even more modularity and to support parallel HPC hardware and accelerators \citep{Engel:2018akg}. It is currently in development and not yet available for physics production. Building on the experience of the Fortran version, it is a community project from the start, open to all collaborations and individuals to contribute. The goal of the project is to build the most comprehensive distribution of models, tools, and techniques, to fill the simulation needs of current and future astroparticle experiments and to make new sophisticated forms of data analyses possible. Authors can replace almost every aspect of the simulation to try out new ideas while relying on the rest of \corsika~8 to work. Through this feature and others, \corsika~8 will make it feasible to investigate systematic uncertainties beyond the existing boundaries of other codes. \corsika~8 simulates electromagnetic interactions with the lepton propagator \proposal \citep{Koehne:2013gpa,Dunsch:2018nsc}. Currently implemented hadronic interaction models are \sibyll{2.3d} and \qgsjet{II.04}.

\item
\cosmos is an air shower program \citep{Kasahara:2007jwa} developed for the Akeno/AGASA experiments \citep{Chiba:1991nf,Takeda:2002at}. It supports the simulation of air showers up to the highest energies with thinning and the simulation of Cherenkov photon emission. It allows users to choose between the high-energy interaction models \qgsjet{} and \dpmjet{III} and low-energy interaction models \textsc{JQMD}, \textsc{JAM} \citep{Koi:2003iq}, and Bertini \citep{Heikkinen:2003sc}. Electromagnetic cascades are simulated with a custom code that includes the LPM effect. \cosmos and \corsika were compared by \cite{Roh:2013ev}. Some differences were found, but the number of muons was consistent within 5\,\%.

\item
\mceq is a program to numerically solve the cascade equations \citep{Fedynitch:2015zma}. It was designed as a fast option to calculate atmospheric lepton fluxes \citep{Fedynitch:2018cbl} for neutrino observatories and can also be used to simulate an average air shower, which requires only a few seconds on modern hardware in both cases. \mceq is completely written in Python and achieves this speed by using sparse matrices and optimised linear algebra libraries. It supports a wide range of interaction models, \sibyll{2.1} to 2.3d, \qgsjet{01c} to \qgsjet{II.04}, \eposlhc, \dpmjet{III~3.0-6} and \dpmjet{III~19.1}, and also provides a range of models of the cosmic-ray flux as input to the atmospheric lepton flux calculations. \mceq can use electromagnetic cross sections from the \emca code \citep{Meighen-Berger:2019cxt}. It is an ideal tool to study the systematic uncertainties that interaction models introduce in the simulation of atmospheric lepton fluxes and air showers \citep{Aartsen:2020fwb}. Like \conex, \mceq does not simulate air showers in four dimensions. Fluctuations and structures in individual events, for example, muon bundles, are not present in the solutions to the cascade equations.

\item
\seneca was a hybrid air shower simulation code \citep{Drescher:2002cr}. The first stages of the shower were fully simulated, then the output was fed into cascade equations, and finally the last stages of the shower were again fully simulated, by sampling particles from the energy distributions produced by the cascade equations. The goal was to greatly reduce the time needed to simulate EeV air showers. While \seneca is not further maintained, the principle ideas have been implemented also in \corsika since then.

\end{itemize}

\subsection{Hadronic interaction models}
\label{sec:generators}

An air shower is mainly driven by the outcomes of relativistic hadron-ion collisions with nitrogen and oxygen atoms under low momentum transfer in the non-perturbative regime of quantum chromodynamics (QCD). Since hadron production under these conditions cannot be calculated directly from first principles, effective theories and phenomenology are used. The codes that simulate hadron production are called generators or hadronic interaction models. They are the largest source of uncertainties in air shower simulations \citep{Engel:2011zzb,Pierog:2017awp}.

Specialised generators are developed for the simulations of extensive air showers (EAS) to address the needs of the astroparticle community, which differ from their pendants in the high-energy physics (HEP) or the heavy-ion collision (HIC) communities. The need to describe interactions at energies beyond the reach of colliders and to handle a variety of projectiles (nuclei, proton, charged pions and kaons) and targets (nitrogen, oxygen, argon) is specific to EAS generators. The production and decay of heavy-flavour (charm, beauty, top) and heavy bosons is important for HEP generators, but largely omitted in EAS generators. The predictive power of the generators is important \citep{dEnterria:2011kw}, since they are used to extrapolate to cms-energies that are one order of magnitude above the LHC and toward forward rapidities that are not well covered by LHC experiments.

The generators \qgsjet{} \citep{Ostapchenko:2005nj,Ostapchenko:2006vr,Ostapchenko:2010gt,Ostapchenko:2010vb} and \sibyll{} \citep{Ahn:2009wx,Riehn:2017mfm,Engel:2019dsg,Riehn:2019jet} are focused on air shower simulation. They have a limited set of parameters (of the order of tens of parameters) and only implement physics which are important for shower development. These generators correspondingly focus on a small data set for tuning. The generators \dpmjet{} \citep{Ranft:1994fd,Ranft:1997pu,Ranft:1999fy,Ranft:2002rj,Roesler:2000he,Fedynitch:2015kcn,Cerutti:2015lcn} and \epos \citep{Werner:2005jf,Pierog:2009zt,Pierog:2013ria} have a more general focus on minimum-bias \pp and heavy-ion collisions and can be used for EAS simulations. Their parameter sets are larger (of the order of 100 parameters), but the data sets used to constrain them are also larger. In principle, all minimum-bias measurements from collider or fixed-target experiments can be used for tuning and validation.

There is also an effort to bridge the divide between HEP and EAS generators from the HEP side. \pythia{} \citep{Sjostrand:2006za,Sjostrand:2007gs} is widely used at electron and proton colliders and recently expanded its focus toward heavy-ion collisions by adding the Argantyr model \citep{Bierlich:2018xfw}. There is also an interest to support EAS simulation. A technical obstacle is the \pythia{}'s initialisation scheme, which was not designed to switch collision energies and colliding particles frequently, but there are plans to mitigate this. Nevertheless, \pythia{} has been used for EAS simulation and compared to other EAS generators in a specialised study by \cite{dEnterria:2018kcz}.

\begin{table*}[t]
\caption{Comparison of the theoretical approaches in commonly used hadronic interaction models.}
\label{tab:generators}
\centering
\setlength{\tabcolsep}{3pt}
\begin{tabular}{p{3.2cm}p{2.9cm}p{2.5cm}p{2.5cm}p{2.6cm}p{2.3cm}}\hline
                & \dpmjet{III.19-1} & \eposlhc  & \qgsjet{II-04} & \sibyll{2.3d} & \pythia{8}
\\ \hline
Domain          & EAS, HEP   & EAS, HIC & EAS & EAS & HEP
\\[0.3em]
Theoretical basis & GRFT + \newline minijet & GRFT + \newline energy sharing & GRFT + \newline resummation & GRFT + \newline minijet & parton model
\\[1.5em]
Nuclear collisions & Glauber & extended \newline GRFT & extended GRFT & extended \newline superposition & Glauber via \newline Argantyr
\\[0.3em]
Pomeron            & soft+hard  & semi-hard & semi-hard & soft+hard & soft+hard
\\[0.3em]
Energy evolution \newline of parton densities & via $Q_0(s)$ cut &  parameterised  & Higher-order \newline Pomeron graphs & via $Q_0(s)$ cut & via $Q_0(s)$ cut
\\[0.3em]
Energy evolution \newline of elasticity & constant & falling & falling & constant & ---
\\[0.3em]
Parton distributions  & CT14  & custom   & custom  & GRV & various
\\[0.3em]
Non-diffractive \newline remnant & ---  &  multi-quark \newline exchange (low \newline to high mass) & one-quark \newline exchange \newline (low mass) & one-quark \newline exchange \newline (low mass) & low mass
\\[1.5em]
Diffractive \newline dissociation \newline (low mass) & 2-channel \newline eikonal  &  diffractive Pomeron & 3-channel eikonal & 2-channel eikonal & Pomeron \newline emission
\\[0.3em]
Diffractive \newline dissocation \newline (high mass) & cut enhanced \newline graphs &  Pomeron \newline exchange & cut enhanced \newline graphs & Pomeron \newline exchange & Pomeron \newline exchange
\\[0.3em]
String fragmentation (fitted data) & Lund (\ee) & area law (\ee) & custom (\pp) & Lund (\pp)  & Lund (\ee)
\\[1.5em]
Forward-central \newline correlation & weak & strong  & strong  & weak & strong
\\[1.5em]
Charm production   & pQCD \newline (incomplete) & --- & --- & parameterised \newline + intrinsic  & pQCD
\\[1.5em]
Collective effects   & string fusion & core-corona \newline (parameterised) & --- & --- & colour reconnection, rope hadronization, string shoving
\\ \hline
\end{tabular}
\end{table*}

The aforementioned generators are compared in \tb{generators} to give an overview of the physics implemented in these models. \pythia{} is included for reference. An explanation of the table content is given in the following.
\begin{itemize}
\item \emph{Gribov-Regge field theory.} To increase their predictive power in particular for the high energy extrapolation, all models used in shower simulation are based on the Gribov-Regge field theory (GRFT) from \cite{Gribov:1968fc}, which links the inelastic cross-section (which can be easily constrained by data) to the particle production using a unique amplitude for the Pomeron exchange. In parton models like \pythia{}, there is no such fundamental link. A Pomeron is a colour-neutral object that can be exchanged between partons, the collective term for quarks and gluons.

A flaw of classic GRFT is that energy conservation is taken into account only at the particle production level, but not for the calculation of the cross-section \citep{Drescher:2000ha}. This is solved only in \epos, where energy sharing is consistently used. This leads to a narrower multiplicity distribution, since events with multiple partonic interactions (MPI) are suppressed, in better agreement with data \citep{Pierog:2017awp} than the Poissonian distribution of MPI given by classical GRFT and the parton model in \pythia{}.

\item \emph{Nuclear collisions.} Generators need to support asymmetric nuclear collisions (projectile masses up to iron are relevant, $A \le 56$; air targets are nitrogen, oxygen, and argon, $A \le 40$). Different approaches are used to predict nuclear collisions from elemental nucleon collisions. The classical Glauber approach \citep{Glauber:1970jm,Miller:2007ri} treats it as independent binary-pair collisions. In the semi-superposition model \citep{Engel:1992vf}, a collision of a nucleus with $A$ nucleons on a nucleus with mass $B$ nucleons is treated as $A \times \pX{B}$ collisions, where each nucleons carries an equal fraction of the energy of the nucleus, while the \pX{B} cross-section is based on a Glauber calculation. The GRFT-motivated calculations consider the reduction the nucleon-nucleon cross-section with respect to \pp via higher-order Pomeron interactions \citep{Drescher:2000ha,Werner:2005jf,Ostapchenko:2010vb}.

\item \emph{Pomeron amplitude.} In GRFT models, the Pomeron amplitude defines the evolution of the model as a function of energy and impact parameter to a large extent. There are two different approaches. In the soft+hard approach, the amplitude is the sum of a purely soft and a purely hard Pomeron based on external parton distribution function (PDF). In the semi-hard approach, the Pomeron is the convolution of a soft and a hard component based on Dokshitzer Gribov Lipatov Altarelli Parisi (DGLAP) equations \citep{Dokshitzer:1977sg,Gribov:1972ri,Altarelli:1977zs}.

In the soft+hard approach, one Pomeron is always connected to valence quarks with a large momentum fraction. The other MPI come from gluons at small momentum fraction $x$. The elasticity in this approach (energy fraction carried by the most energetic particle) is constant with energy and the pseudorapidity distribution is narrower. LHC data suggests that it is too narrow already at the TeV scale \citep{Pierog:2017awp}.

In the semi-hard approach, the parameters of the soft component can be tuned to PDF data, which increases predictive power, and the Fock states that are used as initial partons always carry a significant momentum fraction $x$ of the projectile and target. This leads to a strong correlation between particle production at mid-rapidity due to MPI and beam energy loss which is linked to elasticity \citep{Ostapchenko:2016ytp}. It also leads to a higher average mass for each Pomeron, which broadens the pseudorapidity distribution in better agreement with LHC data.

The Pomeron model (hard or semi-hard) also determines how the non-linear effects (screening or saturation) of the cross-section is handled at high energy. For a purely hard Pomeron, the only parameter which can be adjusted is the minimum integration limit $Q_0$, which is parameterised as a function of energy to reproduce the inelastic cross-section. In the case of a semi-hard Pomeron, $Q_0$ is constant and the non-linearity is introduced via higher-order Pomeron interactions, which are treated explicitly with enhanced graphs in \qgsjet{II} and via an effective amplitude modified with a parametrisation in EPOS, which is obtained by tuning to cross-section and multiplicity data in various systems.

Finally, in the case of the soft+hard approach, there are no beam remnants since the valence quarks (and diquark) of the projectile are always connected to the valence quarks (and diquark) of the target like in \dpmjet{III} and \sibyll{}. In case of semi-hard Pomeron, the beam remnant is used to carry the remaining energy and the parton Fock states. In \qgsjet{II}, only one valence quark is exchanged and a low mass is given to the remnant for hadronization via string fragmentation. In case of \epos, there is no limitation for the number of exchanged quarks, so that high-mass remnants can be produced with more than two or three quarks. These require a special hadronization scheme (micro-canonical). String fragmentation and resonance decay are also implemented, depending on the mass and quark content of the remnant. This scheme is supported by low-energy data on strange baryon production \citep{Bleicher:2001nz,Liu:2002gw}.

\item \emph{Parton distribution functions.} The parton distribution functions (PDFs) are an important ingredient in parton shower models like \pythia{}. They are not a fundamental element of the GRFT models, which are based on the Pomeron amplitude. The leading-order PDFs can be interpreted as the momentum distributions of the partons inside the nucleon. At high momentum transfer, Pomerons can be connected to partons either using the soft+hard approach based on external PDFs and the standard mini-jet calculations of hard scatterings, or the semi-hard approach in which the PDFs are given by the Pomeron amplitude itself. \epos and \qgsjet{II} generate their own PDFs through the semi-hard approach, while \sibyll{2.3d} uses the GRV parametrisation \citep{Gluck:1994uf} as an external PDF. The models which target the HEP domain, \dpmjet{III} and especially \pythia{8}, give more weight to calculating hard scatterings and use external state-of-the-art PDF parametrisations. In the case of \dpmjet{III}, CT14 \citep{Dulat:2015mca} is used, while \pythia{8} allows one to use various leading order and next-to-leading order PDFs \citep{Kasemets:2010sg}, or a mix, where the former is used to simulate parton showers and the latter to simulate hard scatterings.

\item \emph{Diffractive dissociation.} Collisions with diffractice dissociation (which is often shortened to \emph{diffraction}) make up about 20\,\% of the inelastic cross-section at the TeV scale and are an important ingredient for the elasticity of the interaction, which in turn is an important parameter for air shower evolution. No quantum numbers are exchanged between the beam particles in these collisions, but momentum is transferred and new particles are produced. One refers to low-mass diffraction when only a few GeV are transferred. It is handled with \cite{Good:1960ba} theory using a 2- or 3-channel eikonal, depending on whether only one or two diffractive states are considered. The beam particles are simply excited and hadronised. In high-mass diffraction, the momentum transfer is large enough so that the transferred Pomeron can produce new particles or even a jet, leading to much lower elasticity. To remain consistent with the energy sharing scheme in \epos, all types of diffractive dissociation are computed with a special diffractive Pomeron which is added to the semi-hard Pomeron amplitude. The beam remnants from diffractive dissociation are usually hadronised by a simple resonance decay or via string fragmentation. The impact of remnant breakup on muon production was studied by \cite{Drescher:2007hc} and found to be potentially significant, but constant with shower energy.

\item \emph{Collective effects.} The final hadronization of the excited system is very important to get the correct multiplicity and hadron ratios, in particular in regard to the muon production in air showers \citep{Pierog:2006qv}. All GRFT models are based on the fact that a Pomeron has a cylindrical topology which, once it is cut, produces two strings carrying the colour field between the beam remnant and the partons from the jets. The strings are hadronised based on different schemes; Lund \citep{Sjostrand:1986hx} or Area Law \citep{Artru:1974hr,Drescher:2000ha} or some custom process. The scheme is less important than the data on which the parameters are tuned. \sibyll{} and \qgsjet{II} use \pp at various energies and consider only a subset of final-state particles for tuning. \dpmjet{III}, \pythia{}, and \epos use data from \ee collisions to tune the parameters based a larger list of particles. \epos uses also heavy-ion collisions for tuning. Particle multiplicities at high energies and central rapidities are strongly affected by MPI, see \citet{Bartalini:2017jkk} for a review. One of the reasons for some tension of \sibyll{} and \dpmjet{} with high-energy multiplicity distributions are their relatively simple, Poissonian MPI models. Similar models lie at the basis of \pythia{} \citep{Bartalini:2017jkk} that improved the description of multiplicity and $\avg \pt$ by introducing a new mechanism called colour reconnection \citep{Ortiz:2013yxa,Bierlich:2015rha}, a form of string-string interaction. Two other new string-string interaction mechanisms in \pythia{} are string shoving \citep{Bierlich:2017vhg} and rope hadronization \citep{Bierlich:2014xba,Bierlich:2015rha}. \epos keeps string fragmentation universal and uses the core-corona approach to describe high-energy \pp data \citep{Werner:2005jf,Pierog:2013ria}. A core of quark gluon plasma (QGP) droplets is formed in high density regions of the collision. The droplets are hadronised via a micro-canonical statistical decay that produces different particle ratios (more baryons and strangeness), while the corona is the low density region of the collision where strings are fragmented like in \ee collisions. With this scheme, \epos can reproduce all data from \pp to \PbPb and from SPS to LHC energy scales without introducing energy-depend parameters for the hadronization~\citep{Werner:2013tya}.

\item \emph{Charm production and decay.} Some models include charm production, which  (like heavy-flavour production in general) has no impact on conventional air showers and therefore on the Muon Puzzle, but it is important for the prediction of the prompt atmospheric neutrino flux, a major background for neutrino observatories. \dpmjet{} and \pythia{} produce charm based on a pQCD calculation and in fragmentation. In \sibyll{}, central charm production is based on a parametrisation. In addition, intrinsic charm in the nucleon PDF has been introduced to correctly reproduce forward charm production.

\end{itemize}

\subsection{Modern air shower detection}
\label{sec:air-shower-detection}

\begin{figure}[tb]
\centering
\includegraphics[width=\columnwidth]{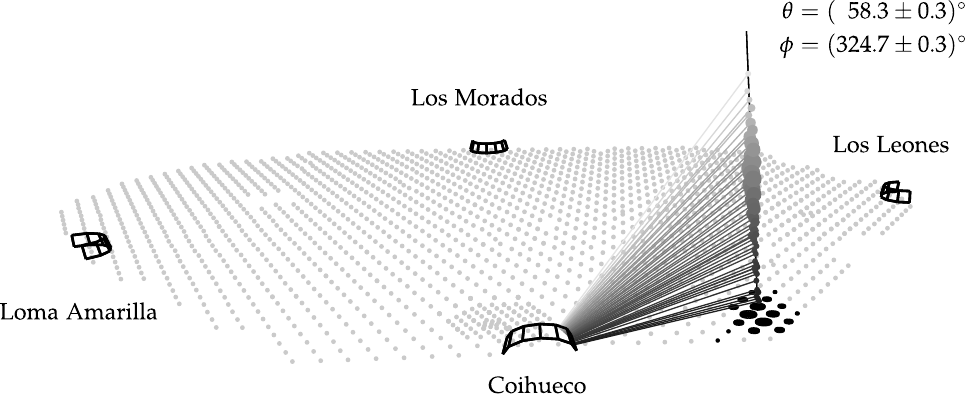}
\caption{Hybrid detection of an air shower illustrated with an event recorded by the Pierre Auger Observatory (from \cite{Aab:2014kda}). Dots represent particle detectors on the ground, the four named structures at the edges of the array are fluorescence telescope stations. The shower axis (solid line) with its light intensity profile (grey dots) is indicated, as well the footprint in the ground array (black cylinders). The sizes of the dots and cylinders indicate the logarithm of the local light and particle density, respectively.}
\label{fig:hybrid-detection}
\end{figure}

A cosmic-ray event is fully characterised by the energy $E$ and mass $A$ of the cosmic ray, and its arrival direction expressed by the zenith and azimuthal angles $(\theta,\phi)$, and its arrival location $(x, y)$ at ground, the shower core. The orientation and location of the cosmic ray directly translates into that of the shower axis. This part of the measurement is relatively straight-forward. The energy is indirectly observed via the number of particles produced in the atmosphere and the mass via the longitudinal evolution of the shower and the number of muons produced.

The latest generation of high-energy air shower experiments are the Pierre Auger Observatory \citep{ThePierreAuger:2015rma} and the Telescope Array \citep{Abbasi:2018fkz}. They use hybrid observations of the longitudinal development of the shower with fluorescence and Cherenkov telescopes and the particle footprint at the ground with particle detectors. An example of a hybrid detection is shown in \fg{hybrid-detection}. Hybrid observation of air showers at ultra-high energy was pivotal for the nearly model-independent observation of the discrepancy in the muon production in air showers.

Air shower observations with distributed detectors at ground were discovered by Kolhörster \citep{Kolhoerster:1938} and further exploited by Auger \citep{Auger:1939sh}. The advantage of ground detection is the duty cycle of nearly 100\,\%, the accurate measurement of the shower arrival direction, and the ability to measure the particle composition at ground, in particular the muon number \nmu. Muons that are measured with ground arrays have typical energies of 10 to 100\si{GeV} \citep{Dembinski:2009jc} and overwhelmingly originate from the last stages of the hadronic cascade. Only a single slice in the longitudinal evolution of the shower is observed. For that reason, measurements of the shower energy are usually fairly model-dependent for ground arrays. This can be compensated by building the ground array at high altitudes so that the slice is close to the shower maximum, which maximises the observable particle density and reduces the systematic uncertainty of the energy measurement.

The observation of showers via fluorescence telescopes was pioneered by the Fly's Eye experiment \citep{Baltrusaitis:1985mx}. It allows for a near-calorimetric measurement of the shower energy since most of the kinetic energy of an ultra-high energy cosmic ray is deposited into the atmosphere and a known fraction is converted into observable light \citep{Aab:2019cwj,Ave:2012ifa}. The observation of the depth of shower maximum \xmax is found to be a very good estimator for the cosmic-ray mass. However, the duty cycle is limited to about 15\,\%, since the detection requires dark nights. The combination of the two complementary techniques offers synergies that made high-precision measurements possible. Observing both \xmax and \nmu allows one to test hadronic interaction models used in air shower simulations, since both are sensitive to the cosmic-ray mass composition and if the implemented physics is correct, the mass composition inferred from \xmax and \nmu must be in agreement.

In summary, the gold standard to study hadronic interactions with air showers is the simultaneous and independent observation of the shower energy $E$, the depth of the shower maximum \xmax to determine the cosmic-ray mass, and the lateral muon density at the ground, $\rho_\mu(r)$, which can be integrated to give the muon number \nmu. The lateral density $\rho_\mu(r)$ depends on the energy $E$ and mass $A$, the zenith angle $\theta$, the shower age (which is the slant depth between the shower maximum and observation level) and the energy threshold $E_{\mu,\text{min}}$ of the detector for muons.

It is not necessary to perform all these measurements simultaneously with the same detector, however, if external measurements in the same energy range are available. The average logarithmic muon number $\avg\lnn$ depends on the average cosmic-ray mass composition quantified by \mlna which can be measured by another experiment in the same energy interval via \xmax observations. If this information is available, it is also possible to measure the muon discrepancy with a detector that can only independently measure the shower energy $E$ and the muon number \nmu, but not \xmax. It is even possible to measure the muon discrepancy with a detector that only measures muons, if both the cosmic-ray mass composition and the energy spectrum are taken from another experiment. We will briefly discuss these different ways in which the muon discrepancy was observed.

\begin{figure*}[tb]
\centering
\includegraphics[width=0.32\textwidth,trim=10 0 40 10,clip]{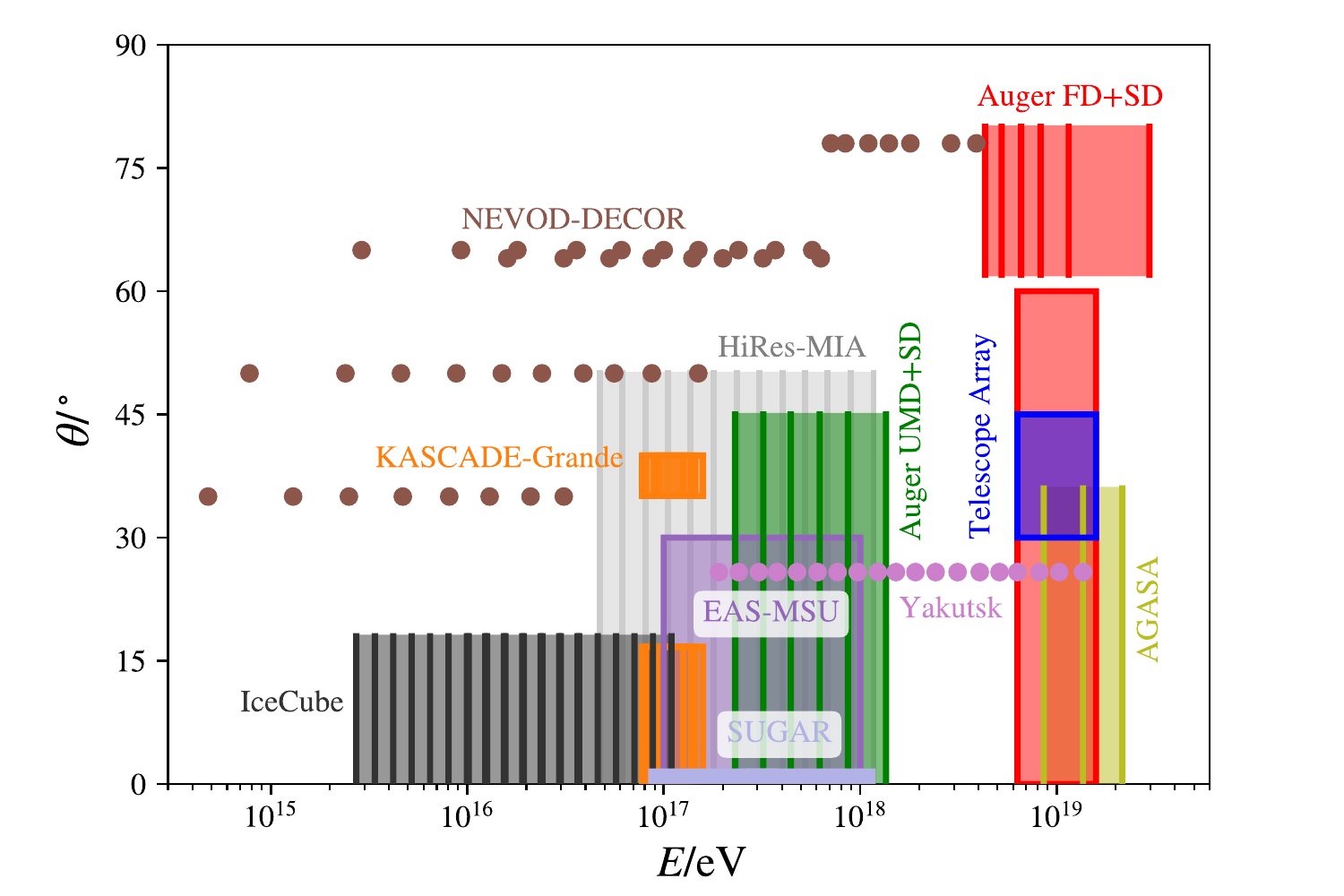}
\includegraphics[width=0.32\textwidth,trim=10 0 40 10,clip]{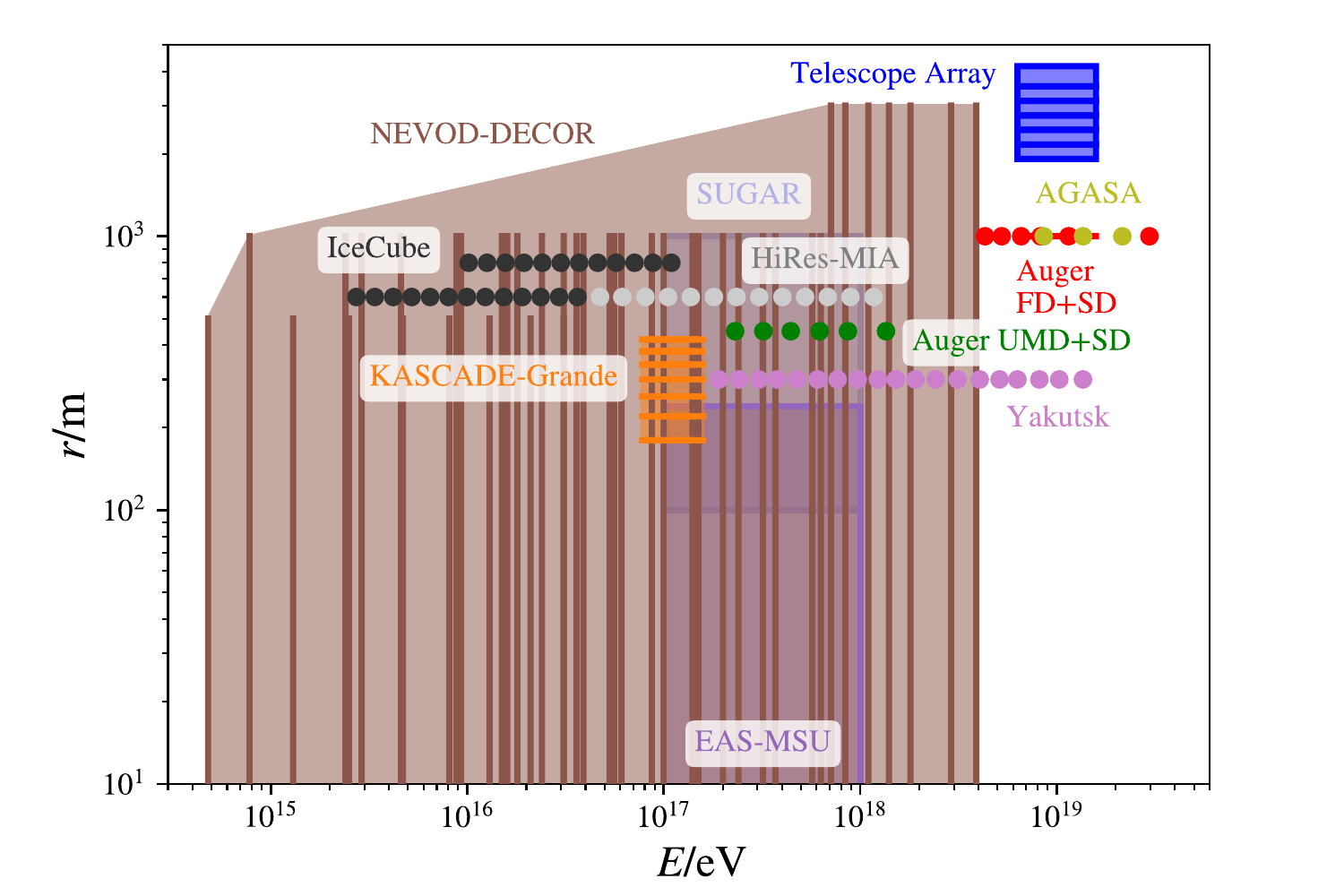}
\includegraphics[width=0.32\textwidth,trim=10 0 40 10,clip]{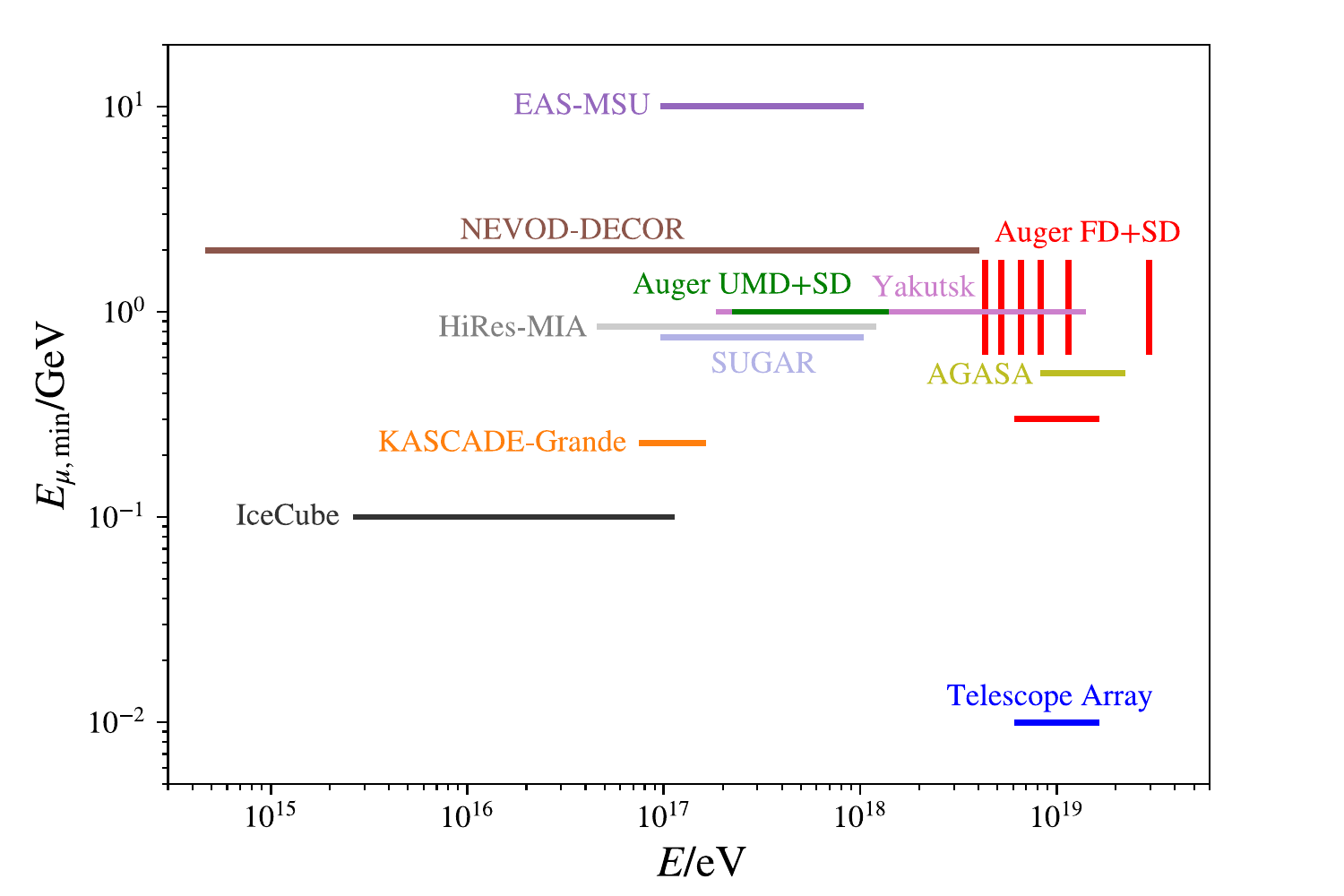}
\caption{Air shower experiments have measured the muon density at ground under various conditions, which are shown here (image from \cite{Soldin:2021wyv}). Points and lines indicate a measurement in a narrow bin of the parameter, while boxes indicate integration over a parameter range. \emph{Left:} Zenith angle of air showers versus shower energy. \emph{Centre:} Lateral distance of the muon density measurement versus shower energy. \emph{Right:} Energy threshold for the muons that are counted in the experiment. Some experiments measure muons below a shielding, which increases the muon energy threshold}
\label{fig:whisp_phase_space}
\end{figure*}

To increase the experimental significance and to uncover more information about the origin of the Muon Puzzle, it is important to combine muon measurements from many experiments. The phase space of air shower parameters covered by the experiments that were used in the meta analysis by the WHISP group (discussed in \sect{meta-analysis} and shown in \fg{whisp_phase_space}) are briefly discussed below.

\begin{itemize}
   \item The \emph{Pierre Auger Observatory} \citep{ThePierreAuger:2015rma} is a hybrid experiment comprising  1660 water-Cherenkov stations positioned on a 1.5\si{km} triangular grid covering an area of 3000\si{km^2} that is overlooked by 27 fluorescence telescopes located at four sites at the periphery of the array. The observatory has the largest collected exposure and the largest collaboration among air shower experiments. The water-Cherenkov stations are very sensitive to muons and, due to their height of 1.2\,m,  also offer acceptance for horizontal air showers. The first and least model-dependent muon measurement at the EeV scale was obtained with highly-inclined air showers \citep{Aab:2014pza} which are naturally muon-rich, followed by a measurement with vertical showers based on shower universality \citep{Aab:2016hkv}. The measurement with highly-inclined showers was recently updated and also the relative shower-to-shower fluctuations in the muon number were measured for the first time in any air shower experiment \citep{Aab:2021zfr}. Since the start of regular operation, a number of enhancements have been installed at the site. AMIGA is a denser infill array with 750\si{m} spacing and buried scintillator detectors which are able to measure the isolated muon component also in vertical showers \citep{PierreAugur:2016fvp}. The infill array is overlooked by high-elevation telescopes optimised for the study of low-energy showers. As part of the on-going AugerPrime upgrade \citep{Castellina:2019irv}, the water-Cherenkov stations are being upgraded with scintillators. The upgraded surface detector array will allow for a model-independent measurement of the muon content in vertical showers with a duty cycle of 100\,\%.

   \item \emph{Telescope Array} \citep{Fukushima:2003ig} is a hybrid experiment that consists of a 700\si{km^2} array of 507 scintillator detectors with 1.2\si{km} spacing, overlooked by three telescope stations. Since scintillators have no acceptance for horizontal air showers, stations at large lateral distance to the shower axis have been used to measure the muon density in which the muon purity reaches 70\,\% \citep{Abbasi:2018fkz}. Telescope Array has a low-energy extension called TALE, which has been used to measure the cosmic-ray flux and composition from 2\si{PeV} to 2\si{EeV} \citep{Abbasi:2018xsn,Abbasi:2021bpq}. A non-imaging optical array called NICHE \citep{Bergman:2020pjh} is currently build, and the main array is upgraded to TA*4 to cover a four times larger area to gain more acceptance for energies above 50\,EeV \citep{Kido:2018dhc}.

   \item \emph{Akeno Giant Air Shower Array (AGASA)} (\cite{Chiba:1991nf}, now decommissioned) was a 100\si{km^2} surface array of 111 scintillation and 27 muon detectors with a spacing about about 1\si{km}. The array measured the cosmic-ray flux from about 3\si{PeV} to about 30\si{EeV}. AGASA data on muons \citep{AGASA:1995kpd,Shinozaki:2004nh} has been re-analysed and compared with modern air shower simulations \citep{Gesualdi:2020ttc,Gesualdi:2021yay} for inclusion in the meta-analysis.

   \item The \emph{Yakutsk EAS Array} is a 8\si{km^2} array of 58 scintillator stations, 48 non-imaging Cherenkov light detectors, six large-area muon detectors, and two partially-imaging Cherenkov light detectors \citep{Anatoly:2013lya}. It has been continuously taking data since 1974 \citep{2018PAN....81..575G} and has published air shower data in the energy range from 1\si{PeV} up to 30\si{EeV}. The small size of the array (varying over time with a peak size of 12\si{km^2}) is somewhat compensated by an exceptionally long exposure time. The combination of data from the Cherenkov detectors, surface scintillators, and muon detectors allows for full hybrid observation of air showers. The array is located close to sea level (0.1\si{km} altitude) and measures air showers well past their maximum. Determining the shower energy independently of the muon content with a surface array at sea-level is challenging. The energy calibration was recently updated by \cite{2018PAN....81..575G} using a combination of shower simulations and data from all Yakutsk detectors. The cosmic-ray flux obtained from surface array data in this way is in agreement with the now excluded AGASA spectrum and incompatible with the flux that Yakutsk measured using only the Cherenkov detectors. This contradiction is unresolved. \cite{Glushkov:2019oft} present measurements of the muon density.

   \item The \emph{IceCube Neutrino Observatory} (IceCube) consists of a cubic-kilometre ice-Cherenkov detector comprised of over 5000 optical sensors in the deep Antarctic ice \citep{Aartsen:2016nxy} shielded by $1350\si{mwe}$ (vertical incidence), and a 1\si{km^2} surface array called IceTop that consists of 81 ice-Cherenkov detector stations with a spacing of 125\si{m} \citep{IceCube:2012nn}. IceCube does not observe air showers with telescopes, but it has the unique capability to simultaneously measure the shower particles at the surface (electrons, photons, and muons) and high-energy ($> 300\si{GeV}$) muon bundles in the deep detector which are sensitive to the first interaction of the cosmic ray and offer unique insights into the hadronic physics of this interaction. The IceTop detector measures air showers at an altitude of 2.8\si{km} above sea level close to the shower maximum, which allows one to infer the shower energy with low model-dependence \citep{Aartsen:2013wda, IceCube:2019hmk}. The muon density at the surface is measured with a statistical technique that distinguishes muon hits far from the shower axis by their characteristic constant signal, while electrons and photons generate a continuum of signals that can be separated and subtracted \citep{Dembinski:2017zkb,Gonzalez:2018IceTop,Gonzalez:2019epd}.

   \item The \emph{KASCADE-Grande} experiment (\cite{Apel:2010zz}, now decommissioned) was a 0.49\si{km^2} array with 252 scintillator stations with 13\si{m} spacing (KASCADE) and additional 37 stations with an average spacing of 137\si{m} (Grande). The denser array contained both unshielded and shielded detectors to separately measure show electrons and muons. The array further contained several muon tracking detectors. The experiment was located close to sea level (0.1\si{km} altitude) and measured air showers well past their maximum. Like in case of Yakutsk, it is challenging to determine the shower energy independently of the muon content at sea-level. A summary of muon results is given by \cite{Apel:2017thr}. The data of the KASCADE-Grande experiment were released to the public \citep{Haungs:2018xpw} and analyses are still forthcoming.

   \item The \emph{EAS-MSU Array} (\cite{Fomin:2016sud}, now decommissioned) was a 0.5\si{km^2} array in its latest configuration with 76 unshielded charged-particle detector stations and additional underground stations which measured atmospheric muons under 40\si{m} water-equivalent with a threshold energy of 10\si{GeV} for vertical incidence. The array is similar in size and capabilities to the KASCADE-Grande array and also close to sea-level (0.15\si{km} altitude). It faced the same challenges in determining the cosmic-ray energy independently from the muon content the shower. Muon data from the array was recently re-analysed by \cite{Fomin:2016kul}.

   \item The \emph{SUGAR Array} (\cite{1968CaJPS..46..259B}, now decommissioned) was a 70\si{km^2} array of 54 muon detectors placed at a depth of $(1.5 \pm 0.3)\si{m}$ with a energy threshold of about 0.75\si{GeV} for vertical muons at an altitude of about 250\si{m} above sea level. The cosmic-ray flux in this experiment was measured by using the muon number \nmu as an energy estimator, using a relationship between energy $E$ and \nmu predicted by air shower simulations. If the simulations suffer from a muon deficit, then the inferred cosmic-ray flux turns to be too high. \cite{Bellido:2018toz} re-analysed the data in this regard by comparing the measured flux with a simulation based on the cosmic-ray flux measured near-calorimetrically by the Pierre Auger Observatory, which is not affected by the muon deficit, and two extreme mass composition assumptions of pure proton and pure iron showers. By attributing the discrepancy between measured and simulated flux solely to the muon deficit in simulations, the muon deficit was calculated.

   \item The \emph{NEVOD-DECOR} experiment \citep{Barbashina:2000bs,Petrukhin:2015kla} is a single large $2000\,\text{m}^3$ water-Cherenkov detector (NEVOD) with a 70\si{m^2} charged-particle detector array on its top (DECOR). The setup measures muon bundles produced by air showers. The density of the bundles and the arrival direction is recorded. Pure muon showers are selected with a cut on highly inclined events with $\theta > 55^\circ$. The shower information collected by the \emph{NEVOD-DECOR} experiment is even further reduced than in case of the SUGAR array. Neither the shower energy nor the shower core location are known event-by-event, since the detector is point-like compared to the lateral shower extension. A muon deficit in simulations has nevertheless been detected in a similar way as in case of the SUGAR array \citep{Bogdanov:2010zz,Bogdanov:2018sfw}. Expected density spectra of muon bundles were simulated based on a model of the cosmic-ray flux fitted to selected world data, which closely follows fluxes that have been measured near-calorimetrically, and two extreme mass composition assumptions of pure proton and pure iron showers. The muon deficit is then inferred from the discrepancy between the simulated and observed muon density spectra. Assigning a shower energy to the observed discrepancy is a challenge, since the measured local muon density spectra are the product of an integral over a wide range of shower energies.
\end{itemize}

In addition to these experiments, other air shower experiments have performed muon measurements or are currently taking data. The HiRes-MIA collaboration performed the first combined measurement of the muon content of air showers together with optical telescopes \citep{AbuZayyad:1999xa}, as a precursor to the Pierre Auger Observatory and Telescope Array. Haverah Park experiment has published a comprehensive series of muon measurements in air showers \citep{Armitage:1987wq,1997APh.....6..263C,Blake:1995ft,Blake:1995kh,Blake:1998fn,Blake:2000bx} which were used to estimate the cosmic-ray mass composition \citep{Ave:2002gc} and put limits on the ultra-high photon flux \citep{Ave:2001xn}. Data from Haverah Park and HiRes-MIA were not included in the meta-analysis, since the data were not re-analysized with modern air shower simulation codes. The GRAPES-3 air shower experiment \citep{GRAPES-3:2005tvd} has a large-area muon tracking detector similar to the KASCADE experiment, but not yet published muon measurements.

\subsection{Observation of the muon discrepancy in air showers}
\label{sec:meta-analysis}

The report by \cite{Aab:2014pza} about a muon deficit in simulations compared to measurements from the Pierre Auger Observatory sparked renewed interest in the muon discrepancy and were followed by new muon measurements and the re-analysis of previously collected data from the EAS-MSU Array \citep{Fomin:2016kul}, the IceCube Neutrino Observatory \citep{Gonzalez:2019epd}, the KASCADE-Grande experiment~\cite{Apel:2017thr}, the NEVOD-DECOR detector~\cite{Bogdanov:2018sfw}, the SUGAR array~\cite{Bellido:2018toz}, Telescope Array~\cite{Abbasi:2018fkz}, and the Yakutsk array~\cite{Glushkov:2019oft}. The Pierre Auger Observatory also followed up with independent measurements using vertical showers, first using shower universality \citep{Aab:2016hkv} and then using direct muon measurements at lower energies with the AMIGA sub-array \citep{Muller:2019dvf,Snchez:2020qeg}. The Pierre Auger Observatory was not the first to report a discrepancy in the number of muons, but it offered the first nearly model-independent measurements with well-controlled systematic effects in comparison with post-LHC models. Discrepancies in the muon number had been reported before by the HiRes/MIA \citep{AbuZayyad:1999xa} and NEVOD-DECOR experiments \citep{Bogdanov:2010zz}, but not by all experiments.  The AGASA experiment, for example, did not report a muon discrepancy \citep{Shinozaki:2004nh}.

The newfound wealth of data created the necessity for a meta-analysis of muon measurements. This led to the foundation of the Working group for Hadronic Interactions and Shower Physics (WHISP) formed by members of the aforementioned experiments (no contact could not be established for the HiRes/MIA experiment) with the goal to develop a common framework to compare all measurements, since a direct comparison is usually not possible. The muon density at the ground depends on many parameters which differ from experiment to experiment:
\begin{itemize}
  \item Cosmic-ray energy $E$,
  \item Zenith angle $\theta$,
  \item Shower age (depends on altitude of the experiment, local atmosphere, and zenith angle of the shower),
  \item Lateral distance $r$ from shower axis,
  \item Energy threshold $E_{\mu,\text{min}}$ of the detectors for muons.
\end{itemize}
The WHISP introduced the abstract muon scale parameter $z$ \citep{Dembinski:2019uta,Cazon:2020zhx,Soldin:2021wyv,Gesualdi:2021ndz} defined as
\beq
z = \frac{\ln \avg{\nmu} - \ln \avg{\nmu}_p}{\ln \avg{\nmu}_\fe - \ln \avg{\nmu}_p},
\label{eq:z-whisp}
\eeq
which can be computed from the data of each experiment,
where $\avg{\nmu}$ is the muon number or anything proportional to it averaged over showers in a narrow shower energy interval, and $\avg{\nmu}_p$ and $\avg{\nmu}_\fe$ are the corresponding values obtained from simulated air showers which undergo a full detector simulation and the same analysis as the real events. This definition cancels potential biases and is insensitive to a mismodelling of the \nmu resolution in the experiment \citep{Dembinski:2017kpa}. The natural range of $z$ in absence of a muon discrepancy is $0 < z < 1$, since proton and iron showers limit the range of observed cosmic-ray masses in practice.

By construction, $z$ depends on the hadronic interaction model used in air shower simulations and therefore different values are obtained for each model. The simulations account for differences in the experimental conditions. This approach is feasible since simulations mainly differ by a global offset in the number of muons and less in other aspects like the lateral density profile or the zenith angle dependence, as shown in several studies, see e.g.~ \cite{Dembinski:2009jc, Aab:2014gua}.

Another important step was to cross-calibrate the energy scales of the participating experiments. Since the muon number \nmu scales almost linearly with shower energy $E$ as described in \sect{heitler}, the measured muon number needs to be compared in \eq{z-whisp} with showers simulated at the exact same energy. Unfortunately, the overall calibration of the energy-scale of each experiment is only known with an accuracy of 10-20\,\%. A 20\,\% energy offset between two experiments translates to a shift of about 0.5 in $z$, half the difference between proton and iron showers. These energy-scale offsets are well-known to affect the cosmic-ray flux measured by different experiments, leading to shifts in the flux that are consistent within the uncertainties of the respective energy scales, see e.g.~\cite{ Hoerandel:2002yg,Berezinsky:2002nc,Berezinsky:2005cq,Dembinski:2017zsh,AbuZayyad:2019tfz}. The WHISP removed the energy-scale offsets in $z$ based on the observed offsets in the measured cosmic-ray fluxes, assuming that the flux offsets are caused entirely by shifts in the respective energy scales. This lead to a remarkable reduction in variance in $z$.

\begin{figure*}[tb]
\centering
\includegraphics[width=0.49\textwidth]{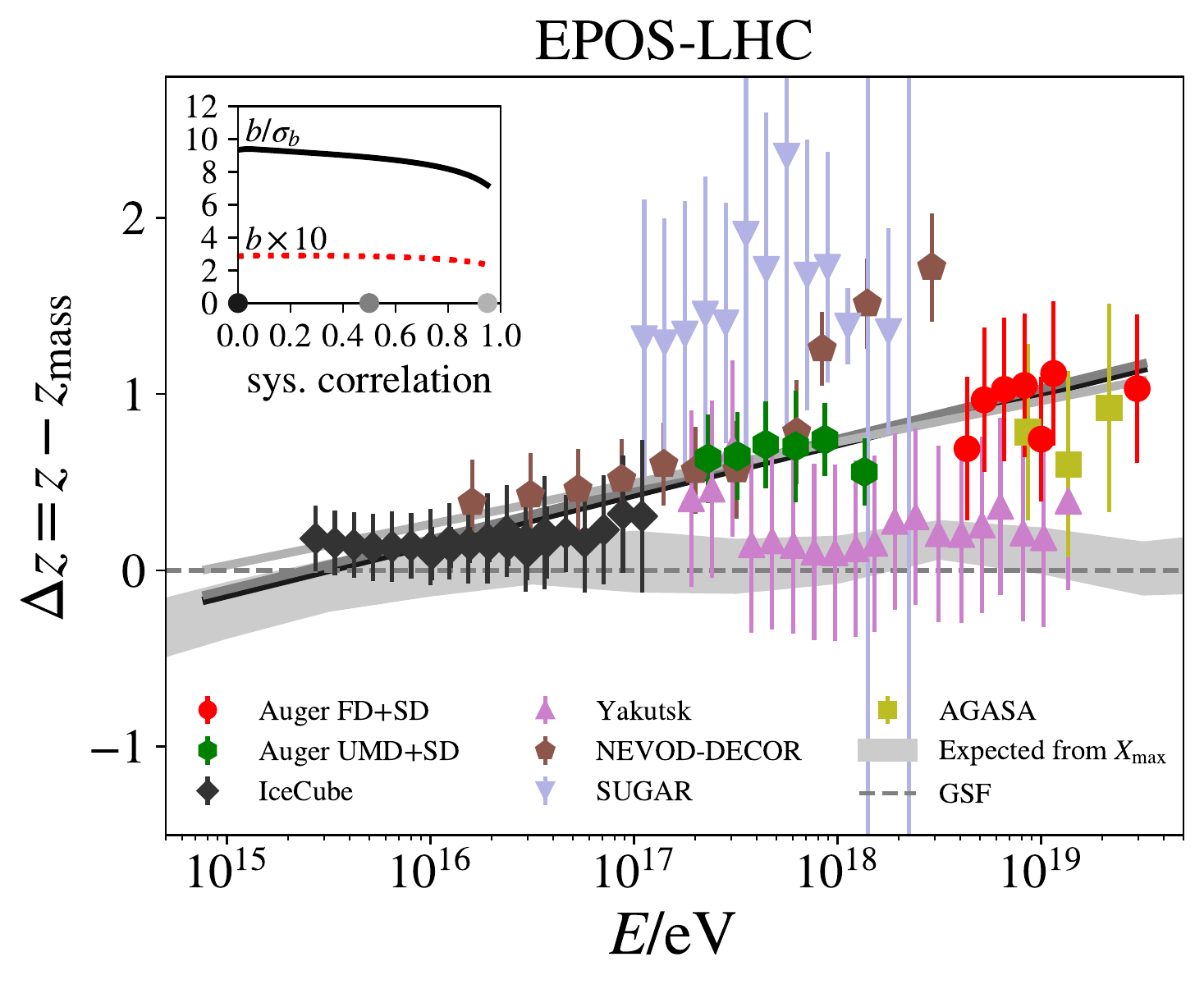}
\includegraphics[width=0.49\textwidth]{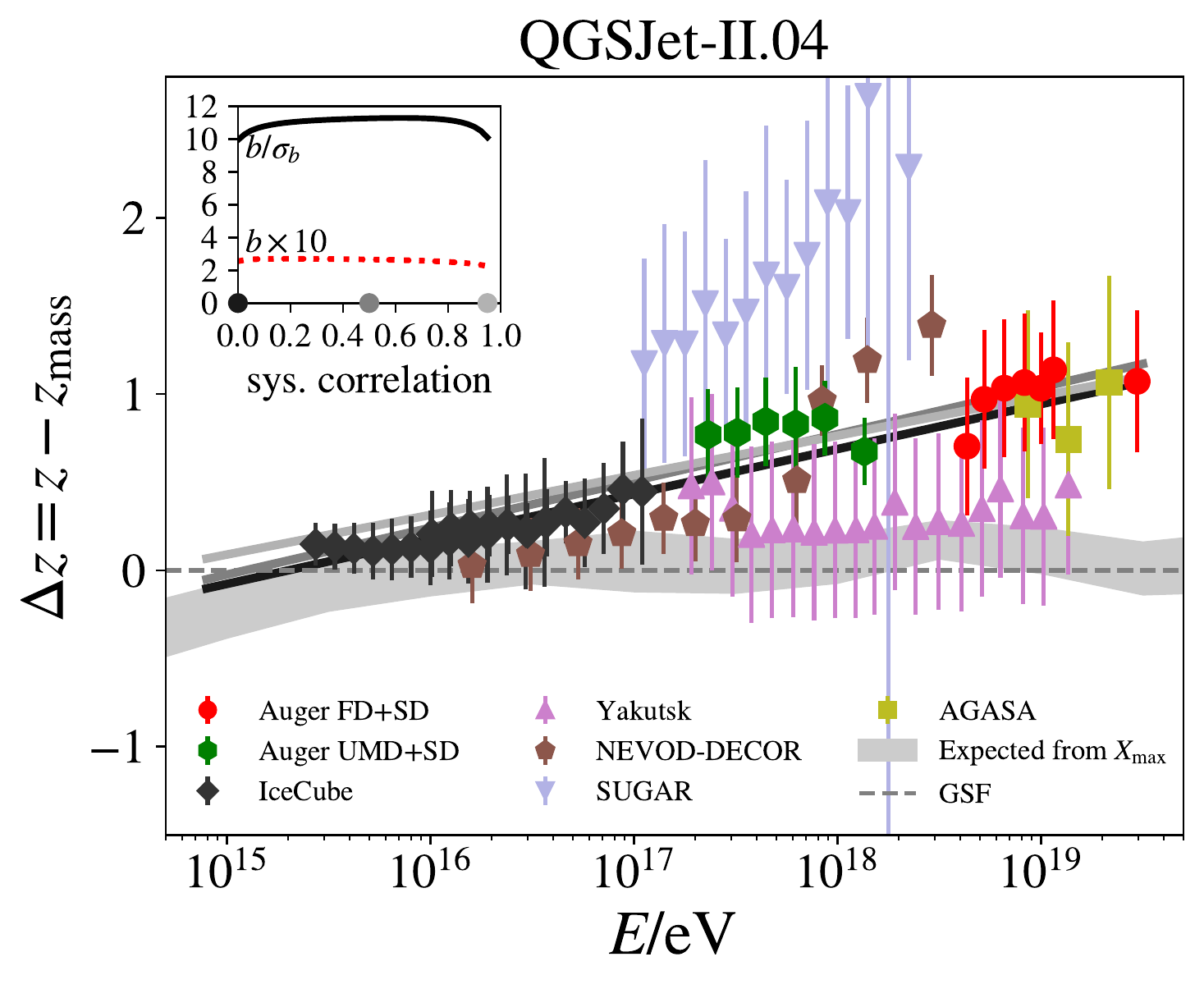}
\caption{Data from \fg{z_rescaled} after subtracting $z_\text{mass}$, the expected variation in $z$ due to changes in the cosmic-ray composition (image from \cite{Soldin:2021wyv}). The expected variations are taken from the GSF model (dashed line), which in turn is derived primarily from \xmax measurements (grey band). Solid lines represent fits that assume different levels of correlated experimental uncertainties. The deviation of the slope $b$ from zero in standard deviations is shown in the inset as function of the assumed correlation. Also shown in the inset is the value of the slope, scaled by a factor of 10.}
\label{fig:muon_offset}
\end{figure*}

The cross-calibrated $z$-values show an upward trend in $z$ with increasing shower energy. The post-LHC models \eposlhc, \qgsjet{II.04}, and \sibyll{2.3} give a better description of air shower data than the pre-LHC models, but the trend is observed for all models. A modulation in $z$ is observed, which follows that of $z_\text{mass}$, the expected value of $z$ for a cosmic-ray mass composition that agrees with the independent $\xmax$ measurements. After subtracting this expected modulation, the difference $\Delta z = z - z_\text{mass}$ shows an approximately linear increase with shower energy $E$, which is shown in \fg{muon_offset}.

The quantity $\Delta z$ can be interpreted as the relative muon deficit in units of proton-iron difference. The value $\Delta z \approx 1$ at 20\si{EeV} corresponds to a muon deficit in simulations of about 40\,\%. However, since the cross-calibration of energy-scales only removes relative offsets, a residual global offset of the energy cannot be excluded, which is roughly estimated at the level of 10\,\% \citep{Dembinski:2019uta}. It follows that all points in \fg{muon_offset} can be moved up and down by about $\pm 0.25$.

A fit of a line model $\Delta z = a + b \log_{10}(E / \text{eV})$ yields a slope $b$, which deviates positively from zero with 8 standard deviations or more. The slope is independent of the previously mentioned global shifts. Its fitted value depends only weakly on the model and the assumed correlation coefficient for the reported experimental uncertainties. The latter are known to be largely positively correlated, but the correlation coefficients for each experiment are not generally known. Therefore, a scan is performed over all values of the correlation coefficients from zero to one. To account for residual discrepancies in the data, the raw covariance matrix of the fitted parameters in this analysis is scaled by the $\chi^2$ of the fit divided by the degrees of freedom $n_\text{dof}$, see \cite{Dembinski:2019uta} for details. This is procedure effectively scales the uncertainties of all data points with a common factor so that $\chi^2 = n_\text{dof}$, the same technique is used in similar cases by the Particle Data Group \cite{ParticleDataGroup:2020ssz}. Data points which

The result of the meta-analysis is remarkable in several ways. It is the first unified statement of this kind from nine air shower experiments and it demonstrates that results are fairly consistent if the well-known systematic effect of the energy-scale uncertainties is removed. We list the main conclusions:
\begin{itemize}
\item The origin of the muon discrepancy can be studied at the LHC. The deviation starts around $40\si{PeV}$, which corresponds to a cms-energy $\sqrtsnn \approx 8\tev$ within the reach of the LHC. Below this threshold data and simulations are consistent within uncertainties.
\item The previous point implies that late shower stages in which the hadron energy is below 40\si{PeV} are sufficiently well described by simulations. The origin of the muon puzzle therefore is likely to be found in the first stages of the shower, not in the late stages.
\item The relative muon deficit increases approximately linearly with $\ln\! E$ above this threshold. The fact that the number of shower stages above this energy also increases linearly with $\ln\! E$ points toward a compounding effect, as described in \sect{heitler}.
\end{itemize}
The observed behaviour makes a non-exotic explanation more likely in which a comparably small change in the hadronic interactions causes large changes in \nmu over several shower stages, see \sect{heitler}, \sect{impact}, and \sect{solutions} for more details.

Although the data taken as a whole shows a muon discrepancy with high significance, one can distinguish two groups of experiments. The experiments which either measure the shower energy near-calorimetrically (Auger, AMIGA, IceCube) or do not measure the shower energy at all (NEVOD-DECOR, SUGAR) show a strong deficit, while the experiments which use an energy estimator that is correlated to the muon number (KASCADE-Grande, Yakutsk, EAS-MSU) show a much weaker deficit. This is still under study, but the discrepancy could be caused by this correlation. In case of AGASA, no muon discrepancy was originally reported, since the original energy scale of the experiment was shifted, but it appeared after the energy-scale offset was adjusted. In regard to the Yakutsk data, there is an unresolved discrepancy between the Cherenkov and surface detector measurements, which may affect these results.

\subsection{Results connected to the Muon Puzzle}
\label{sec:connected-results}

The Muon Puzzle specifically refers to a deficit in GeV muons that are produced near the end of the hadronic cascade in a simulated air shower. This is not the only discrepancy between muon measurements and simulations, however. We list here other measurements that found discrepancies and results potentially connected to the Muon Puzzle.
\begin{itemize}
\item \emph{Fluctuation of the muon number.} The shower-to-shower fluctuations of the muon number \nmu are sensitive to the first interactions in the hadronic cascade and to the cosmic-ray mass composition \citep{10.1143/PTPS.16.1,Cazon:2018gww}. They provide important evidence toward the source of the muon deficit in simulations. The fluctuations have been measured recently by the Pierre Auger Observatory \citep{Aab:2021zfr} for the first time and reasonable agreement was found between the measurement and the post-LHC models \eposlhc, \qgsjet{II.04}, and \sibyll{2.3d}. This limits exotic explanations of the Muon Puzzle, in which an extreme change to the physics of the first interaction is proposed, followed by ordinary QCD for the rest of the shower development. For example, the Chiral Symmetry Restauration toy model by \cite{Farrar:2013sfa} would drastically reduce the muon number fluctuations. A very large sample of \nmu-measurements could be used to measure the \piz production cross-section  \citep{Cazon:2020jla}.

\item \emph{Muon production depth.}  The longitudinal muon density profile cannot be observed optically, since the small muon signal is overwhelmed by the electron signal, the related profile of muon production depths has been observed based on the signal arrival times in the surface detectors of the Pierre Auger Observatory \citep{Aab:2014dua} and via the muon tracking detectors of the KASCADE-Grande experiment \citep{Apel:2011zz}. The profile contains sensitive information about the hadronic cascade and about the cosmic-ray mass composition. The muon production depth is correlated to the depth \xmax of the electromagnetic shower maximum, but they are not identical and arise from different shower physics. At the EeV scale, \qgsjet{II.04} describes both quantities consistently, while \eposlhc does not. Deviations were also found at the PeV scale in comparison to simulations with the older model \qgsjet{II.02}.

\item \emph{Muon attenuation with zenith angle and mass overburden.}
Several experiments have reported deviations between simulated and measured the number of muons as a function of the zenith angle and mass overburden. KASCADE-Grande expressed the measurement in form of the effective attenuation length for muons in air showers \citep{Apel:2017thr}. The measurement is independent of assumptions regarding the cosmic-ray flux, since the cosmic-ray flux is isotropic and therefore the intensity is known in each zenith angle interval. Based on this, the attenuation of the muon flux can be computed. Simulations with the models \qgsjet{II.02}, \sibyll{2.1}, \qgsjet{II.04}, and \eposlhc consistently show smaller attenuation lengths than the experimental result, although the deviations is weaker for the two post-LHC models. In other words, the muon number in simulations decreases more rapidly with zenith angle than in the measurement, which could indicate that the muon energy spectrum in simulations is steeper than in data. The measured attenuation also shows a dependence on the lateral distance to the shower axis, which is not reproduced correctly by current air shower simulations.

Some underground experiments observed the opposite effect for TeV muons that originate primarily from the first interaction in contrast to GeV muons in air shower experiments like KASCADE-Grande. The Fréjus and AMANDA experiments have measured the muon rate as a function of the mass overburden which is computed from the zenith angle \citep{2001ICRC....3..985D,Berger:1989hs}. Simulations were provided by \cite[chapter 5]{rhode_habil} and \cite[chapter 6]{schroeder_diss}. A muon deficit is seen in simulated showers with vertical incidence which either disappears or turns into an excess at larger zenith angles with large mass overburden, depending on the simulation code. A re-interpretation of these old measurements would require new simulations with recent models of the cosmic-ray flux and composition and hadronic interaction models.

\item \emph{High-energy atmospheric muon flux.} Only muons with energies above 300\si{GeV} penetrate the Antarctic ice and reach the deep-ice detector of the IceCube Neutrino Observatory. Shower muons arrive in bundles that appear like a single track. Although individual muons cannot be resolved, the experiment can distinguish between bundles with and without high-energy muons based on the presence of stochastic energy losses from bremsstrahlung. IceCube has measured the high-energy muon flux up to PeV energies in this way \citep{Aartsen:2015nss} which provides evidence for a prompt muon flux in air showers. This flux mainly originates from decays of charmed and unflavoured hadrons \citep{Fedynitch:2018cbl}. While the conventional high-energy muon flux from decays of light hadrons above 10\tev is well reproduced by \sibyll{2.1}, discrepancies between simulations and experimental data are found in the zenith angle distribution of muons that are unresolved \citep{Soldin:2018vak}.

\item \emph{Simultaneous measurements of GeV and TeV muons.} The IceCube Neutrino Observatory is capable of simultaneously measuring an air shower in the deep-ice detector and in the IceTop array and to study the low-energy (GeV) and high-energy (TeV) muon component event-by-event and their correlation. The measurement of muons at two vastly different energies provides information about the energy sharing between low- (late) and high-energy (early) interactions during the shower development and is of particular interest in regard to the Muon Puzzle. Preliminary studies by \cite{DeRidder:2017alk} indicate that the yield of low- and high-energy muons differs among hadronic interaction models. A consistent interpretation of the experimental data is obtained for \sibyll{2.3} and \qgsjet{II.04}, while \sibyll{2.1} and \eposlhc show discrepancies. As described by \cite{Engel:2019dsg}, these measurements can be used to constrain hadronic interaction models and provide unique tests of muon production models in EAS.

\item \emph{Lateral separation of TeV muons.} The IceCube Neutrino Observatory observes events with well-separated pairs of near-parallel tracks in the deep-ice detector with lateral distances between 135\si{m} to 400\si{m} \citep{Abbasi:2012kza}. Simulations have shown that these events are dominantly caused by the production and decay of hadrons with large transverse momentum $\pt$ in the first interaction of an air shower \citep{Soldin:2018vak}. The observed distribution of lateral separations (LS distribution) can be related in a model-dependent way to the $\pt$ distribution of the decaying mesons. The momentum threshold $p_\text{T,min}\simeq 2\,\gevc$ for the observed mesons is large, which means that their production rates can be calculated in perturbative QCD and since the $\pt$ distribution is also sensitive to the mass of the cosmic ray, it may offer an alternative way to measure the mass composition of cosmic rays with independent systematic uncertainties \citep{KLEIN2008346, Gerhardt2009, Soldin:2014ouk, Soldin:2016GH}. The LS distribution for PeV showers are compatible with predictions by \sibyll{2.1}, while \qgsjet{II.04} and \eposlhc show deviations. \sibyll{2.1} does not describe the zenith-dependency of the LS distribution, however. The tensions between experimental data and simulations are not yet understood. Underground detectors like MACRO \citep{Ambrosio:1999qu} and Fréjus \citep{Berger:1989hs} have previously measured the LS distribution at greater depths, but these measurements have not been studied in the context of modern hadronic interaction models.

\item \emph{Seasonal variations of atmospheric muon and neutrino fluxes.} It has been proposed long ago to use the correlation between the variation in atmospheric temperature and muon flux as a probe of hadronic interactions \citep{Barrett:1952woo}. Seasonal temperature variations cause density variations that modify the mean-free path for mesons in the atmosphere. If the density is lower, mesons are more likely to decay than to hit another target and produce more mesons, which reduces the muon flux. Measurements of the flux variations yield important information about the dynamics in the central stratosphere, such as ozone hole dynamics and the temporal behaviour of the stratospheric temperatures \citep{Tilav:2019xmf}. Furthermore, the size of this effect depends on the life-time of the meson. Charged pions have the largest life-times and are more affected by density variations than kaons. This can be exploited to infer the $K/\pi$-ratio from the size of the flux variation for a given temperature variation \citep{Grashorn:2009ey,Desiati:2010wt}. Model-dependent measurements of the $K/\pi$-ratio via variations in the muon flux have been performed by MINOS \citep{Adamson:2009zf,Adamson:2014xga}, MACRO \citep{Ambrosio:1997tc}, and IceCube \citep{Tilav:2010hj,Desiati:2011hea}. The flux-based measurements are compatible with direct measurements of both $pp$ and in heavy-ion collisions, due to the comparably large uncertainties.

Another approach to measure the $K/\pi$ ratio is the study of the electron neutrino and muon neutrino flux as a function of energy and zenith angle \citep{Fedynitch:2019bbp}. At energies below $\sim 80\gev$ (for vertical incidence), muon neutrinos follow the flux of parent pions, while at higher energies that of charged kaons. Low-energy electron neutrinos come from muon decays and at higher energy from charged and neutral kaon decays. The features of the observed neutrino spectrum and angular distribution provide constraints on the $K/\pi$-ratio. IceCube recently reached sufficient exposure to also observe seasonal variations in the atmospheric neutrino flux \citep{Heix:2019jib} that provide an additional probe of the $K/\pi$-ratio.

\begin{figure}[tb]
\centering
\includegraphics[width=\columnwidth]{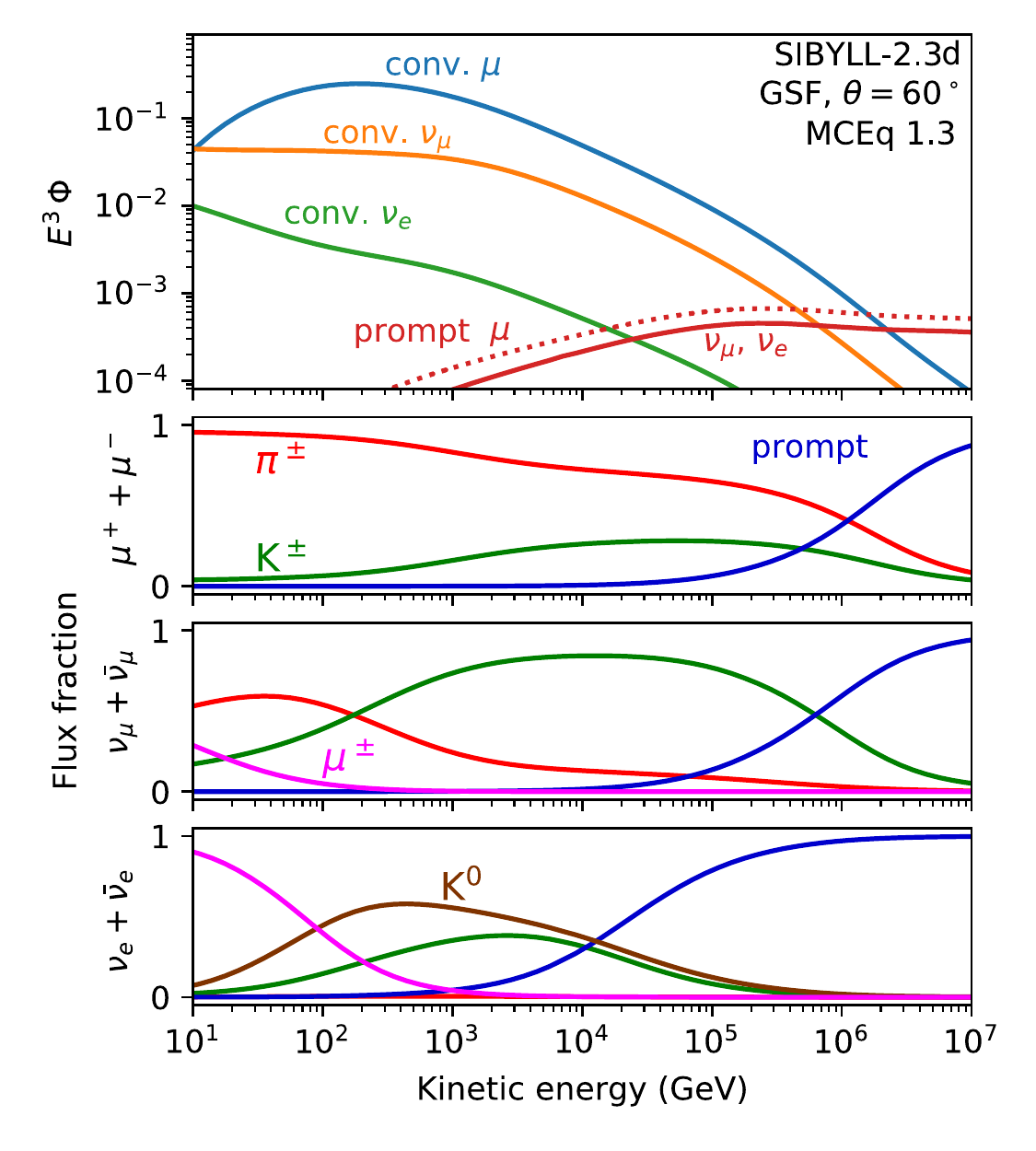}
\caption{Top panel: Conventional and prompt lepton fluxes in $\text{GeV}^{2}/(\text{cm}^2\, \text{s}\, \text{sr})$ and fractional contributions from parent mesons for $\theta = 60^\circ$. The calculation uses \mceq, the \sibyll{2.3d} hadronic interaction model and the GSF cosmic-ray flux model as described in the text. Rare decays of unflavoured vector mesons and electromagnetic pairs contribute to the difference between prompt muons (dotted line) and neutrinos (solid line). Lower panels: flux fraction by the type of lepton ancestor. Prompt leptons originate predominantly from decays of charm mesons.}
\label{fig:inclusive-flux}
\end{figure}

\item \emph{Inclusive muon and neutrino fluxes}. Muons and atmospheric neutrinos are closely related since they are produced in decays of the same ancestors \citep{Gaisser:2002jj}, but the relative contributions of ancestor particles to the different lepton fluxes vary, as shown in \fg{inclusive-flux}. A muon deficit in simulated air showers is therefore expected to be connected to a corresponding deficit in the conventional atmospheric lepton flux, but the correspondence is not trivial. The atmospheric lepton fluxes are integrals over the contributions from individual showers weighted by the flux of the primary cosmic rays, as discussed in \sect{cascade-equations}. The steep cosmic-ray spectrum emphasises the particle production phase space at $x_\text{lab} > 0.1$, and suppresses the importance of pion-air and kaon-air interactions \citep{Gaisser:2016cr,Fedynitch:2018cbl}. If the muon deficit in simulated air showers is caused by a compounded effect over several steps of the hadronic cascade, then the atmospheric lepton fluxes should be less affected. Above PeV energies, the prompt component takes over which is not directly linked to the muon deficit in air showers.

Since intrinsic shower fluctuations do not play a significant role in the calculation of atmospheric lepton fluxes, the cascade equations \eq{cascade-equations} can be solved directly, either with semi-analytical approximations \citep{Zatsepin:1962ta,Volkova:1980sw,Gaisser:1983vc,Naumov:1993yr,Lipari:1993hd}, iterative semi-analytical approaches \citep{Kochanov:2008pt,Sinegovskaya:2014pia} or iterative numerical solvers such as \mceq{} \citep{Fedynitch:2015zma}. The accuracy of these methods is comparable \citep{Gaisser:2019xlw,Morozova:2017fof}, but challenged by the increasing precision of muon flux measurements.

An alternative approach is to simulate many air showers over a wide range of energies and zenith angles and weight the results with the cosmic-ray flux \citep{Rancati:1999vb,Barr:2004br,Honda:2006qj,Fedynitch:2012fs}. \cite{Honda:2006qj} showed that simulations of the atmospheric muon flux with the \dpmjet{III} model underestimate measurements by up to 20\,\% at 1\tev, while showing no discrepancy at 10\gev. Recent computations with post-LHC interaction models confirm this observed deficit \citep{Fedynitch:2018cbl}, which remains after considering systematic uncertainties of modern muon flux and charge-ratio measurements \citep{Yanez:2019bnw}. As in the case of the interpretation of air shower data, significant uncertainties arise from a lack of data on forward particle production and in general the hadronic phase-space coverage at high energies \citep{Barr:2006it,Honda:2019ymh}. At high energies ($> 100\gev$), uncertainties of the models for the cosmic-ray spectrum and mass composition \citep{Gaisser:2012zz,Gaisser:2013bla,Fedynitch:2012fs,Dembinski:2017zsh, Evans:2016obt} introduce errors up to several tens of percent \citep{Barr:2006it,Fedynitch:2018vfe}. The GSF model \citep{Dembinski:2017zsh} helps in breaking the degeneracy between uncertainties in the cosmic-ray flux and hadronic interactions. It is an nearly model-independent parametrization (GSF stands for Global Spline Fit) of the world data on cosmic rays and provides a parameter covariance matrix that accounts for experimental systematic uncertainties through cross-calibration between experiments.

A viable strategy to improve the accuracy of the calculated atmospheric neutrino flux is to exploit the link between muon and neutrino fluxes by calibrating against muon flux data \citep{Honda:2006qj,Honda:2019ymh}. However, this ansatz is challenged by a lack of accurate muon flux data above PeV energies in the range relevant for IceCube and future astrophysical neutrino observatories. A further complication in this energy range is the onset of the prompt component \citep{Volkova:1983yf,Gaisser:1983vc,Inazawa:1984rk,Volkova:1985vx,Bugaev:1985bs,Gondolo:1995fq,Pasquali:1998ji,Costa:2000jw,Martin:2003us,Enberg:2008te}. Prompt leptons originate from semi-leptonic decays of charm and bottom mesons. In case of muons, additional prompt components are expected from decays of unflavored vector mesons \citep{Illana:2009qv} and from electromagnetic muon pairs \citep{Gamez:2019dex,Meighen-Berger:2019lld}. The prompt atmospheric neutrino flux has not yet been conclusively measured. The best limit is set by IceCube \citep{Aartsen:2016xlq} to 1.04 times the central prediction of \cite{Enberg:2008te} based on a dipole model. The prompt atmospheric flux is a significant background for the measurement of astrophysical neutrinos in track-based methods \citep{Aartsen:2016xlq,Stettner:2019tok}. The veto-based analyses by IceCube \citep{Abbasi:2020jmh,Aartsen:2020aqd} are less affected, since atmospheric backgrounds including prompt ones are efficiently suppressed.

Modern calculations \citep{Garzelli:2015psa,Bhattacharya:2015jpa,Gauld:2015yia,Garzelli:2016xmx,Bhattacharya:2016jce,Sinegovsky:2018vju,Zenaiev:2019ktw} aim to reduce the theoretical uncertainties of the prompt neutrino flux with LHC data. The production cross-sections for charm hadrons are not as tightly constrained theoretically as they are experimentally by direct measurements. The theoretical calculations are limited by the uncertainties in the scale, charm mass, and the nuclear PDFs. A non-perturbative intrinsic charm component proposed by \cite{Brodsky:1980pb} may also contribute \citep{Bhattacharya:2018tbc}. Prompt lepton production is not directly linked to the Muon Puzzle in air showers and primarily constrained by measurements of heavy-flavour production, but there is an overlapping interest in the analysis of proton-oxygen collisions at the LHC to better understand potential nuclear effects in the production of light and heavy flavour.

\end{itemize}

\subsection{Impact of changes in hadronic interaction features on air shower features}
\label{sec:impact}

A first connection between features of microscopic hadronic interactions and air shower features was made in \sect{heitler} with the Heitler-Matthews model. Due to the involved simplifications and approximations, these can only guide the research. It is essential to study these connections with full air shower simulations, but such studies are challenging due to the complexity of air showers and the computational cost of simulating a large number showers with varied parameters in the hadronic interaction model. Major accomplishments in connecting microscopic and macroscopic shower physics are works on the impact on emission of Cherenkov light from a gamma shower \citep{Fortson:1999nw}, on the asymmetries of muon footprints on the ground \citep{Ave:2000xs}, on the time structure of signals in the water-Cherenkov detectors of the Pierre Auger Observatory \citep{Aglietta:2007yx}, on the impact of basic features of hadron production in a hadronic shower on the moments of \xmax and \nmu \citep{Ulrich:2010rg}, and on the corresponding impact on atmospheric lepton fluxes \citep{Fedynitch:2012fs}.

The Heitler-Matthews model has shown the following basic parameters of interest with a large impact on air showers: the inelastic hadronic cross-section $\sigma_\text{inel}$, the multiplicity $\nmult$ of secondary particles, the elasticity $E_\text{leading}/E_0$ defined as the energy fraction carried by the most energetic particle, and the energy ratio $R = E_\gamma/E_\text{long-lived hadrons}$ of energy carried by photons from short-lived hadrons like the \piz to the energy in long-lived hadrons \citep{Baur:2019cpv}.

These parameters need to be well-known especially for the most common \piX{air} interaction in air shower. The interactions of other long-lived particles such as the $K$ meson are less important for the air shower development and the total number of muons produced in air showers \citep{Maris:2009} due to their lower mean multiplicity, also compare \fg{inclusive-flux}. For forward production in high-energy collisions, the quark flavour in the projectile is not important, since most quarks are produced from the vacuum and strong interactions are flavour-blind. This also means that high-energy \piX{air} interactions can be constrained with collider measurements of \pX{air}.

\begin{figure*}[tb]
\centering
\includegraphics[width=0.99\textwidth]{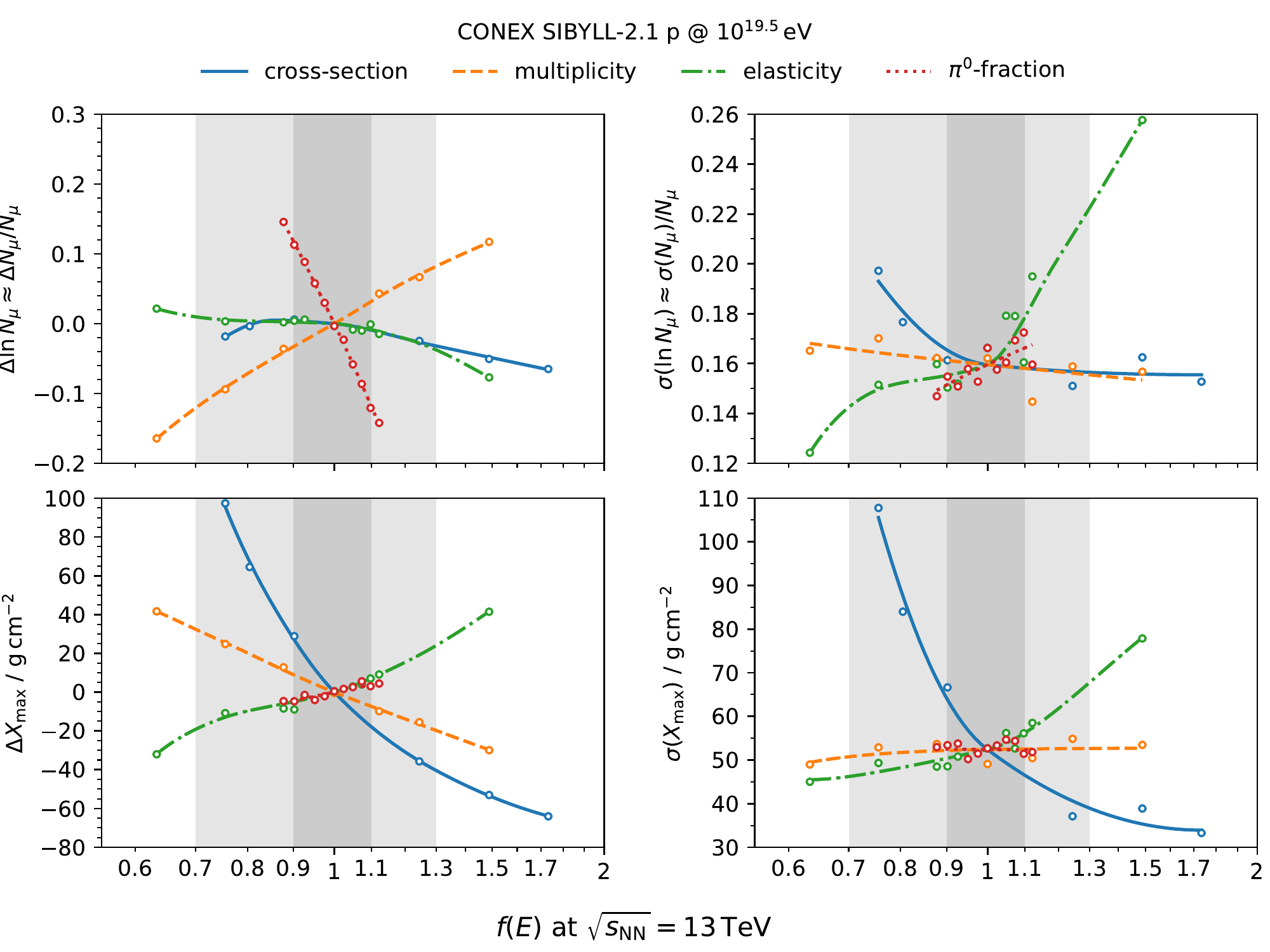}
\caption{Impact of changing basic parameters of hadronic interactions (see text for details) on the means and standard deviations of the logarithm of the muon number \nmu (top row) and the depth \xmax of the shower maximum (bottom row) for a $10^{19.5}\si{eV}$ proton shower simulated with \conex using \sibyll{2.1} as the baseline model, as described in the text. Relative shifts to the mean values are shown on the left-hand side. Fluctuations are shown on the right-hand side. The original data from \cite{Ulrich:2010rg} was refitted for this plot with monotonic cubic splines and are shown as a function of the modification in the nucleon-nucleon system at a cms-energy $\sqrtsnn = 13\tev$, which is extrapolated logarithmically towards higher energies as described in the text. The shaded bands highlight a $\pm 10\,\%$ and $\pm 30\,\%$ modification, respectively.}
\label{fig:adhoc}
\end{figure*}

\begin{figure}[t]
\includegraphics[width=0.48\textwidth]{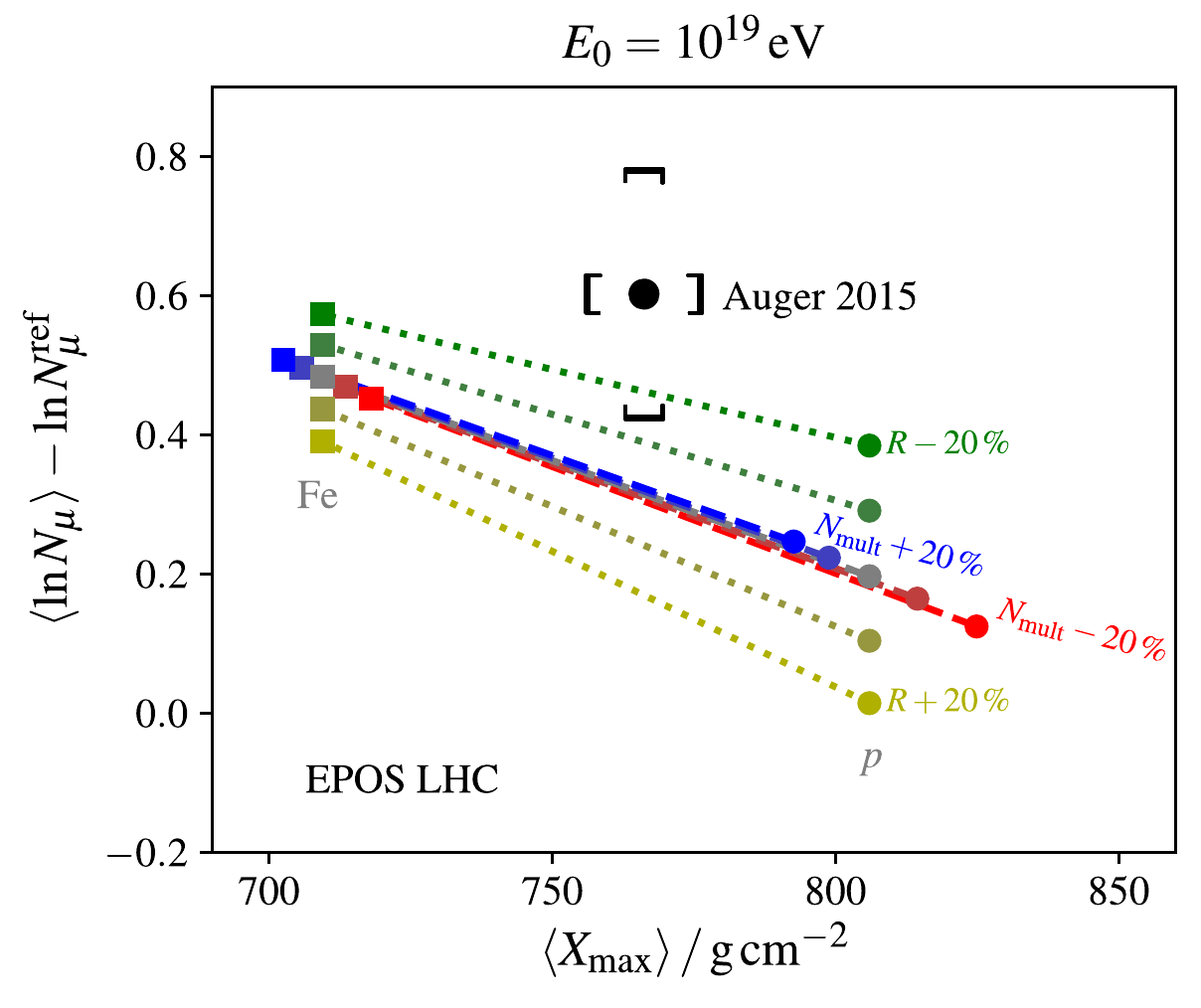}
\caption{Impact of modifying the average hadron multiplicity $\nmult$ and the ratio $R$ of electromagnetic to hadronic energy in an air shower on predictions for the depth of shower maximum \xmax and the number of muons \nmu for $10^{10}$\,\si{GeV} showers (image from \cite{Baur:2019cpv}). The representative data point is taken from the Pierre Auger Observatory \citep{Aab:2014pza}.}
\label{fig:impact_muons_nmult_r}
\end{figure}

\cite{Ulrich:2010rg} introduced an \emph{ad-hoc} model for air shower simulations in which these parameters are changed during the run-time of an air shower simulation as a function of the energy $E$ of the colliding hadron in the frame where the target is at rest. It uses the original predictions of a particular event generator as the baseline (in this case \sibyll{2.1}, but any generator can be used), which are scaled with an energy-dependent factor $f(E)$. The factor is 1 below a chosen energy threshold of 1\pev and grows logarithmically above. This is motivated by the fact that generators are fairly constrained by accelerator data at low energies, but diverge logarithmically when they are extrapolated to higher energy where accelerator data is missing. The size of the modification is governed by the parameter $f_{19}$,
\begin{equation}
    f(E) = 1 + (f_{19}-1) \cdot
    \begin{cases}
    0 & E < 1\pev \\
    \frac{\log_{10}\left(\frac{E}{1\pev}\right)}{\log_{10}\left(\frac{10\si{EeV}}{1\pev}\right)} & E \ge 1\pev
    \end{cases},
    \label{eq:mod}
\end{equation}
which is the size of the modification for a hadron with $10\eev = 10^{19}\si{eV}$ (an arbitrary scale). Equation\,\ref{eq:mod} is applied in an air shower simulation to each individual hadron collision to modify the respective parameter, as listed above. Very large distortions of $f(E)$ would correspond to exotic modifications of QCD and may conflict with more recent LHC measurements, while small deviations may be within the realm of conventional scenarios and compatible with LHC data.

The results of the study by \cite{Ulrich:2010rg} for the mean and standard deviation of the muon number \nmu and the depth of shower maximum \xmax for a $10^{19.5}\si{eV}$ proton shower are shown in \fg{adhoc} as a function of the modification factor at the LHC energy scale of nucleon-nucleon collisions at 13\tev which corresponds to a projectile energy of 90\pev in the fixed-target system. We note that the fraction of neutral pions among all pions was modified in the \emph{ad-hoc} model in the original study and not the energy ratio $R$, but the effect of modifying this fraction and modifying $R$ are numerically similar. The most effective way to increase the muon number in air showers is to decrease the \piz-fraction. A 10\,\% reduction increases \nmu by 13\,\%. This is less than predicted by the Heitler-Matthews in \sect{heitler}, but can be understood by the fact that the modification in this study only affects the shower evolution above 1\pev, while in \sect{heitler} a modification of the whole hadronic cascasde was considered. The muon number also increases with the multiplicity, but the effect is much weaker. A 30\,\% increase would increase the muon number by only 9\,\%. Changes to the inelastic cross-section and elasticity have negligible impact on the muon number.

The impact on the standard deviation of the muon number is also important, which has been measured recently for the first time by the Pierre Auger Observatory \citep{Aab:2021zfr}. Reasonable agreement between the measurement and the post-LHC models \eposlhc, \qgsjet{II.04}, and \sibyll{2.3d} was found. This puts strong constraints on changes to the elasticity, which is the only one of the four considered parameters with a large impact on the \nmu-fluctuations. The measured \nmu-fluctuations could be used to severely constrain the elasticity. A reduction of the \piz-fraction by 10\,\% would only change the \nmu-fluctuations by one percentage point.

Since air shower simulations with post-LHC models give a reasonable description of the depth of the shower maximum, \xmax, it is important to also consider the impact of changes on \xmax. Air shower simulations for proton and iron showers bracket the measurements over a wide range of shower energies and the mass composition inferred from \xmax is astrophysically plausible. This suggests that the parameter values that influence \xmax cannot deviate too much from those in current models without destroying the consistency. The depth of the shower maximum is most sensitive to the inelastic cross-section which has been measured very precisely in proton-proton collisions at the LHC. A remaining theoretical uncertainty arises from the extrapolation of these data to the \pX{air} and \piX{air} cross-sections. Modifications of the multiplicity, elasticity, and \piz-fraction all have a similar impact on \xmax.

The standard deviation of \xmax is even more sensitive to the inelastic cross-section than its average. It weakly depends on changes to the elasticity and is unaffected by changes to multiplicity and \piz-fraction. The standard deviation has been measured precisely by the Pierre Auger Observatory \citep{Aab:2014kda}, and it is found that is challenging for \qgsjet{II.04} to simultaneously describe the mean and the fluctuation of \xmax, while \eposlhc gives a consistent interpretation. The sensitivity of the \xmax-fluctuations to the inelastic cross-section has been further exploited to infer the inelastic \pX{air} cross-section from air shower measurements by the Pierre Auger Observatory \citep{Collaboration:2012wt}, although the analysis uses the shape of the tail towards large \xmax values instead of the standard deviation, which is beneficial in presence of a mixed composition of cosmic rays.

These individual results can be combined to reveal an interesting point illustrated in \fg{impact_muons_nmult_r} \citep{Citron:2018lsq,Baur:2019cpv}, where the impact of changes in the hadron multiplicity $\nmult$ and the energy ratio $R$ on the means of the logarithm of the muon number and \xmax is shown for $10^{19}\si{eV}$ showers and compared to a measurement by the Pierre Auger Observatory. In this double-logarithmic scale, any possible mass composition of cosmic rays between the two extremes of pure proton showers (bottom right) and pure iron showers (top left) produces a point on a straight line. The standard prediction by EPOS-LHC is indicated by the grey line (hard to see under the other colored lines). The muon deficit in simulations is the reason why this line does not overlap with the data point.

In addition to the standard prediction by EPOS-LHC, also \emph{ad-hoc} modified predictions are shown. The multiplicity $\nmult$ (blue and red lines) and the ratio of energy going into neutral pions $R$ (yellow and green lines ) are changed, respectively, by up to $\pm 20\%$. While the former shifts the lines along themselves and has no power to resolve the muon mystery, the change of $R$ has an effect perpendicular to the lines and has a large impact on the interpretation of the data. This strongly suggests that the solution to the Muon Puzzle lies in a modification of the energy ratio $R$.

\section{Possible solutions for the Muon Puzzle}
\label{sec:solutions}

The muon deficit in air shower simulations starts to appear when the cms-energy in the nucleon-nucleon system reaches $\sqrtsnn \approx 8\tev$. Since there is no large deficit observed in post-LHC models at lower energies, it seems that a new phenomenon in unbiased hadron collisions becomes important at this energy scale that can be neglected at lower energies. As discussed in \sect{heitler} and \sect{impact}, the only plausible way to increase the muon number sufficiently is to decrease the energy fraction lost to photon production (mainly from $\piz$ and $\eta$ decay) in hadron collisions.

A comprehensive summary of rejected attempts to explain the Muon Puzzle in some other way is given by \cite{Farrar:2013sfa}. The effect must come from soft-QCD, since hard scatterings that produce new heavy particles are too rare to significantly change the hadron composition. The standard model of soft low-energy hadronic interactions leaves no room to change the hadron composition. It uses string fragmentation and remnant excitation to produce hadrons. Both effects are believed to produce an universal hadron composition independent of the collision energy. A string is a colour-neutral flux tube made of virtual quarks and gluons that spans between two colour charges. As the colour charges move away from each other, the string gains potential energy, which is subsequently converted into quark/anti-quark pairs created from the vacuum. String fragmentation produces the same composition of hadrons however the strings are created \citep{Andersson:1998tv}. However, basic arguments suggest that this picture has to be modified at high $\sqrtsnn$ and in nuclear collisions with high string density.

Before following this thought further, we want to mention two earlier proposals for solving the Muon Puzzle that are disfavoured by the early onset of the muon discrepancy around $\sqrtsnn \approx 8\tev$. The proposals have in common that an extreme change in the first or the first few generations of the hadronic cascade is introduced at cms-energies outside the reach of colliders, while the rest of the shower develops normally. \cite{Farrar:2013sfa} proposed a toy model based on Chiral Symmetry Restoration to suppress pion production, which effectively replaces mesons with baryons in a fraction of the high energy events generated by EPOS and some minor modifications of the average multiplicity and the inelastic cross-section. The modified model reproduces \xmax measurements with proton cosmic rays and a range of other measurements available at the time while increasing the muon number by a factor 2. \cite{Anchordoqui:2016oxy} assume that cosmic rays at the highest energies are heavy or at least medium mass nuclei, and proposes that the first interaction creates quark gluon plasma (QGP), which we discuss further below. They then argue that the high baryochemical potential of this initial state (the excess of $u$ and $d$ quarks) leads to a shifted equilibrium with enhanced strangeness. This correspondingly leads to a reduction of the $\piz$-fraction and an increase in the muon content by 40\,\%. To achieve this with a modification of only the first interaction requires an extreme combination of temperature and baryochemical potential not reproduced in heavy ion collisions at the LHC. \cite{LaHurd:2017pmb} investigated the impact of QGP formation with conventional parameters if it affects only the first interaction and found only a small increase in the muon number, not sufficient to explain the Muon Puzzle.

\begin{figure}[tb]
\includegraphics[width=\columnwidth]{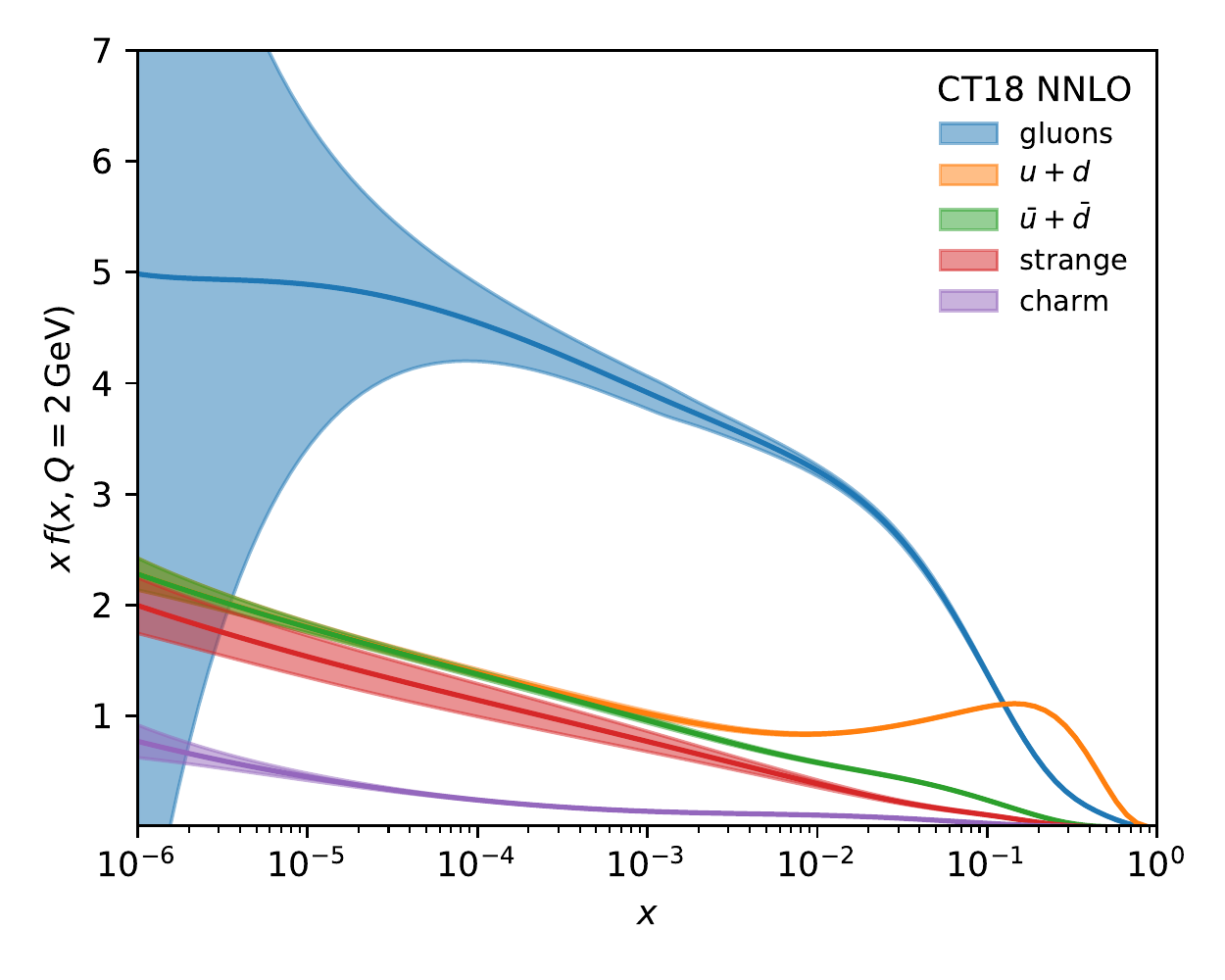}
\caption{Parton density functions (PDF) times $x$ of quarks and gluons in the proton as obtained from CT18 global analysis by \cite{Hou:2019efy}, evaluated at the scale $Q = 2\gev$. Bands indicate one standard deviation. Values were taken from APFEL Web by \cite{Carrazza:2014gfa}.}
\label{fig:proton_pdf}
\end{figure}

The issues with these approaches are avoided if the responsible effect is weaker, but potentiates over several steps in the hadronic cascade and starts well below the EeV scale. Recent LHC data and theoretical considerations suggest a mechanism which fits this description qualitatively. We start by noting that the parton density function (PDF) of the gluon inside the nucleon rises rapidly with decreasing momentum fraction $x$, as shown in \fg{proton_pdf}. Producing hadrons requires a minimum 4-momentum squared $Q_\text{min}^2 \propto x_1 \, x_2 \, s$, which means that the growth in hadron multiplicity as a function of $\sqrt{s}$ is driven by the increase in gluon density at small $x$. Partons in high-energy collisions are dominantly produced by gluon fusion, while forward produced hadrons originate dominantly from quark-gluon scattering.

At large $\sqrt{s}$, the high gluon density leads to simultaneous multi-parton interactions and high string densities. Since strings are color-neutral, they do not interact at a distance, but lattice calculations show that strings are flux tubes with a finite radius \citep{Cea:2014uja,Cea:2015wjd}. It follows that strings must eventually overlap in space, since the string density grows much faster with $\sqrt{s}$ than the radii of the colliding hadron discs. In hadron-nucleus collisions this effect happens at lower $\sqrt{s}$ due to the larger parton densities in the nuclei, but it is expected to happen in any system at sufficiently high energy. If the string interactions lead to a sufficiently strong modification in the hadron composition of forward produced particles, it can solve the Muon Puzzle.

Heavy-ion collisions have already demonstrated a modifications of the hadron composition with respect to classic string models. Strangeness and baryon enhancement have been observed at the SPS, RHIC, and the LHC \citep{Capella:1994pr,Arsene:2004fa,Adcox:2004mh,Adams:2005dq,Becattini:2008yn}. These collisions are very successfully described with models that assume the formation of a new state of matter, the QGP, which hadronises collectively after reaching thermal equilibrium.

It was commonly believed that QGP does not form in small systems like \pp and \pX{A} collisions, since these systems do not seem to offer enough time for the colliding partons to thermalize into a QGP. However, several QGP-like signatures were also found in these systems. Recent overviews on these collective phenomena and theories to explain them are given by \cite{Dusling:2015gta,Nagle:2018nvi,Citron:2018lsq}. A range of models have been developed that explain these effects with and without the formation of QGP. The latter explain the observations through new properties of the initial state of the colliding hadrons at high $\sqrtsnn$, which can be described as a colour glass condensate (CGC), or by string interactions. Mixed forms are also explored, for example, a CGC that decays into a QGP.

\begin{figure}[tb]
\includegraphics[width=\columnwidth]{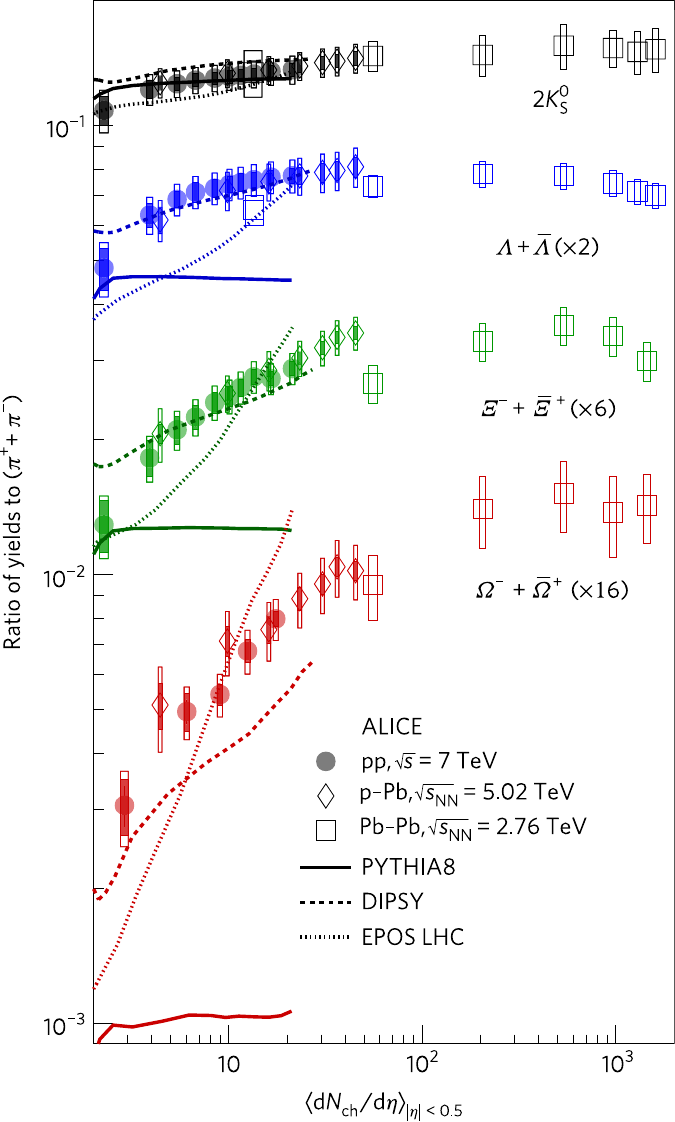}
\caption{Yield ratios of strange hadrons and pions in \pp, \pPb, and \PbPb as a function of the multiplicity at mid-rapidity (image from \cite{ALICE:2016fzo}).}
\label{fig:alice-strangeness}
\end{figure}

The potential key to the Muon Puzzle was discovered by the ALICE collaboration, which found a universal strangeness and baryon enhancement in \pp, \pPb and \PbPb \citep{ALICE:2013wgn,Adam:2015vsf,ALICE:2016fzo,Vasileiou:2020rov}. As shown in \fg{alice-strangeness}, the enhancement only depends on the multiplicity of the event at mid-rapidity, and not on the details of the collision system like its size or $\sqrtsnn$. This is highly remarkable: if this universality holds, it allows one to predict the hadron composition in average collisions of \pX{air} at $\sqrtsnn \gg 10\si{TeV}$ well beyond the reach of colliders based on reference measurements in central \PbPb collisions at current LHC energies.

Most of the data on collective effects was obtained with particles emitted at mid-rapidity $|y| < 2$. This region is not directly relevant for air showers as shown in \fg{eta-spectra}. It has not been shown experimentally yet that these effects can also be seen in hadrons produced at forward rapidities. Theoretical calculations suggest that hadrons produced from QGP decay reach well into the forward region, in \pp at 13\si{TeV} to $y \simeq 6.5$ \citep{Busza:2018rrf,Baur:2019cpv}. It is  therefore important to search for collective effects also at forward rapidities $y > 2$ as a function of multiplicity in small systems, \pp, \pPb, \pO, and \OO. The most important effect to search for is a strangeness and baryon enhancement.

We briefly summarise the other known signatures of QGP and why do they do not relate to the Muon Puzzle directly.
\begin{itemize}
\item \emph{Short-range and long-range correlations of hadron momenta and azimuthal modulation of multiplicities.} Because of collective flow in the QGP, hadron momenta become correlated and modulated along the azimuthal angle. This has been observed experimentally in systems ranging from \pp to \PbPb \citep{Khachatryan:2010gv,CMS:2012qk,Khachatryan:2016txc,Aaboud:2018syf,ALICE:2021nir}. The correlations are a key signature of collective hadronization, but there is a wide range of models with and without QGP formation that describe the data. Flow has no direct consequences for the Muon Puzzle or for the air shower development in general. The effects would average out over the shower development and are negligible compared to intrinsic shower fluctuations.
\item \emph{Jet quenching.} A key signature of QGP formation is the energy loss of high-$\pt$ partons traversing the medium. Jets are produced by partons generated in a hard scattering. If a parton has to traverse a QGP medium, it looses energy compared to free propagation. This effect was first discovered at RHIC in \AuAu collisions and has been extensively studied in \PbPb and \XeXe at the LHC \citep{Adcox:2001jp,Arsene:2012uj,Balek:2014uha,Jung:2014qta,Sirunyan:2017xss,Sirunyan:2017oug,Aaboud:2018twu,CMS:2018yyx,Sirunyan:2018nsz,CMS:2016xef,Acharya:2018eaq,ALICE:2018vuu,Adam:2015jda,Adam:2015kca,Festanti:2014foa,Grelli:2012yv}. It is measured with the nuclear modification factor $R_{AA}$, which compares the $\pt$-dependent yields for particle production in \XX{A}{A} collisions with the corresponding \pp collision. This effect is absent in \pPb \citep{ALICE:2012mj,Balek:2014uha,Aad:2014bxa,Jung:2014qta,Li:2014dha,Adam:2015iga,Adam:2015hoa,CMS:2016xef,Sirunyan:2016fcs,Balek:2017man,Aaij:2017cqq,Sun:2019eaa,Acharya:2020rvc,ALICE:2018vuu,Acharya:2018yud}, which implies it is also absent in \pX{air} and therefore irrelevant for the Muon Puzzle. Some modification is observed at low $\pt$ in \pPb collisions, but that is attributed to the initial state, such as modifications of the parton density functions of the nucleon inside a nucleus compared to free protons. These so-called cold-nuclear matter effects are relevant for air showers and need to be measured for oxygen in \pO collisions.
\item \emph{Enhancement of $\avg{\pt}$ for hadrons with higher mass}. When hadrons form in the rest frame of a moving fluid, they receive a shift towards higher momentum as a function of their mass that can be observed in their $\pt$ distribution. This effect is not relevant for the Muon Puzzle, since the $\pt$ distribution in the first few interactions has no impact on the later shower development.
\end{itemize}

In the following, we give an overview of different approaches to explain QGP-like phenomena. Most theoretical works focus on the mid-rapidity region. For a potential solution to the Muon Puzzle, more theoretical investigations into the forward hadron production in hadron-ion collisions at energies $\sqrt{s} > 10\tev$ are needed. So far, the only predictions for air showers come from the EPOS model which assumes QGP formation in small systems.

\subsection{Quark gluon plasma in small systems}

Quark gluon plasma (QGP) is a high temperature state of quark and gluon matter in which the partons are no longer bound into colour-neutral hadrons. The formation of QGP is the standard model for high-energy heavy-ion collisions in very good agreement with observations. Recent overviews on QGP and related phenomena are given by \cite{Dusling:2015gta} and \cite{Busza:2018rrf}. The QGP medium behaves like a super-fluid with zero viscosity; its equation of state has been computed with lattice QCD. The expansion is described hydrodynamically. QGP evolution is therefore theoretically well-understood and the thermalization (at least in heavy-ion collisions) hides details in the initial state which are less well understood. Both aspects give QGP models high predictive power. Because of the universal aspects, the remaining parameters of QGP models can be tuned in reference systems measured at colliders and then should work for any other collision system.

QGP models enjoy great success in describing heavy-ion collisions, but there is a controversy whether QGP also forms in small systems. To illustrate this we describe the standard picture of a heavy-ion collision. When two nuclei collide, they appear to each other as Lorentz-contracted thin discs, which have a short traversal time $\ll 1\si{fm}/c$. So the initial state of the collision is a pre-equilibrium with a very inhomogeneous distribution of deposited energy in the plane perpendicular to the beam direction. Some time is required to equilibrate into the QGP fluid, during which the matter expands in the longitudinal direction and radially in the transverse plane. The QGP fluid expands and cools until it reaches the QGP freeze-out temperature $T \simeq 170\si{MeV}$ (see \cite{Aoki:2006we}) and then breaks up into hadrons. The hadron density is initially large and scattering is frequent. This phase is called a hadron gas. As the gas expands and the density drops, it experiences a chemical freeze-out when inelastic interactions stop and, shortly after, a kinetic freeze-out when elastic interactions stop. At this point the hadrons have the final-state momenta that are measured experimentally.

It has been argued that small systems expand too fast for the initial state to equilibrate into a QGP. While small systems display several phenomena that are QGP-like, there are also other plausible mechanisms that produce these signatures. One of the key signatures of QGP formation is jet quenching, which is particularly difficult to explain without QGP, but jet quenching is not observed in \pPb collisions. This seems at variance with QGP formation. \cite{Sievert:2019zjr} have simulated \OO collisions (a small system) and found that the initial parton density and thus temperature fluctuations in \OO are larger than in \PbPb. Locally, very high temperatures can be reached. This suggests that only local QGP droplets form in small systems instead of a large QGP medium and could explain the lack of jet quenching. How small QGP droplets can become has been investigated theoretically by \cite{Chesler:2015bba,Chesler:2016ceu}, with the conclusion that the minimum radius is given by the inverse of the temperature. QGP droplets as small as the size of a proton in \pA collisions seem theoretically possible.

EPOS \citep{Pierog:2009zt,Pierog:2013ria} was the first generator that combined a string model with QGP formation to describe both small and large collision systems. In EPOS, strings are formed in all systems. If the string density surpasses a threshold, strings are merged to form QGP, while they hadronise classically below that threshold. In general, this gives rise to two hadronization regimes, called core and corona \citep{Becattini:2008yn}. The core consists of particles from QGP hadronization, while the corona is dominated by string fragmentation and contributions from the excited remnant. The core dominates at mid-rapidity but extends well into the forward region which is important for air showers. \cite{Baur:2019cpv} showed that there is still a significant contribution at $\eta \simeq 7$ in \pp collisions at 13\si{TeV}.

How large the core contribution is depends on tuning parameters in the EPOS model. In the widely used version EPOS-LHC, the core contribution seems to be underestimated in light of the recent ALICE data on strangeness enhancement. Larger contributions would be compatible with LHC data and quantitative projections from \cite{Baur:2019cpv} have shown that an enhancement of the core is a potential solution to the Muon Puzzle. Measurements of strangeness and baryon enhancement at forward rapidities as a function of multiplicity are a stringent test for this theory, as well as measurements of the ratio of electromagnetic to hadronic energy as a function of the multiplicity in the forward region, which have been performed in \pp collisions at 13\si{TeV} by CMS~\cite{CMS:2019kap} and should be repeated for \pPb collisions.

\subsection{Color glass condensate and glasma}

The colour glass condensate (CGC) model is an effective field theory which offers an alternative explanation for many phenomena attributed to QGP formation. For introductions to the CGC see \cite{Iancu:2003xm,Gelis:2010nm,Dusling:2015gta}.

Colliding hadrons in a high-energy collision appear to each other as flat Lorentz-contracted discs that are densely packed with gluons. Gluons with a small momentum fraction $x$ are much more numerous and have short life-times compared to gluons at large $x$ whose life-times are time dilated. The small $x$ gluons can be approximated by static classical fields for which the large $x$ gluons act as sources. The large $x$ gluons then appear weakly coupled because of the screening provided by the small $x$ gluon fields. The fields have the form of Lorentz-boosted Coulomb fields of electrodynamics with random colour, polarisation and density. The spectrum of fluctuations can be computed in the CGC framework and has a universal solution for small $x$ for any hadron. This gives CGC models high predictive power.

When two sheets of coloured glass collide, the classical colour fields change from transverse orientation and, being confined to the thin sheets, to longitudinal colour electric and colour magnetic fields that span longitudinally along the direction of motion between the sheets. The initial distribution of these longitudinal fields and their evolution until thermalization is referred to as the Glasma. The glasma eventually decays into quarks and gluons which are close in description to that of a QGP.

CGC calculations can reproduce the flow in heavy ion collisions observed at the LHC and RHIC \citep{Dumitru:2010iy,Dusling:2012iga,Schenke:2012wb,Mantysaari:2017cni}, and recently have been shown to also reproduce strangeness enhancement in \pp collisions \citep{Siddikov:2021cgd}.

\subsection{Enhanced parton models}

QGP-like phenomena have been also reproduced by parton shower models that are extended with string-string interactions or with parton scattering. The former idea was already described in the introduction to this section. Strings are colour-neutral and thus do not interact at a distance, but they have a finite radius and therefore should influence each other when they overlap. The overlap can be neglected in low-multiplicity events, but not in high-multiplicity events. Parton transport models instead implement parton-parton scattering to achieve a similar effect, gluons and quarks are treated as quasi-particles which scatter during the collision and effectively produce hydrodynamic-like flow patterns. It is a weakly interacting scenario in contrast to the strongly coupled hydrodynamics. Notable parton transport models are BAMPS \citep{Xu:2004mz} and AMPT \citep{Lin:2004en}. Whether these ideas are a dual picture to hydrodynamics or a real alternative is an open question. The details of these interactions are not well understood, which leaves freedom for model builders.

Several kinds of string interactions have been proposed. Colour reconnection \citep{Ortiz:2013yxa,Bierlich:2015rha} is implemented in \pythia{8} and produces flow-like effects. It can occur in events with multi-parton interactions, when several hard scatterings produce parton pairs simultaneously. Partons that cross existing strings created by other partons are colour reconnected in such a way that the total string length becomes as short as possible. This introduces momentum correlations for the hadrons produced by the two independent hard scatterings and enhances baryon production, but not strangeness. String shoving in \pythia{8} \citep{Bierlich:2017vhg} is another approach to produce correlated flow. A repulsive force is assumed between overlapping strings which introduces a transverse velocity that is transferred to the hadrons after string fragmentation.

Rope hadronization \citep{Bierlich:2014xba,Bierlich:2015rha} is a mechanism that produces strangeness and baryon enhancement. It assumes that parton pairs which form next to each other in geometric space act coherently to form a colour rope instead of two independent strings. The rope has a higher effective string tension, which results in more strange quarks and diquarks produced in its fragmentation. Rope hadronization is implemented in the DIPSY model that partially reproduces the strangeness enhancement in ALICE data \citep{ALICE:2016fzo}.

\section{LHC measurements: Status and Prospects}
\label{sec:lhc}

\begin{figure}[tb]
\includegraphics[width=0.9\columnwidth]{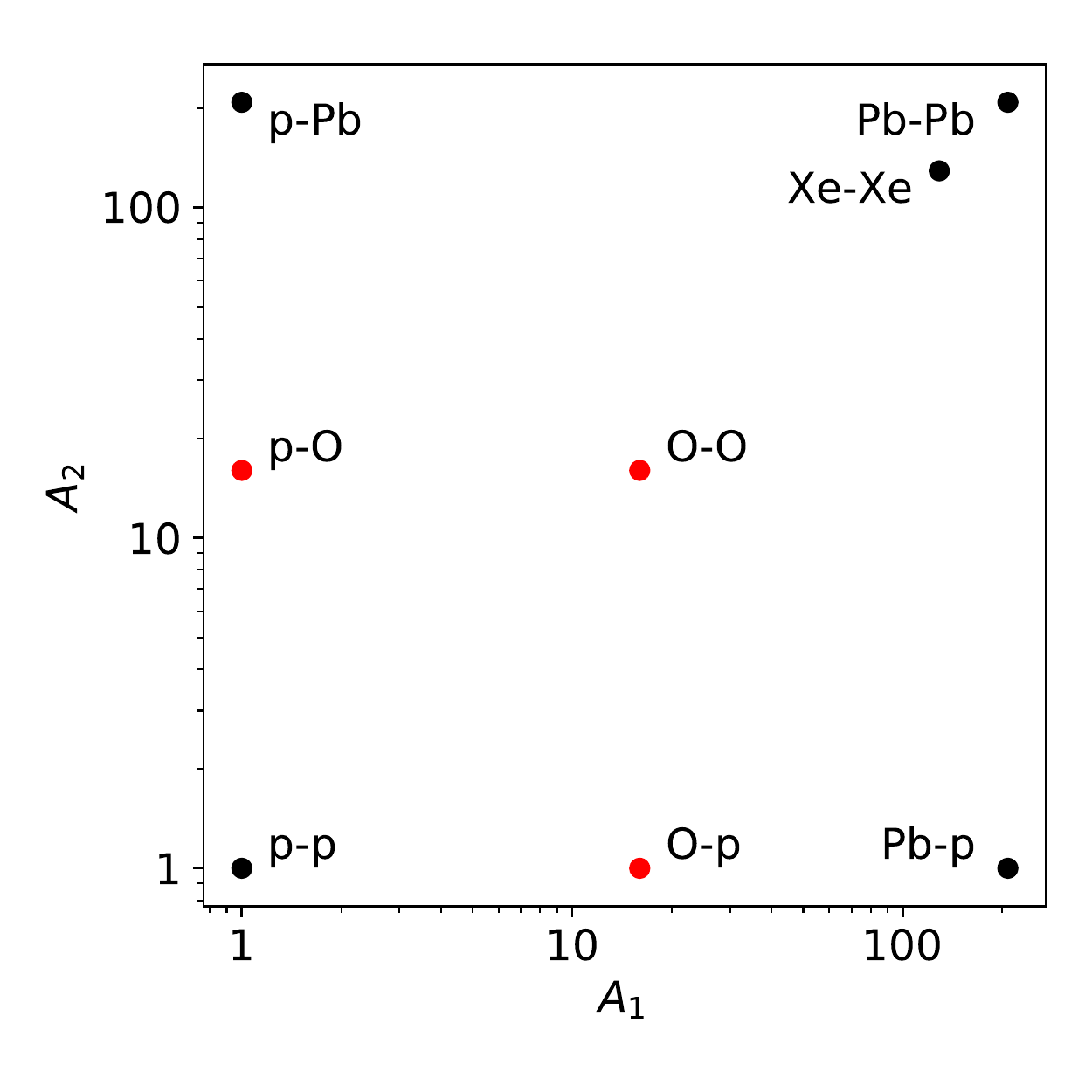}
\caption{Collision systems with data recorded at the LHC in Run 1 and 2 (black). Also shown are possible configurations with oxygen beams that have been proposed for Run 3 (red). The nucleon number is shown in log-scale since typically the relevant physics varies on a logarithmic scale.}
\label{fig:collision_systems}
\end{figure}

The Large Hadron Collider (LHC) accelerated protons up to 13\tev and lead ios up to 5.02\tev per nucleon during Run~1 and 2 to study proton-proton, proton-lead, and lead-lead collisions. A pilot run with \XeXe collisions followed in 2017 and not-fully-ionised lead nuclei in 2018.
Due to the success of the LHC heavy ion program in Run~1 and Run~2, an extension towards lighter ions has been proposed in~\cite{Citron:2018lsq} for the upcoming Run~3 and Run~4. A pilot run with oxygen beams has been proposed for the end of the upcoming Run~3 in 2023. These collision systems are visualised in \fg{collision_systems}.

\begin{figure}[tb]
\includegraphics[width=\columnwidth]{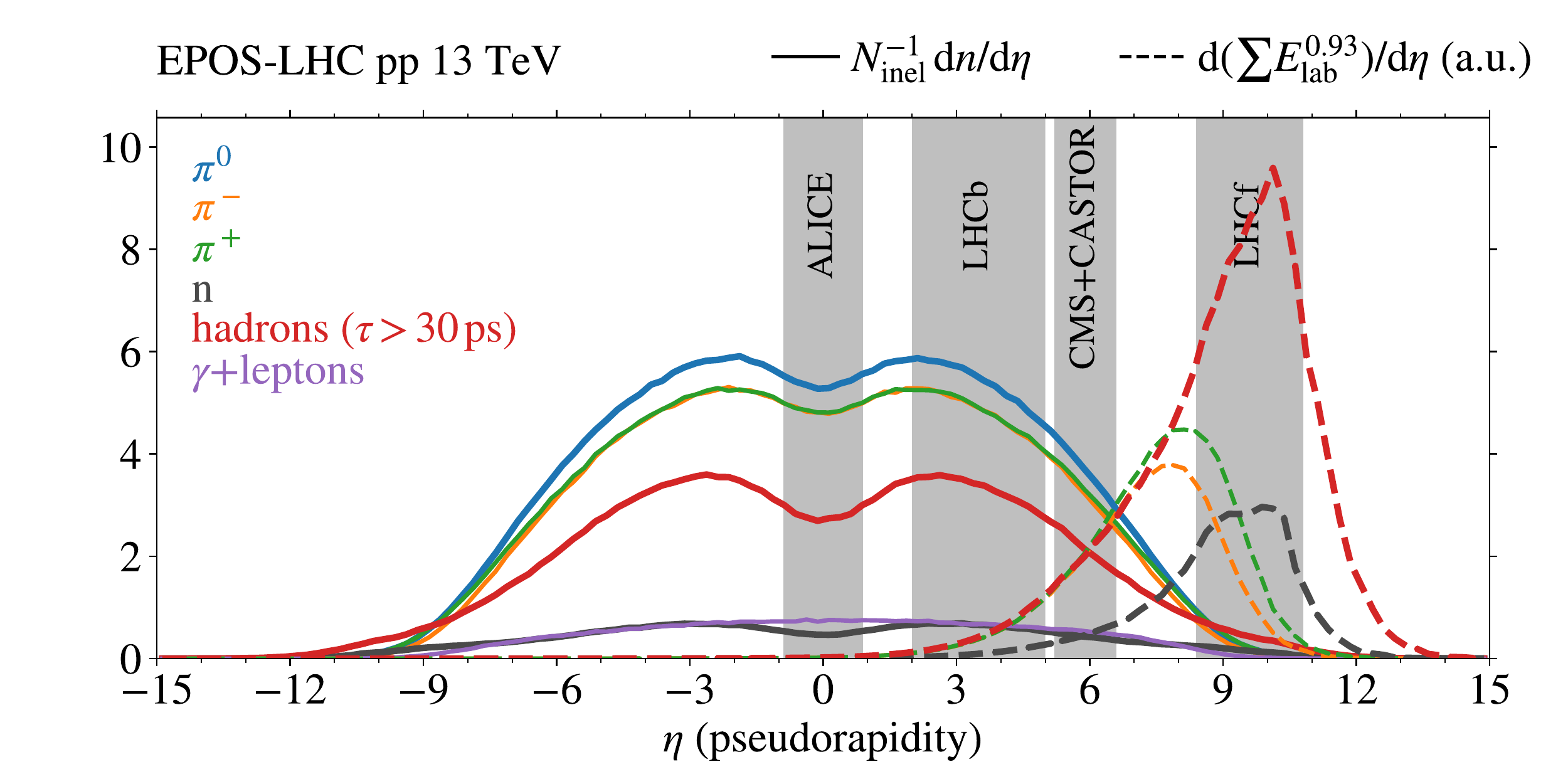}
\includegraphics[width=\columnwidth]{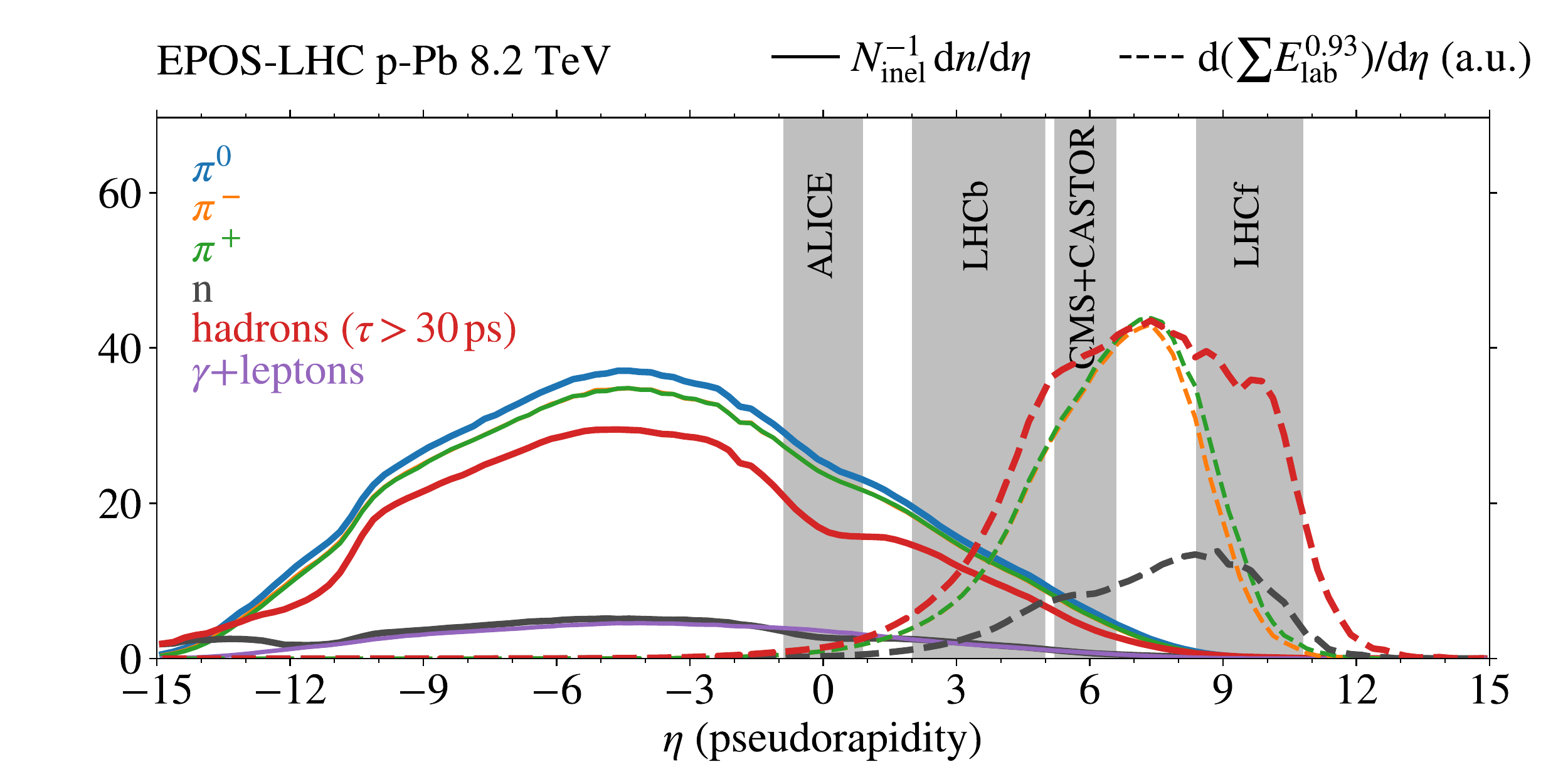}
\includegraphics[width=\columnwidth]{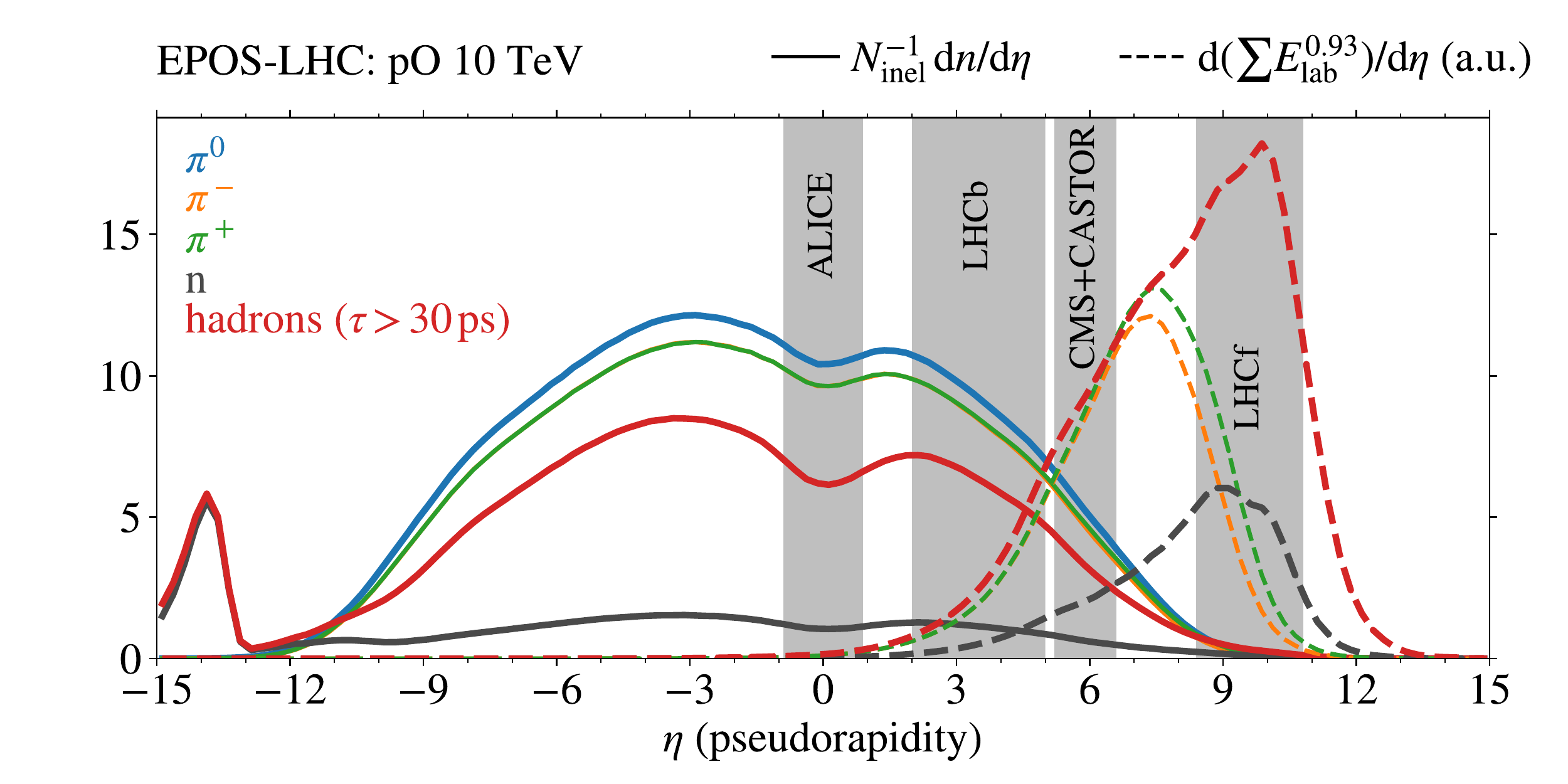}
\caption{Simulated densities of prompt particles (solid lines) in high-energy \pp, \pPb, and \pO collisions. Dashed lines show the estimated number of muons produced by the secondaries if they were propagated through the atmosphere, assuming $\nmu \propto E_\text{lab}^{0.93}$, where $E_\text{lab}$ is the energy of the secondaries in the boosted system.}
\label{fig:eta-spectra}
\end{figure}

The LHC as a \pp and \pX{A} collider with the highest available cms-energies $\sqrtsnn$ in the nucleon-nucleon system from 0.9 to 14\tev is an essential source of data for the modelling and understanding of extensive air showers. For a proton cosmic ray with energy $E_0$  hitting an air nucleus at rest, the conversion between $E$ and $\sqrtsnn$ is in the ultra-relativistic limit
\beq
\sqrtsnn = \sqrt{2 E_0 m} \quad \leftrightarrow \quad E_0 = \frac{s_\text{NN}}{2m},
\eeq
where $m$ is the nucleon mass. LHC data is used to provide anchor points for the tuning of parameters of hadronic interaction models used for air shower simulations. The models are essential to extrapolate into the uncharted phase-space of typical hadronic interactions in air showers since LHC data does not directly mimic these interactions. The following extrapolations are involved.
\begin{itemize}
\item \emph{Towards higher centre-of-mass energies}. The LHC energies from 0.9 to 14\tev cover projectile lab energies in air showers from 0.4 to 104\pev. The cosmic-ray energy spectrum extends at least three orders of magnitude further. The highest-energy cosmic-ray event ever recorded by the Fly's Eye experiment had an energy of $(320\pm 90)\eev$, see \cite{Bird:1994uy}, corresponding to $\sqrtsnn = (780 \pm 110)\tev$. The Pierre Auger Observatory also has recorded events that exceed 100\eev \citep{Aab:2020gxe}.
\item \emph{Towards hadron-nuclear collision systems}. The ability to describe hadron-nuclear collisions is very important for generators used in air shower simulations. The most common interaction in an air shower is \piX{N} and the most important first interaction is $\pX{N}$, since nuclear projectiles in air showers behave in good approximation like a superposition of elementary nucleon interactions as far as the projectile is concerned (the situation is kinematically different for the target, which cannot be approximated by a superposition of nucleons). These systems are far away from both \pp and \pPb, as indicated in \fg{collision_systems}. Generators mostly extrapolate from \hX{p} to \hX{N} without using the \pPb data. \sibyll{2.3d} is strict about this limitation and rejects projectiles heavier than iron and targets heavier than argon.
\item \emph{Towards forward rapidities}. The mid-rapidity region $|\eta| < 2$, which is most precisely measured at LHC, is only indirectly relevant for air showers. This is illustrated in \fg{eta-spectra}, which shows that the muons in an air shower are dominantly produced by long-lived hadrons emitted in the forward region at $\eta > 2$. Except for LHCb, the LHC experiments were not designed to perform precision tracking and PID at forward rapidities $\eta > 2$, also since there are considerable technical challenges for forward measurements at high luminosity conditions at the LHC. The radiation damage can become severe.
\end{itemize}

Because of the limited ability of most current hadronic interaction models to describe heavy-ion collisions, the LHC configurations with \pPb, \PbPb, and \XeXe currently cannot be fully used for parameter tuning and model validation.  This is a severe drawback in regard to the rich results obtained with the heavy ion program of the LHC and the recent discovery of QGP-like effects in light systems including \pPb from \pp. LHC collisions with lighter nuclei are needed to resolve this limitation.

\begin{figure*}[tb]
\includegraphics[width=0.32\textwidth,trim=10 0 10 0,clip]{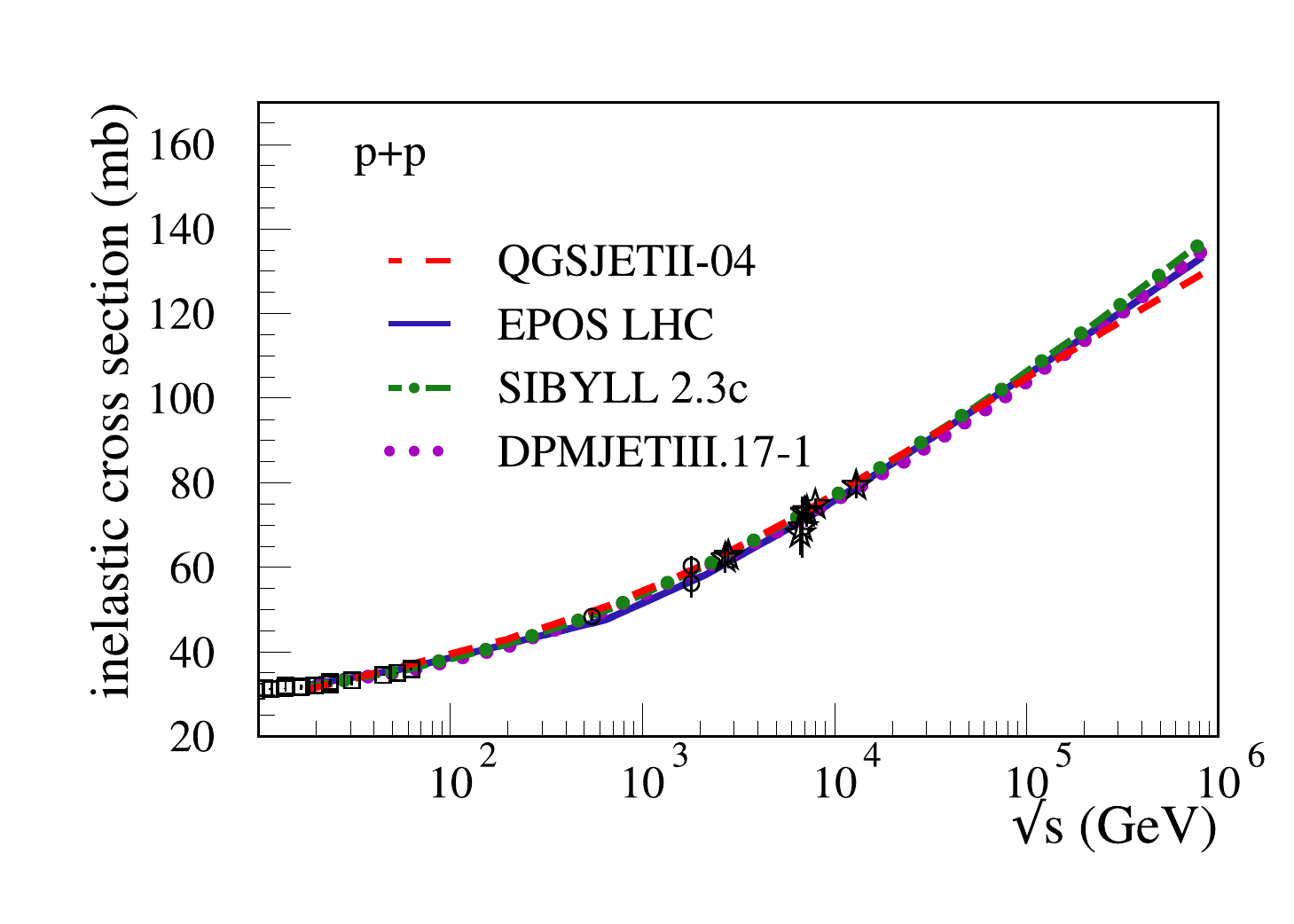}
\includegraphics[width=0.32\textwidth,trim=10 0 10 0,clip]{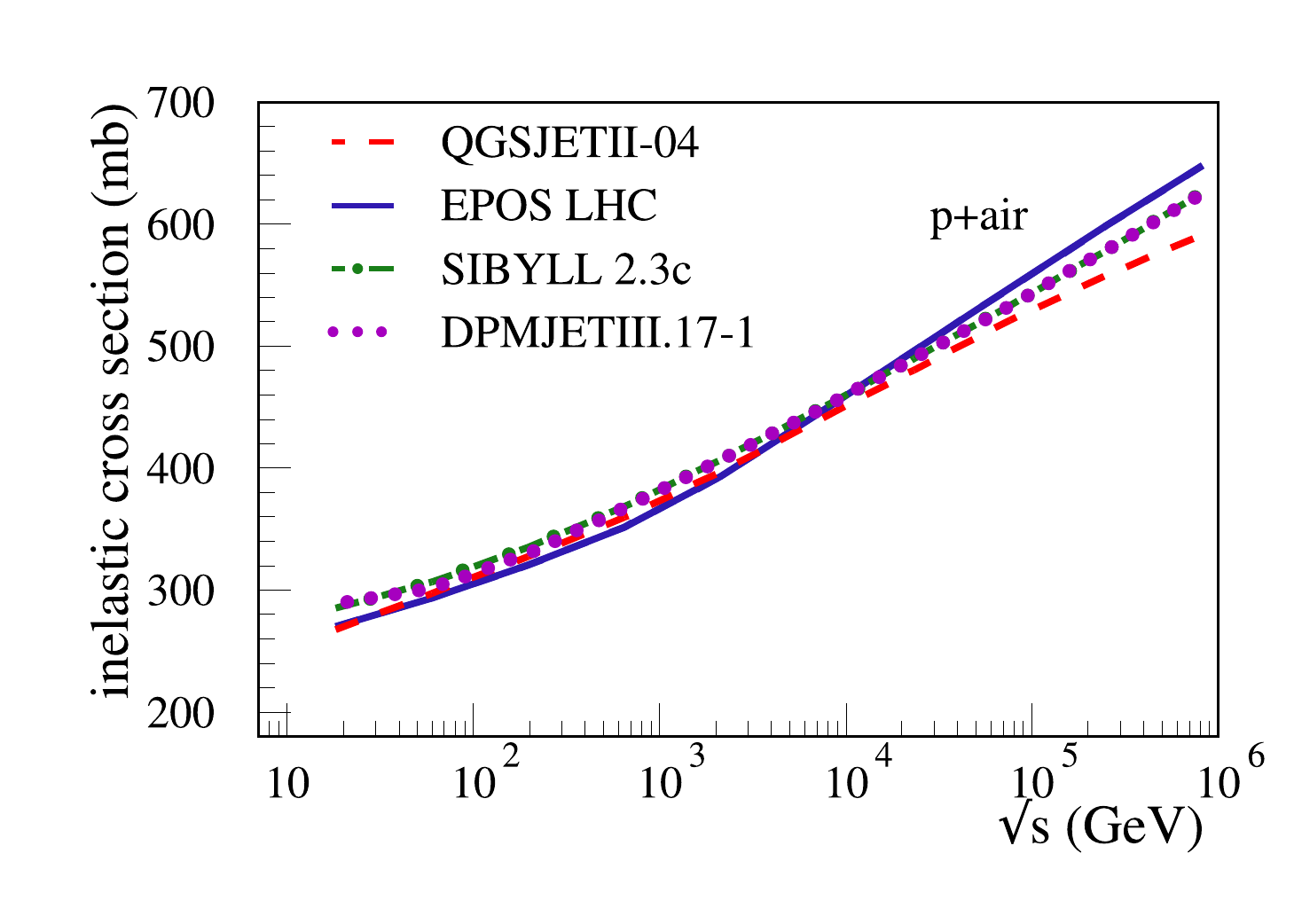}
\includegraphics[width=0.32\textwidth,trim=10 0 10 0,clip]{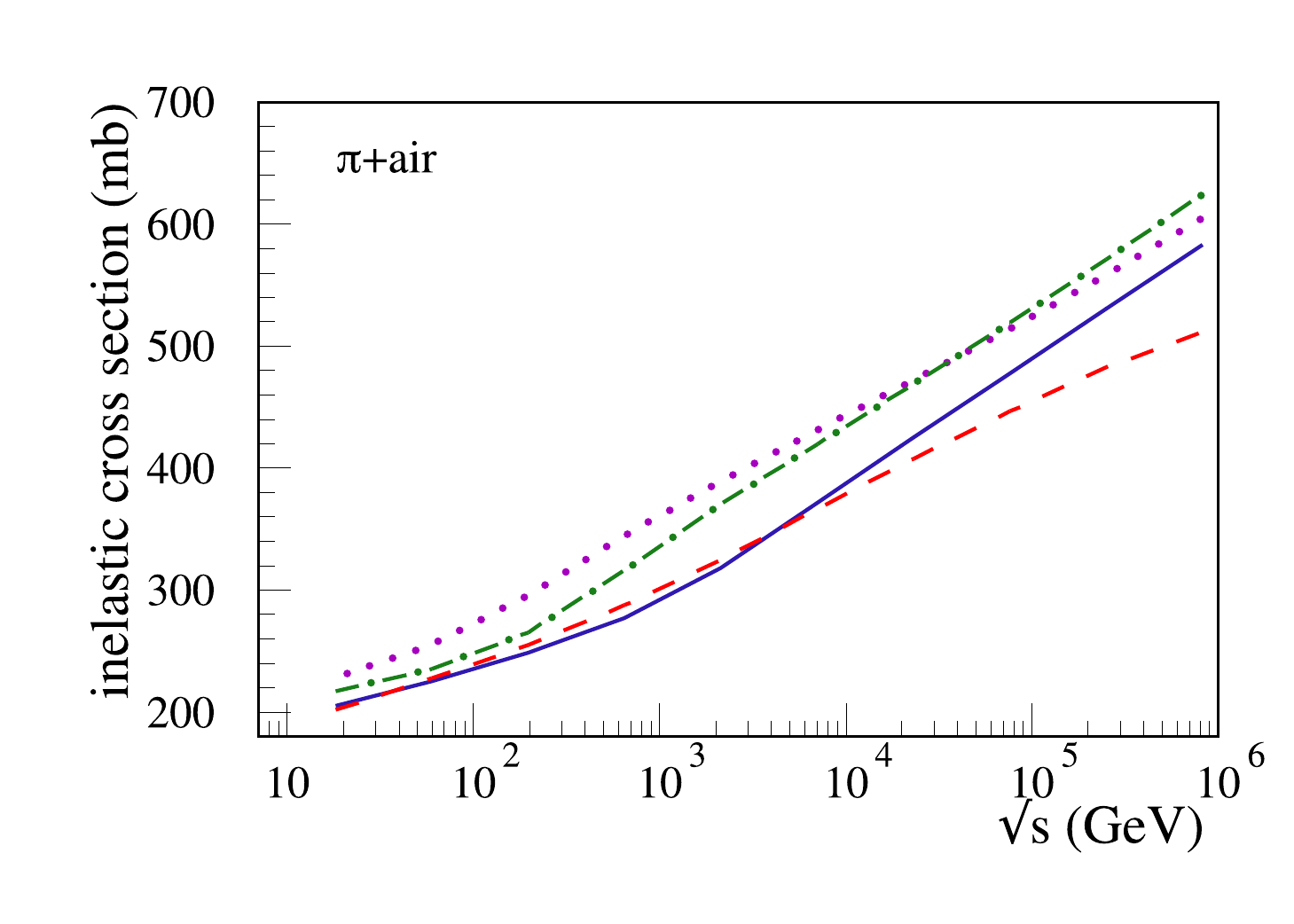}
\\
\includegraphics[width=0.32\textwidth,trim=10 0 10 0,clip]{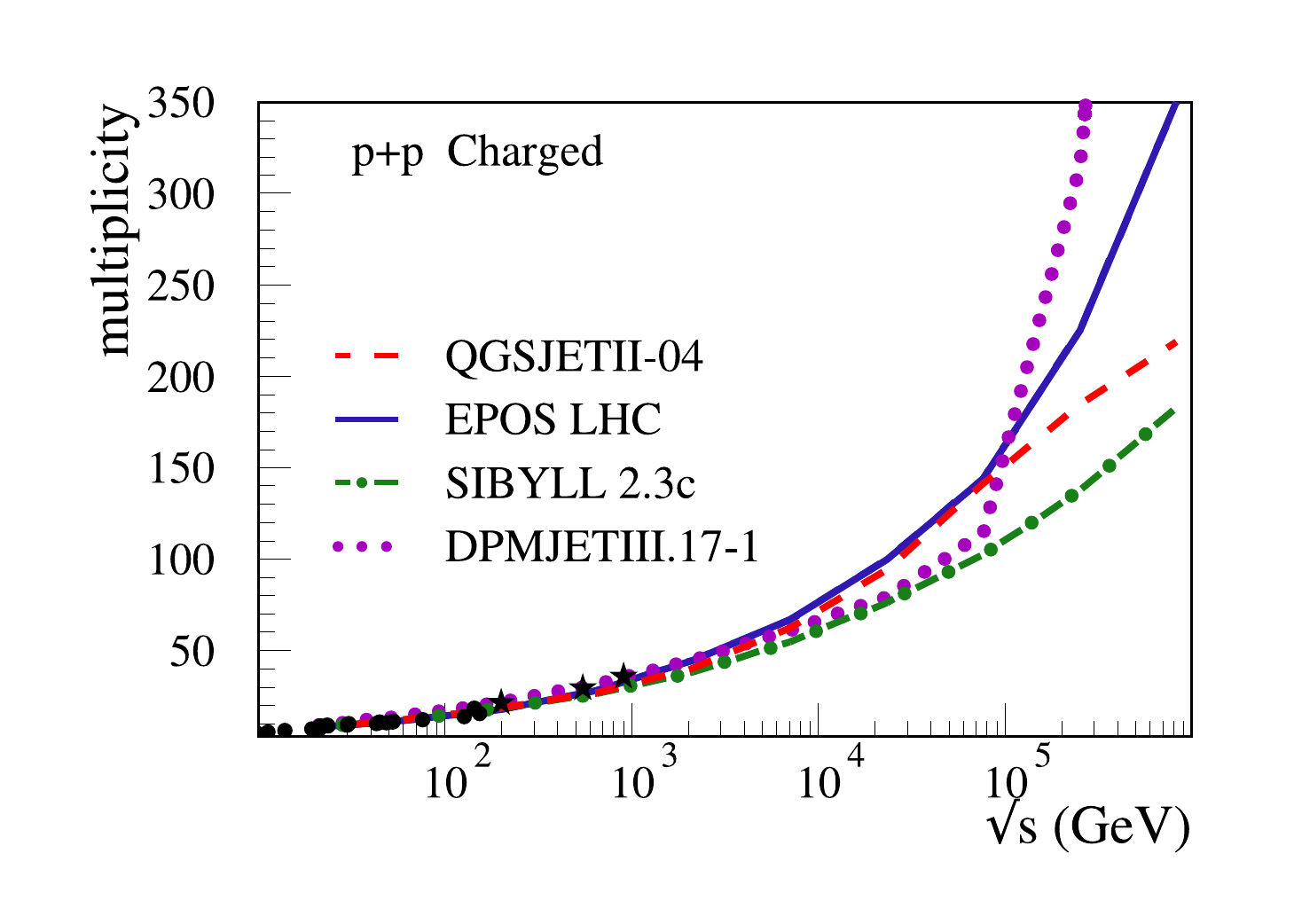}
\includegraphics[width=0.32\textwidth,trim=10 0 10 0,clip]{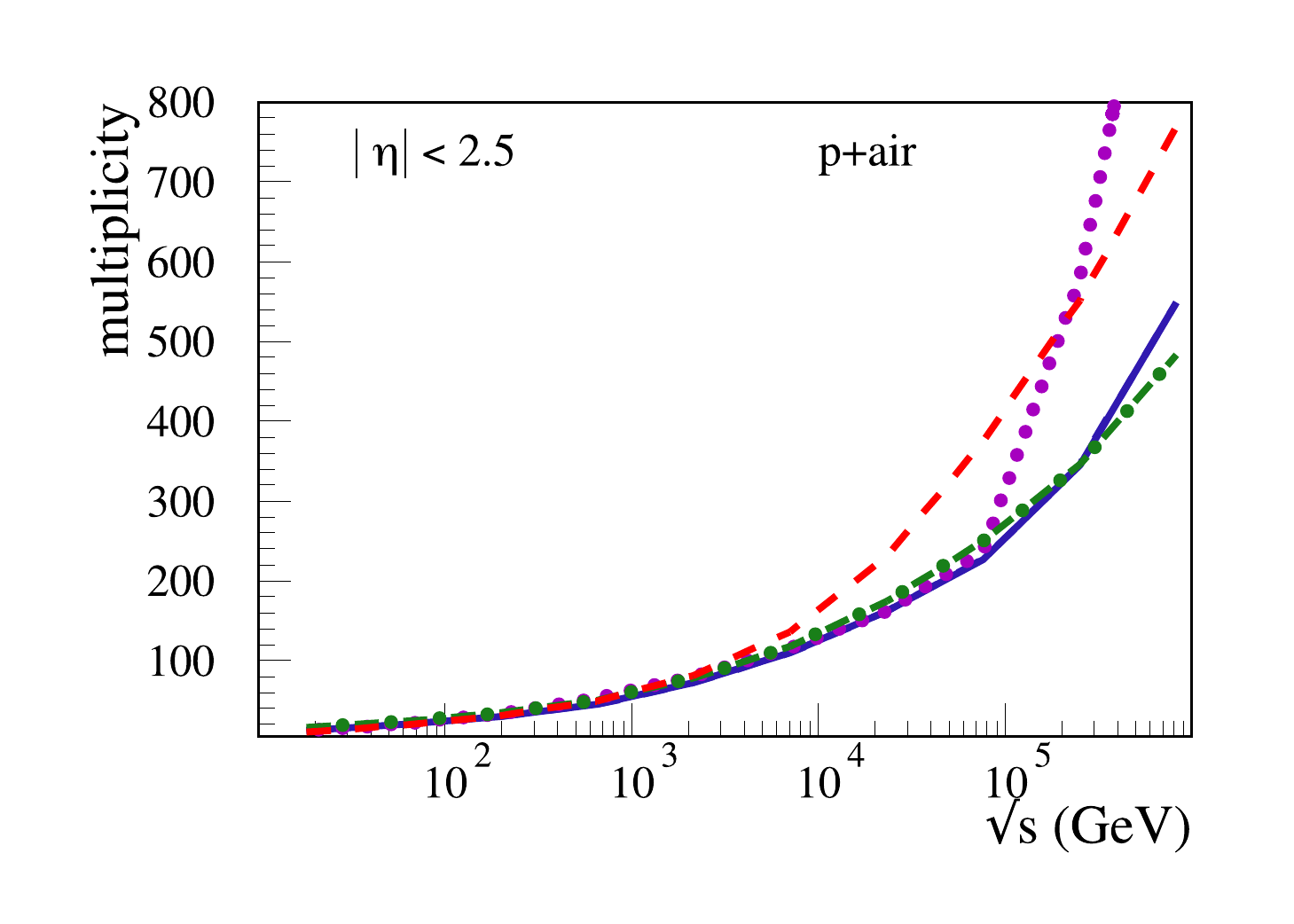}
\includegraphics[width=0.32\textwidth,trim=10 0 10 0,clip]{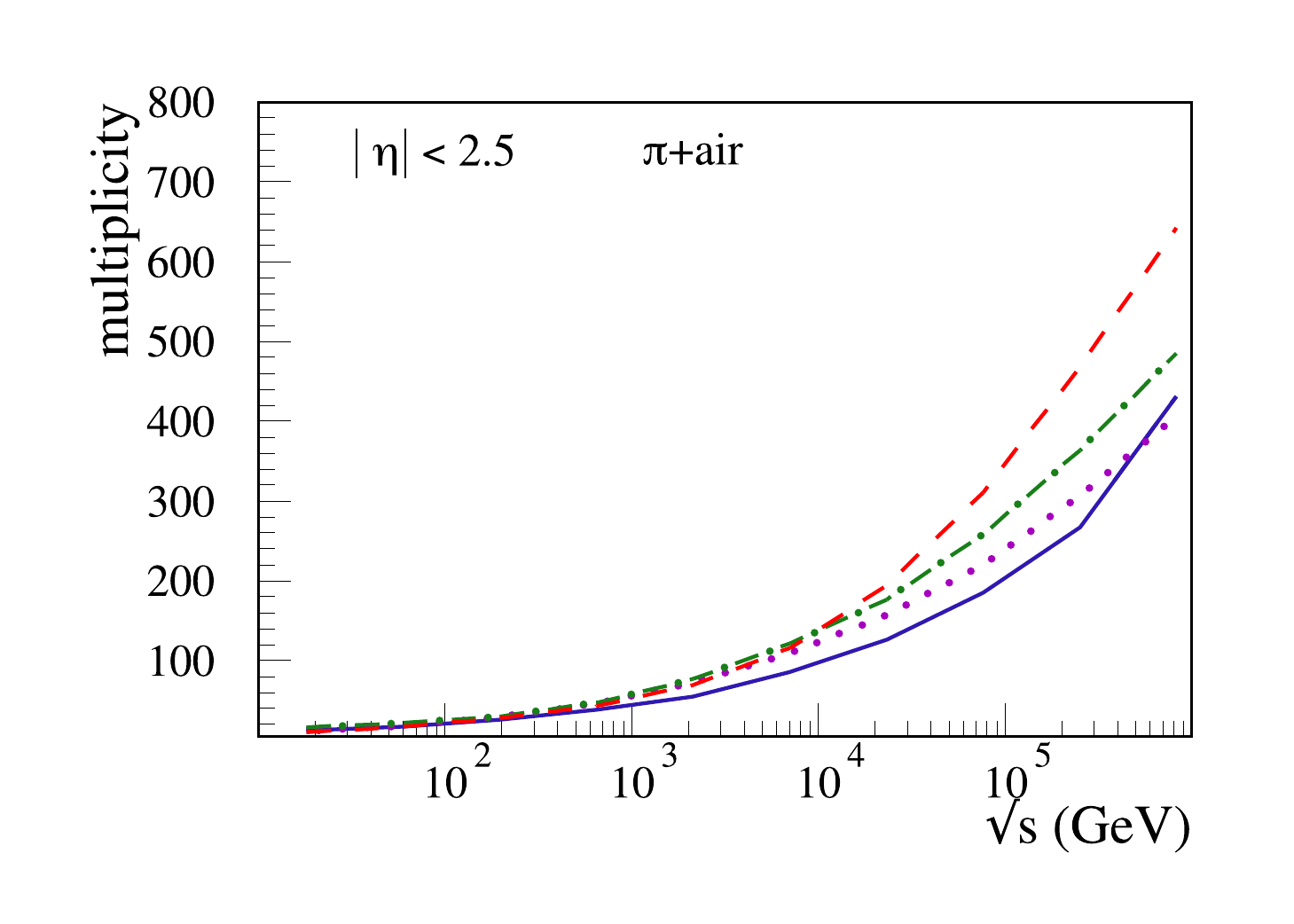}
\\
\includegraphics[width=0.32\textwidth,trim=10 0 10 0,clip]{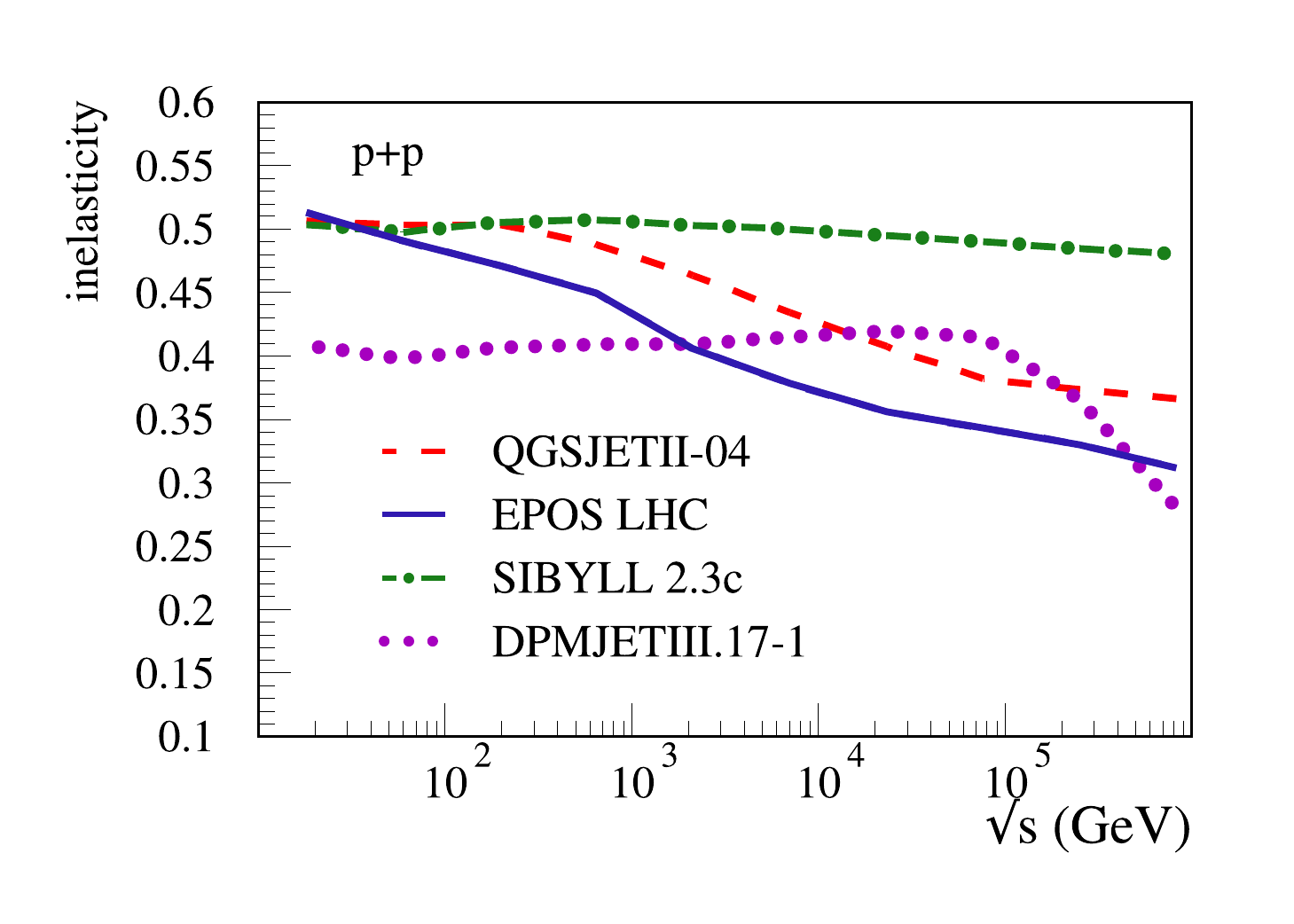}
\includegraphics[width=0.32\textwidth,trim=10 0 10 0,clip]{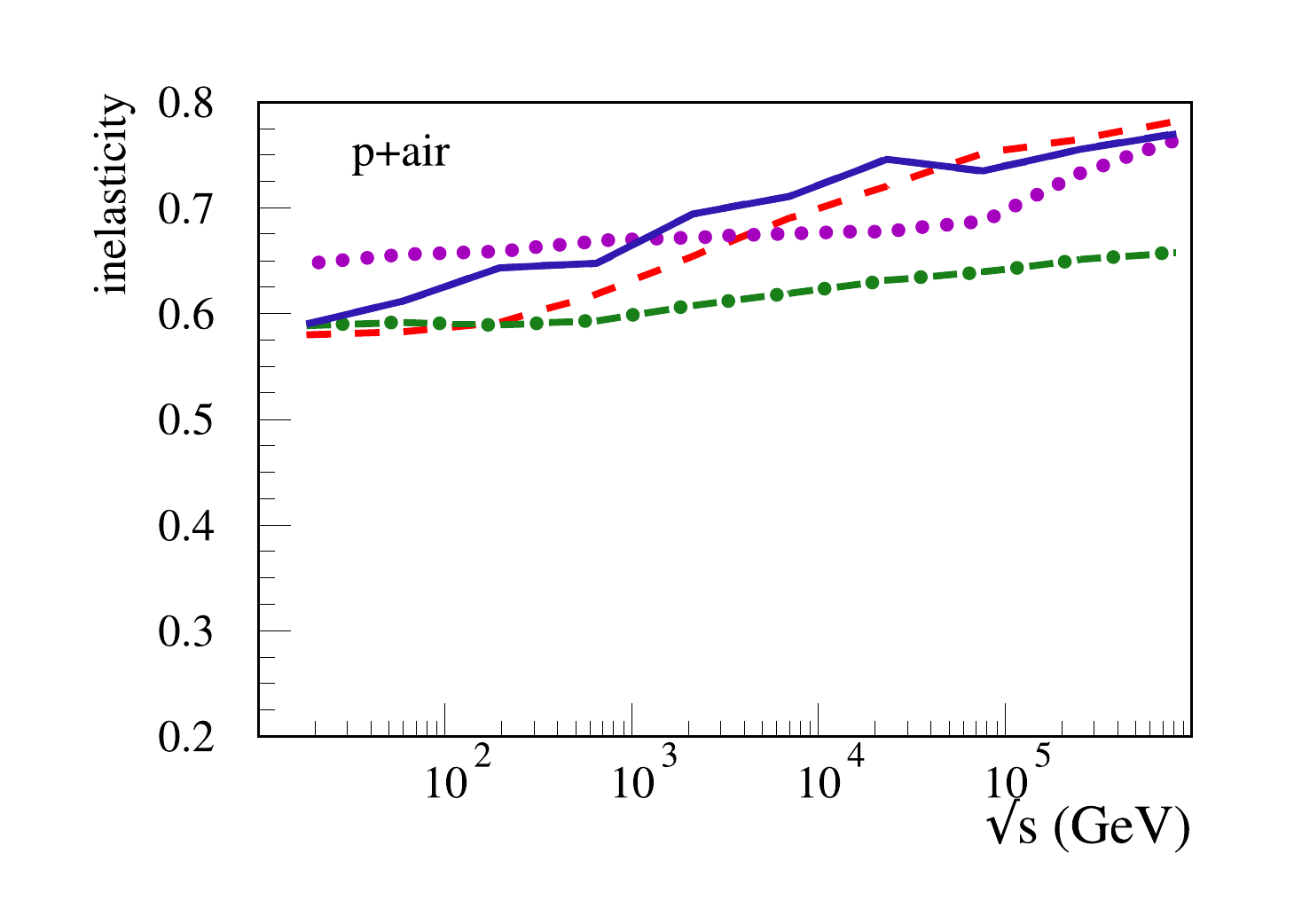}
\includegraphics[width=0.32\textwidth,trim=10 0 10 0,clip]{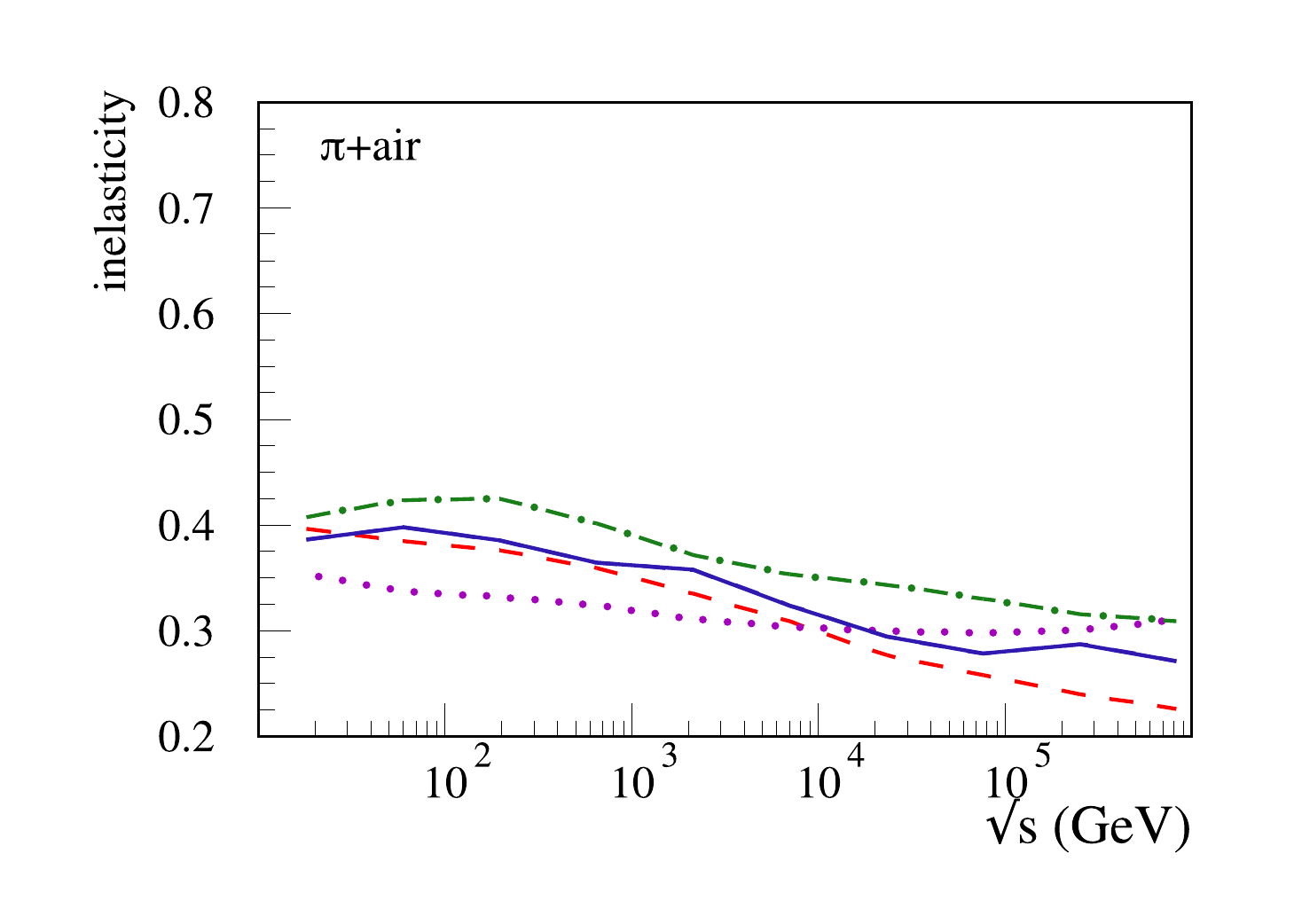}
\caption{Basic features of hadronic interactions as a function of $\sqrtsnn$ (images from \cite{Pierog:2017awp}). From left to right: \pp, \pX{air}, and \piX{air}. From top to bottom: total inelastic cross-section, hadron multiplicity at mid-rapidity $|\eta| < 2.5$, and inelasticity (complement of the energy fraction carried by the most energetic particle). The data points shown in the two upper left plots are taken from several experiments, see \cite{Pierog:2017awp} for details.}
\label{fig:basic_model_features}
\end{figure*}

Basic features of hadronic interactions which are important for air shower simulation are shown in \fg{basic_model_features}, compare with \sect{impact}. The inelastic cross-section has a high impact on the predicted value of the depth of shower maximum \xmax but not on the muon number \nmu. LHC measurements have constrained the inelastic \pp cross-section to very high precision and resolved the $1.9\sigma$ ambiguity in earlier Tevatron data \citep{Abe:1993xy,Amos:1991bp}. The measurement of the \pPb inelastic cross section \citep{CMS:2015nfb} at 5.02\tev is also important, since it validated the standard Glauber model to better than $\approx 10\,\%$. This had a noted impact on the systematic uncertainty of \xmax predictions. There is still a remaining uncertainty in the extrapolation of the inelastic cross-section from \pp to \pX{air}, which could be reduced with future data from \pO collisions. The \piX{air} cross-section remains weakly constrained.

The charged particle multiplicity has a significant impact on both \xmax and \nmu. The multiplicity in \pp collisions at mid-rapidity is well constrained by LHC data up to 13\tev to the level of a few percent; the most recent data is not shown in \fg{basic_model_features}. We emphasise that the multiplicity at mid-rapidity is important for tuning and a benchmark point for models, but has no direct impact on air shower development. Of direct interest is the model variance of the hadron multiplicity in the forward region $\eta \gg 2$, since forward-produced particles carry the highest energies, in particular if the system is boosted to the fixed target air shower frame, thus, they produce the largest sub-showers in the next step of the hadronic air shower cascades. This forward region can be constrained with LHC data up to $\eta < 6.4$ with TOTEM.

The inelasticity is the complement of the energy carried away by the most energetic particle in an inelastic collision. The inelasticity is an important quantity in air shower physics with a high impact on \xmax, but small impact on \nmu. It is not directly measurable at the LHC, but measurements related to inelasticity are the diffractive cross-section and the far-forward production cross-section of photons and neutrons.

\subsection{LHC experiments}
\label{sec:lhc-experiments}

\begin{figure}[tb!]
\centering
\includegraphics[width=\columnwidth]{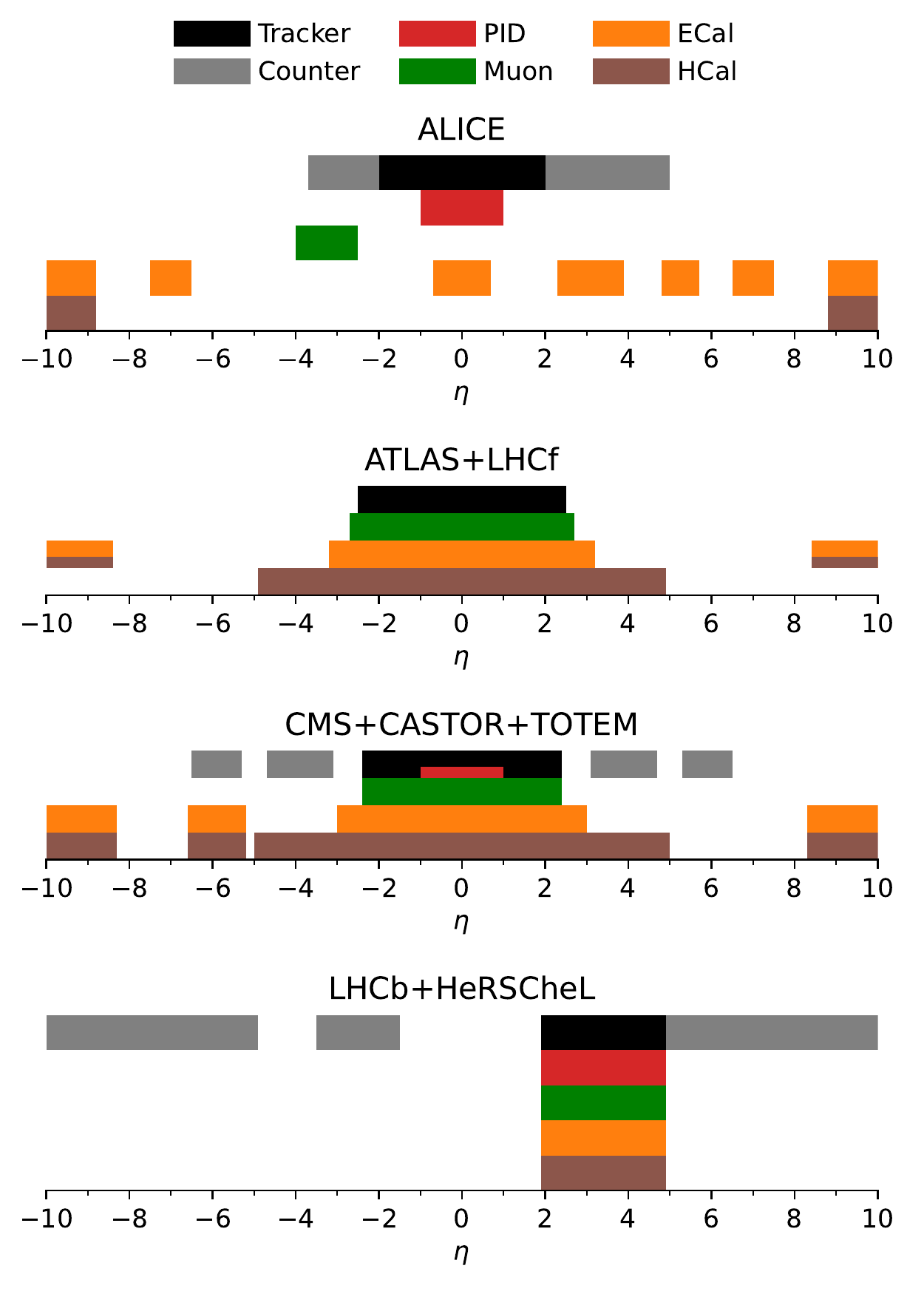}
\caption{Acceptances of the LHC experiments ALICE, ATLAS, CMS, LHCb, TOTEM, and LHCf, see \cite{Abelev:2014ffa,Aad:2008zzm,Aad:2009wy,Bayatian:2006nff,Suranyi:2021ssd,Alves:2008zz,Aaij:2014jba,Khachatryan:2020mpd,Adriani:2008zz,Anelli:2008zza} for details. In the legend, a \emph{tracker} follows individual particles in a magnetic field, while a \emph{counter} measures particle densities in $\eta$-intervals. \emph{Muon} refers to a special muon tracker, \emph{PID} refers to the ability to identify individual particles, \emph{ECal} and \emph{HCal} refer to electromagnetic and hadronic calorimeters, respectively. The inner part $|\eta| < 1$ of the CMS tracker is partially marked red to indicate its PID capabilities for particles with $\pt \lesssim 2\gevc$. Similarly, the acceptance $|\eta| > 8.4$ of the LHCf experiment around ATLAS is partially marked brown to indicate its capability to measure neutrons.}
\label{fig:lhc_detector_acceptance}
\end{figure}

The acceptances of the LHC experiments discussed here are compared in \fg{lhc_detector_acceptance}. We will briefly discuss the advantages of each experiment in regard to measurements for air shower physics in the following.

\begin{itemize}

\item \emph{ALICE} was designed for heavy-ion physics at the LHC \citep{Abelev:2014ffa}. Its strength is the high-resolution tracking system at mid-rapidity with excellent hadron particle identification (PID) capabilities. The tracking system can handle even central \PbPb collisions,
which can have more than 10000 tracks at $|\eta| < 0.5$. The electromagnetic calorimeter (ECal) partially covers the same region as the PID system, it captures photons and provides electron identification. ALICE has no hadronic calorimeter (HCal), since it would not provide significant additional information. A small single-arm muon system in the backward region capture muons from decays of charm and beauty. A system of charged-particle counters without tracking capabilities extends the acceptance of ALICE into the forward and backward region. A sophisticated system of zero-degree calorimeters (ZDCs) detects protons, neutrons, and photons emitted at small beam angles.

ALICE has provided a wealth of high-precision data of identified hadron spectra at mid-rapidity, differential production cross-sections as a function of $\eta$, $\pt$, and multiplicity, which have been one of the main sources for parameter tuning and validation of hadronic interaction models. The models are tightly constrained at mid-rapidity by these measurements; they fix the multiplicities of various light hadrons, the average charged particle multiplicity, the multiplicity spectrum. Various QGP-like effects at mid-rapidity have also been discovered, the strangeness enhancement is most important for air showers which may have a profound impact on the models beyond tuning. 
The backward muon system has been used to measure the differential production cross-sections of $D$ and $B$ mesons, which are inputs to constrain the heavy-quark parton density functions of the free proton and the bound nucleon. These in turn are used to predict the prompt atmospheric lepton flux which forms the principal background for high-energy neutrino observatories.

\item \emph{ATLAS} is a general purpose symmetric spectrometer with wide acceptance \citep{Aad:2008zzm,Aad:2009wy}. The systems ordered by increasing pseudorapidity are: the central tracker with $|\eta| < 2.5$, the muon system with $|\eta| < 2.7$, the ECal with $|\eta| < 3.2$, and the HCal with $|\eta| < 4.9$. Notable is the ALFA Roman Pot system to precisely measure the total and elastic cross-section via the optical theorem. ATLAS profits from combined measurements with the LHCf experiment, which is described further below.

Relevant strengths of ATLAS are its wide acceptance, the precisely measured luminosity, and its ability to combine measurements with LHCf. The collaboration has measured charged particle spectra in the central region as a function of $\pt$ with high precision at the level of 1\,\% at mid-rapidity and the forward energy flow, which constrain the hadron multiplicity in models. The wide acceptance was used to measure the cross-section for diffractive events, interactions in which the beam particles exchange no colour. The inelasticity of a hadronic interaction is sensitive to these interactions, which produce very few but  high-energy particles far forward. Thanks to its ALFA detectors, ATLAS provided the most precise measurements of the inelastic \pp cross-section so far, a direct input for hadronic models which extrapolate this to \pX{air} and higher energies.

\item \emph{CMS} is the second general purpose symmetric spectrometer at LHC \citep{Bayatian:2006nff}.  The systems ordered by increasing pseudorapidity are: the central tracker and the muon system with $|\eta|< 2.4$, the ECal with $|\eta| < 3$, and the HCal with $|\eta| < 5$. A system of zero-degree calorimeters \citep{Suranyi:2021ssd} detects neutral particles emitted at extremely small beam angles. Notable is also the single-arm very forward calorimeter system CASTOR \citep{Khachatryan:2020mpd}, which covers $-6.6 < \eta < -5.2$. Moreover, CMS profits from combined measurements with the TOTEM experiment, described further below.

Particular strengths of CMS are its wide acceptance covering parts of the forward phase-space, single-particle identification at central rapidities, as well as high-accuracy luminosity measurements. CMS has measured, for example, charged particle spectra and forward energy flow to constrain the hadron multiplicity in models, the cross-section for diffractive dissociation and total inelastic cross-sections. CMS has used particle identification capabilities of its tracker (via specific energy loss $\dd E/\dd x$, see \citet{CMS:2017eoq} for a recent analysis) for low momentum particles at mid-rapidity. The method is technically limited to small momenta $p \lesssim 2\gevc$, measurements are therefore restricted to the mid-rapidity region $|y| < 1$ and $\pt \lesssim 2\gevc$. This acceptance for  particle identification overlaps with ALICE.

Thanks to the CASTOR system, CMS has the widest nearly continuously covered acceptance in $\eta$ over 13.2 units of rapidity. CASTOR has only a single-arm, but that is not a limitation for most measurements; the recorded collisions are either symmetric or both beam configurations were recorded like in the case of \pPb. CASTOR is especially powerful for the measurement of diffractive events, low-$x$ parton physics, and forward energy flow. The ability of the CASTOR system to distinguish between hadronic and electromagnetic energy deposit is particularly useful. CASTOR is currently decommissioned as part of the LHC-wide upgrade for Run 3 due to changes in the beam pipe geometry and shielding at CMS.

\item \emph{LHCb} is a single-arm general purpose forward spectrometer designed for the study of flavour physics \citep{Alves:2008zz,Aaij:2014jba}. It features a tracking and muon system, an ECal and HCal system, and a ring-imaging Cherenkov detector for particle identification over an acceptance $1.9 < \eta < 4.9$. The tracker has acceptance also at $-3.5 < \eta < -1.5$, but without magnetic field. The single-arm geometry is not a caveat for most measurements, since the recorded collisions are either symmetric or both beam configurations are recorded. The LHCb experiment also features a set of forward scintillators around the beam pipe called HeRSCheL \citep{CarvalhoAkiba:2018paq} located upstream and downstream of the main experiment. The HeRSCheL system is designed as a veto for diffractive and inelastic events for the study of central-exclusive production $p+p \to p + X + p$. This system is decommissioned as part of the LHC-wide upgrade for Run~3. LHCb features an integrated gas target to study fixed-target collisions at $\sqrtsnn = 68 \text{ to } 110\gev$ with the LHC beams, which described in more detail in \sect{fixed-target}.

LHCb is the only fully instrumented general-purpose spectrometer in the forward region $1.9 < \eta < 4.9$ with tracking and particle identification. This range is of direct interest for air shower physics since it covers the onset of the forward region where most of the energy is directed to.
LHCb was designed to measure the forward production cross-sections of prompt $D$ and $B$ mesons. The time-dilation that forward produced particles experience together with the high resolution tracker allows one to distinguish prompt and non-prompt production and to tag $b$-decays. LHCb data uniquely allows one to constrain heavy-quark parton density functions of the free proton and the bound nucleon at very low $x$, which are used in turn to predict the prompt atmospheric lepton flux which forms the principal background for high-energy neutrino observatories. Further opportunities in the forward region are the study of identified charged hadron spectra, energy flow (separated by electromagnetic and hadronic flow), production cross-sections for photons and \piz, and production cross-sections for \vzero particles important for air shower physics like $\rho^0$, $K^0$, and $\Lambda$. The HeRSCheL system could be used to measure the diffractive cross-section and improve the LHCb measurement of the inelastic cross-section.
\end{itemize}

In addition to the four large general purpose LHC experiments there two specialised experiments with a high impact on air showers, LHCf and TOTEM. Both experiments can measure particles emitted in the very-forward region.
\begin{itemize}
\item \emph{LHCf} is a system of electromagnetic zero-degree sampling calorimeters located up- and downstream of the ATLAS experiment \citep{Adriani:2008zz}. It was designed to measure the production cross-sections of photons and \piz in the far forward region $|\eta| > 8.4$ with energy resolution of a few percent. It is also capable of measuring neutrons with an energy resolution of about $40\,\%$ and detection efficiency of better than $\approx50\,\%$ \citep{LHCf:2018gbv}. It is an independent experiment with its own trigger, but the time-stamps are partially synchronised with ATLAS so that some events may be merged offline. This has been used to study diffractive events, which had no activity in the ATLAS tracker.
A highlight of LHCf is the measurement of neutron spectra at zero-degree, which is beyond the design goals of the calorimeter \citep{LHCf:2015nel,LHCf:2018gbv}.

\item \emph{TOTEM} is a system of charged particle trackers surrounding the CMS experiment, designed to measure the total \pp cross-section at the TeV-scale to very high precision via the optical theorem and to study elastic scattering and diffractive dissociation \citep{Anelli:2008zza}. The experiment consists two large trackers T1 and T2 which are installed near the CMS experiment to provide particle tracking in the forward region $3.1 < |\eta| < 4.7$ and $5.3 < |\eta| < 6.5$. These trackers are used by TOTEM to tag inelastic collisions and offer the most forward data on charged particle densities at the LHC. Further small trackers are installed in special movable beam-pipe insertions (called Roman Pots) at several locations in large distances from the interaction point. The trackers inside the Roman Pots are moved within 1\si{mm} of the beam to measure elastic and diffractive scattering under extremely small angles. The TOTEM measurements of the inelastic cross-section of \pp collisions are among the most accurate to date. Also the shape of differential $|t|$-distribution for elastic events has been measured at unprecedented accuracy, illustrating higher-order correction to the assumed Gaussian shape of the proton-proton overlap region \citep{Antchev:2018edk}. Another unique highlight is the measurement of the Pomeron-photon interference region at extremely small scattering angles \citep{Antchev:2016vpy}, which was only done by TOTEM at LHC energies.
\end{itemize}

\subsection{Relevant LHC measurements}
\label{sec:lhc-measurements}

The quantities that need to be measured at the LHC to improve the simulation of air showers are primarily average properties of light-flavor hadron production at low momentum transfer in the realm of semi-hard- and soft-QCD. Rare events like the production of heavy and/or high-$\pt$ particles do not significantly influence air shower development and are not of prime interest, with the exception of forward heavy-flavour production, which is the source of the so-called prompt atmospheric neutrino flux. The most important measurements at LHC for air shower physics are
\begin{itemize}
\item the inelastic cross-section, $\sigma_\text{inel}$,
\item the hadron multiplicity over a wide range of rapidity,
\item the diffractive cross-section (cross-section for events with rapidity gaps to occur),
\item the composition and spectra of light hadrons: pions, kaons, protons, \kzero, and $\Lambda$,
\item the forward production cross-sections of certain key particles: $\piz$, $\eta$, $\rhoz$, the lightest $D$ mesons and $B$ mesons,
\item and the forward energy flow $\dd E/\dd \eta$.
\end{itemize}
The total energy flow is a more effective quantity, while the ratio of hadronic to electromagnetic energy flow is a direct measurement of the energy fraction $\alpha$ which remains in the hadronic cascade to which the number of low energy muons produced in an air shower is very sensitive to. The energy fraction $1-\alpha$ is lost via photon production in each step through decays of short-lived mesons. These are mostly \piz and $\eta$ mesons. In \piX{air} interactions, the forward production of \rhozero mesons plays an important role in increasing $\alpha$ \citep{Drescher:2007hc,Ostapchenko:2013pia,Engel:2019dsg} and generally the production of baryons \citep{1973ICRC....4.2467G,Pierog:2006qv,Engel:2019dsg}. The relevant process for \rhozero production in pion-projectile interactions cannot be directly measured with the LHC beams, but the ratio of \rhozero to \piz production in \pp and \pPb collisions is an important benchmark for the hadronic generators. Forward baryon production can be measured, and anti-baryons are excellent proxies for baryons newly formed in interactions as opposed to being created in the disintegration/fragmentation of the projectile.

The measurement of $D$ and $B$ is not motivated by the Muon Puzzle, but by the wish to accurately predict inclusive atmospheric lepton fluxes above 1\pev, which is the main background for high-energy neutrino observatories. Air showers provide the conventional component of this flux, the amount is directly linked to the muon production and thus the Muon Puzzle, since most muons and neutrinos are produced in pairs. At neutrino energies above 1\pev, the prompt component becomes dominant which arises from the production and decay of charm and beauty in the first interactions of a cosmic-ray nucleus with air. The production cross-sections for $D$ and $B$ mesons in \pp, \pPb, and \pO collisions constrain the corresponding nuclear parton density functions, which are key ingredients for the flux calculations.

\begin{table*}
\caption{Peer-reviewed papers on LHC measurements related to air shower physics for ALICE (A), ATLAS (T), CMS (C), LHCb (B), TOTEM (O), and LHCf (F). Collision energies are in $\sqrtsnn/\tev$.}
\label{tab:lhc-measurements}
\footnotesize
\setlength{\tabcolsep}{1.4pt}
\begin{tabular}{p{4.5cm}p{0.75cm}p{0.75cm}p{0.75cm}p{0.75cm}p{0.75cm}p{0.75cm}p{0.75cm}p{0.75cm}p{0.75cm}p{0.75cm}p{0.75cm}p{0.75cm}p{0.75cm}p{0.75cm}p{0.75cm}}
\hline
                                                             & pp 0.9         & pp 2.36     & pp 2.76    & pp 5.02     & pp 7\phantom{XX}                            & pp 8\phantom{XX}   & pp 13\phantom{XX}   & pPb 5.02             & pPb 8.16   & PbPb 2.76   & PbPb 5.02   & XeXe 5.44   & pHe 0.09   & pHe 0.11   & pAr 0.11   \\
\hline
 Inelastic cross-section                                     & A1             &             & A1         &             & A1 T1 C1 B1 O1                              & T2                 & T3 C2 B2 O2         & C3                   &            &             &             &             &            &            &            \\
 Charged particle spectra $|\eta| < 2.5$                     & A2 T4 C4 C5 C6 & A2 T4 C4 C6 & A2 A3      & A3 A4       & A2 A4 T4 C4 C5 C7                           & A2 T5 C8           & A4 T6 T7 C9 C10     & A3 T8 T9 C11 C12 C13 & C13        & A3 C14      & A3 C12      & C15 C16     &            &            &            \\
 Charged particles spectra $\eta > 2.5$                      & A5             &             &            & B3          & A5 B4                                       & A5 O3              & B5                  & B3                   &            &             &             & A6          &            &            &            \\
 Forward energy flow $\eta > 2$                              & C17            &             &            &             & C17 B6                                      &                    & C18 C19 C20         &                      &            &             &             &             &            &            &            \\
 Identified hadron spectra                                   & A7 C21 B7      &             & C21        &             & A8 C21 B7                                   &                    & C22                 & A9 A10 C23           &            & A11 A12     &             &             &            & B8         &            \\
 Inclusive photon spectra, \newline neutral pions and $\eta$ & A13 A14 F1     &             & A15 A14 F2 &             & A13 A14 F3 F2                               &                    & F4                  & A16 F5               &            & A15 A17     &             &             &            &            &            \\
 Forward neutron spectra                                     &                &             &            &             & F6                                          &                    & F7 F8               &                      &            &             &             &             &            &            &            \\
 Strangeness                                                 & T10 C24        &             &            & C25         & A18 T10 T11 C24 C26 B9                      &                    &                     & A18 C26 C25          &            & A18 C26     &             &             &            &            &            \\
 Diffractive cross-section                                   & A1             &             & A1         &             & A1 C27 O4                                   &                    &                     &                      &            &             &             &             &            &            &            \\
 D and B meson production                                    &                &             & A19 A20    & A21 A22 B10 & A23 A24 T12 T13 C28 C29 B11 B12 B13 B14 B15 &                    & C30 B16 B15 B17     & A25 A26 B18          & B19        & A20 A27     &             &             & B20        &            & B20        \\
\hline
\end{tabular}
\tiny
A1: \cite{ALICE:2012fjm}, A2: \cite{ALICE:2015olq}, A3: \cite{ALICE:2018vuu}, A4: \cite{ALICE:2020swj}, A5: \cite{ALICE:2017pcy}, A6: \cite{ALICE:2018cpu}, A7: \cite{ALICE:2011gmo}, A8: \cite{ALICE:2015ial}, A9: \cite{ALICE:2013wgn}, A10: \cite{ALICE:2016dei}, A11: \cite{ALICE:2012ovd}, A12: \cite{ALICE:2013mez}, A13: \cite{ALICE:2012wos}, A14: \cite{ALICE:2014rma}, A15: \cite{ALICE:2014hpa}, A16: \cite{ALICE:2018vhm}, A17: \cite{ALICE:2018mdl}, A18: \cite{ALICE:2016fzo}, A19: \cite{ALICE:2012inj}, A20: \cite{ALICE:2012sxy}, A21: \cite{ALICE:2019nxm}, A22: \cite{ALICE:2021mgk}, A23: \cite{ALICE:2015ikl}, A24: \cite{ALICE:2017olh}, A25: \cite{ALICE:2014xjz}, A26: \cite{ALICE:2019fhe}, A27: \cite{ALICE:2015ccw}, T1: \cite{ATLAS:2011zrx}, T2: \cite{ATLAS:2016ikn}, T3: \cite{ATLAS:2016ygv}, T4: \cite{ATLAS:2010jvh}, T5: \cite{ATLAS:2016qux}, T6: \cite{ATLAS:2016zba}, T7: \cite{ATLAS:2016zkp}, T8: \cite{ATLAS:2015hkr}, T9: \cite{ATLAS:2016xpn}, T10: \cite{ATLAS:2011xhu}, T11: \cite{ATLAS:2014pju}, T12: \cite{ATLAS:2013cia}, T13: \cite{ATLAS:2015igt}, C1: \cite{CMS:2012gek}, C2: \cite{CMS:2018mlc}, C3: \cite{CMS:2015nfb}, C4: \cite{CMS:2010qvf}, C5: \cite{CMS:2011mry}, C6: \cite{CMS:2010wcx}, C7: \cite{CMS:2010tjh}, C8: \cite{CMS:2014kix}, C9: \cite{CMS:2018nhd}, C10: \cite{CMS:2015zrm}, C11: \cite{CMS:2015ved}, C12: \cite{CMS:2016xef}, C13: \cite{CMS:2017shj}, C14: \cite{CMS:2011aqh}, C15: \cite{CMS:2018yyx}, C16: \cite{CMS:2019gzk}, C17: \cite{CMS:2011xjg}, C18: \cite{CMS:2017dou}, C19: \cite{CMS:2018lqt}, C20: \cite{CMS:2019kap}, C21: \cite{CMS:2012xvn}, C22: \cite{CMS:2017eoq}, C23: \cite{CMS:2013pdl}, C24: \cite{CMS:2011jlm}, C25: \cite{CMS:2019isl}, C26: \cite{CMS:2016zzh}, C27: \cite{CMS:2015inp}, C28: \cite{CMS:2011pdu}, C29: \cite{CMS:2011oft}, C30: \cite{CMS:2016plw}, B1: \cite{LHCb:2014llk}, B2: \cite{LHCb:2018ehw}, B3: \cite{LHCb:2021vww}, B4: \cite{LHCb:2014wmv}, B5: \cite{LHCb:2021abm}, B6: \cite{LHCb:2012gpm}, B7: \cite{LHCb:2012lfk}, B8: \cite{LHCb:2018ygc}, B9: \cite{LHCb:2011ijs}, B10: \cite{LHCb:2016ikn}, B11: \cite{LHCb:2010wqx}, B12: \cite{LHCb:2011leg}, B13: \cite{LHCb:2013xam}, B14: \cite{LHCb:2013vjr}, B15: \cite{LHCb:2016qpe}, B16: \cite{LHCb:2015swx}, B17: \cite{LHCb:2019fns}, B18: \cite{LHCb:2017yua}, B19: \cite{LHCb:2019avm}, B20: \cite{LHCb:2018jry}, O1: \cite{TOTEM:2013vij}, O2: \cite{TOTEM:2017asr}, O3: \cite{CMS:2014kix}, O4: \cite{TOTEM:2013pio}, F1: \cite{LHCf:2012stt}, F2: \cite{LHCf:2015rcj}, F3: \cite{LHCf:2012mtr}, F4: \cite{LHCf:2017fnw}, F5: \cite{LHCf:2014gqm}, F6: \cite{LHCf:2015nel}, F7: \cite{LHCf:2018gbv}, F8: \cite{LHCf:2020hjf}
\end{table*}

On the one hand, the published data already constrain well the inelastic cross-sections, the hadron multiplicity, the diffractive cross-section, and the production cross-sections for $D$ and $B$ mesons, also to some part in the forward region. On the other hand, we identify a lack of data on identified hadron spectra and on strangeness production in the forward region especially in \pX{A} collisions, which are expected to have a high impact on the Muon Puzzle. The relevant measurements that have been published in refereed journals so far are listed in \tb{lhc-measurements} and are further discussed in the following.
\begin{itemize}
\item \emph{Inelastic and elastic cross-sections}.
Two complementary techniques are used to measure the inelastic cross-section. One is based on counting empty events; the probability to observe empty events decreases as the inelastic cross-section increases. It requires wide acceptance, a precise measurement of the beam luminosity, and theory input to extrapolate the measured fiducial cross-section to the full inelastic cross-section. The other technique is employed by TOTEM and ATLAS/ALPHA and is based on observing elastic scattering down to very low momentum transfer. From the forward amplitude of elastic scattering, the total cross section can be calculated using the optical theorem, which is based on conservation of probability. This also requires some small model-dependent phase-space corrections as well as an independent measurement of luminosity. However, when inelastic event counting is combined with the elastic measurement, it is possible to perform a luminosity-independent cross section measurement. This is a particular advantage of the experimental setup of TOTEM and also the ATLAS/ALPHA system. Furthermore, from elastic scattering further important physics parameters can be extracted. The most important one is the shape of the differential elastic cross section in $|t|$, which corresponds to the size and shape of the actual collision region in impact parameter space. For small $|t|$ the shape of this distribution is exponential to very good approximation with small higher-order corrections~\citep{TOTEM:2018hki}. Another quantity that is measured by TOTEM in dedicated data taking and special analysis is the ratio of the real to imaginary part of the elastic forward scattering amplitude. This can be extracted from very high-$\beta^*$ data by observing the Coulomb-nuclear interference region. These parameters are further important ingredients in modelling of cross section, e.g. via Glauber or Gribov-Regge theory.

Measurements of the inelastic cross-section of \pp collisions have been performed from 0.9 to 13\tev by multiple experiments. The most precise measurements have been obtained by TOTEM. CMS has also measured the inelastic cross-section of \pPb collisions at 5.02\tev, which is very interesting to test predictions based on Glauber models directly. There are currently no published measurements for \PbPb and \XeXe.

These \pp measurements are very precise, and significantly improve the extrapolation toward the highest energies as they occur in air showers. This increase in accuracy had a significant impact on the predictions of the depth of shower maximum \xmax in air shower simulations, but it is less relevant for the number of muons \nmu produced in the shower.

\item \emph{Diffractive cross-sections and rapidity gaps}. Diffractive collisions are a subclass of inelastic events (about 20\,\% of the inelastic cross-section at the TeV scale) in which no colour is exchanged between the beam particles, but momentum is transferred and new particles are produced. The experimental signature of a diffractive event are large rapidity gaps, regions in rapidity devoid of particles. Rapidity gaps were first studied at HERA \citep{ZEUS:1993vio,H1:1994ahk}. Important for air showers are single diffractive events, in which only one of the beam particles is dissociated, and double diffractive event, in which both are dissociated. The measurement is ideally performed with a detector with wide acceptance and very-forward detectors to detect the dissociation.

The differential rapidity-gap cross section measurement by ATLAS and CMS are particularly useful. In this case the diffractive cross-section is given as a function of the size of the gap in pseudorapidity, which allows one to directly study the transition from inelastic to diffractive event topologies. This new type of measurement had an impact on the development of EPOS and revealed tension to air shower data that could only be overcome by an individual tuning of pion-projectiles in EPOS~\citep{Pierog:2015ifw}. This underlined the importance of pion-projectiles in air showers, and the related potential modelling uncertainties.

Diffractive cross-sections were measured in \pp collisions from 0.9 to 7\tev by ALICE, CMS, and TOTEM. Measurements in \pPb collisions are under study by CMS \citep{Sosnov:2020gaq} and would help to understand how diffractive dissociation is modified in an ion collision.

\item \emph{Charged particle spectra}. Charged particle spectra can be measured with particle tracking as a function of $\eta$ and $\pt$ or without tracking as a function of $\eta$ only. Both types of measurements are of interest for air shower physics, the loss of the $\pt$ information is acceptable at the TeV scale, since most of the lateral spread of particles perpendicular to the air shower axis is generated at lower energies. At high energies, the boost factor $1/\gamma$ suppresses the transverse spread of the particles.

Measurements at mid-rapidity $|\eta| < 2.5$ have been performed in \pp collisions from 0.9 to 13\tev, in \pPb from 2.76 to 8.16\tev, in \PbPb from 2.76 to 5.02\tev, and in \XeXe collisions at 5.44\tev. Forward measurements that cover the relevant $\eta$-region for air showers have also been performed up to $\eta = 6.4$ in \pp collisions from 0.9 to 8\tev and in \XeXe collisions at 5.44\tev by ALICE, TOTEM, and LHCb. There are currently no published forward measurements for \pPb and \PbPb. The forward measurements are especially valuable for air shower simulations, since the evolution of the hadronic cascade is dominated by forward-produced hadrons.

\item \emph{Forward energy flow}. The energy deposits of particles in a electromagnetic or hadronic calorimeter are measured instead of tracking or counting charged particles. The advantage is that also  neutral particles are captured. The disadvantage is the loss of the $\pt$ information, but as mentioned previously this is not a major concern at the TeV scale. Calorimeters are preferred in the forward region over charged-particle counters or even trackers, since they better withstand high radiation levels closer to the beam. The CASTOR calorimeter of the CMS experiment has the most forward acceptance up to $\eta = 6.6$, apart from zero-degree calorimeters (ZDCs), which cover $\eta > 8.2$ but can only measure neutral particles. A particular ZDC is the LHCf experiment, which is specifically designed for tuning models employed in air shower simulations.

Forward measurements up to $|\eta| < 6.6$ were performed in \pp collisions from 0.9 to 13\tev by CMS and LHCb. And by CMS also in \pPb collisions at 5.02\tev \citep{Sirunyan:2018nqr} as well as in \PbPb collisions at 2.76\tev \citep{Chatrchyan:2012mb}.

Of particular interest for air shower physics is also the ratio $R$ of electromagnetic and hadronic energy flows, which so far has been only measured once in \pp collisions at 13\tev by CMS using the CASTOR calorimeter. Of great interest would be energy flow ratio in \pPb collisions as a function of charged particle multiplicity.

\item \emph{Identified hadron spectra.} Experiment with a tracker and a particle identification system (based on specific energy loss, time-of-flight, Cherenkov cone angle, etc.) can discriminate individual hadron species track-by-track. Measured are then either yield ratios or identified hadron spectra, the product of relative yields and charged particle spectra. Only ALICE and LHCb have been specifically designed with particle identification capabilities, but CMS has successfully used the energy deposits in its tracker to identify particles with $p < 2\gevc$.

Pions, kaons, and protons were measured at mid-rapidity $|\eta| < 1$ in \pp, \pPb, and \PbPb collisions from 0.9 to 13\tev by CMS and ALICE. Measurements in the forward region $2.5 < \eta < 4.5$ were performed in \pp collisions at 0.9 and 7\tev by LHCb. The anti-proton flux was measured in \pX{He} collisions at $\sqrtsnn = 110\gev$ with LHCb in fixed-target mode. We emphasise that there are currently no forward data available for \pPb collisions. The hadron composition in the forward region and its potential nuclear modification plays an important role for the muon production in an air shower.

\item \emph{Inclusive photon, neutral pion, and $\eta$ spectra}. The inclusive flux of photons, produced dominantly from the decay of \piz and $\eta$ mesons, is the electromagnetic complement to the hadronic energy produced in a collision. The muon production in air showers is very sensitive to this ratio and therefore measurements are of great interest in regard to the Muon Puzzle. The spectra of neutral and charged pions are linked to first order by isospin symmetry, which means that measurements of charged pions also constrain neutral pions, but the yields are not exactly identical because of additional effects, for example, pions produced in decays of short-lived particles.

ALICE has measured inclusive photon, neutral pion, and $\eta$ production in \pp collisions from 0.9 to 7\tev, in \pPb at 5.02\tev and in \PbPb collisions at 2.76\tev. The LHCf experiment has measured photon and neutral pion spectra in the very forward rapidity range $8.4 < \eta$ in \pp collisions from 0.9 to 13\tev, and in \pPb at 5.02\tev. The LHCf measurements are particularly important for the Muon Puzzle, since an unexpectedly low flux of neutral pions in the very forward region would have been a sign of new physics at the LHC energy scale with a direct impact on air showers. Since large deviations were not observed in the very forward region nor at mid-rapidity, the search is now focused on a smaller modification that applies to a wide rapidity range.

\item \emph{Very forward neutron spectra and inelasticity.} The LHCf experiment has measured forward neutron spectra in \pp collisions at 7 and 13\tev. While the actual inelasticity remains a more theoretical concept not directly accessible by measurements, an analysis based on the forward neutrons can be  performed \citep{LHCf:2020hjf} exploiting the fact that the forward neutron is often the most energetic particle, and subsequently the inelasticity is the complement of the energy fraction carried by the neutrons. The inelasticity as a conceptual parameter has a large impact on the depth of shower maximum \xmax, while the number of produced muons in the shower depends only weakly on it. The ratio of the very forward neutral pion and neutron spectra is an important parameter for muon production in air showers and are constrained by these measurements. An analysis of \pPb collisions at 5.02\tev with LHCf is needed to further complete the picture.

\item \emph{Strangeness production}. The enhanced strangeness production in events with high multiplicities is understood as a high-density QCD effect (related to QGP, or colour glass phases, etc.). Measurements by ALICE suggest that strangeness enhancement is to a large degree independent of the collision system or the collision energy and even appears in small collision systems like \pp collisions -- it mainly depends on hadron multiplicity. This is a potential hint on how QCD physics could be extended to resolve the Muon Puzzle in air showers, as described in \sect{solutions}. Experimentally, strangeness production is measured by reconstructing decays of strange mesons and baryons, $K_S^0$, $\phi$, $\Lambda$, $\Xi^-$, $\Omega^-$ and their antiparticles. This requires a high-precision tracker, since the momenta of the decay products need to be known and the impact parameter of candidate tracks to reduce combinatorial background. Particle identification is not required but beneficial to reduce combinatorial background. Multi-strange hadrons are particularly sensitive probes for a strangeness enhancement.

ALICE has measured a strangeness enhancement at mid-rapidity $|y| < 1$ as a function of multiplicity at mid-rapidity $|y|<0.5$ using data from \pp collisions at 7\tev, \pPb at 5.02\tev, and \PbPb at 2\tev.  CMS has measured strangeness in \pp, \pPb, and \PbPb collisions up to $|y|<2.4$. Earlier measurements were performed without splitting the data into multiplicity classes. ATLAS has measured \kshort and $\Lambda$ in \pp collisions at 0.9 and 7\tev up to $|y| < 2.5$. LHCb has measured $\phi$ production in \pp collisions at 7\tev in the range $2.44 < y < 4.06$.

The strangeness enhancement was observed at mid-rapidity, but it has not been experimentally demonstrated in the forward region. In order to solve the Muon Puzzle in air showers, the strangeness enhancement has to be present also in the forward region and it has to be sufficiently large to make an impact on the shower development. In order to search for this, high-precision measurements of strangeness production as a function of the particle multiplicity need to be performed in the forward region $y > 2.5$ in \pp and \pPb, and optionally \PbPb collisions with LHCb. This will address the key question whether the universal strangeness enhancement observed at mid-rapidity is also present in the forward region.

\item \emph{D and B meson production cross-section}. The cross-sections for $D$ and $B$ meson production have no impact on air shower development, but their decays give rise to the prompt component of the atmospheric neutrino flux, which dominates the overall flux above about 10\pev, see \fg{inclusive-flux}. The production via charm decays is dominant. The neutrino yield from decays of $D$ and $B$ mesons is similar, but the production cross-section for $c\bar c$ is an order of magnitude higher than for $b\bar b$ \citep{Martin:2003us,Bhattacharya:2016jce}. $B$ decays contribute less than 10\,\% of the prompt component. The production cross-sections for $D$ and $B$ mesons are also important inputs for the (nuclear) parton density density functions (PDF) at high momentum transfer \citep{Ball:2014uwa,Hou:2019efy,Zenaiev:2019ktw}. The neutrino flux calculations are sensitive to the PDFs at small $x \lesssim 10^{-5}$, which can only be constrained by forward measurements at large $\eta$.

Production cross-sections for $D$ and $B$ mesons have been measured at mid-rapidity in \pp collisions from 2.76\gev to 13\tev, in \pPb at 5.02 and 8.16\tev, and in \PbPb at 2.76\tev with ALICE, ATLAS, and CMS. Forward measurements with LHCb have been performed in \pp collisions from 7 to 13\tev and in \pPb collisions at 5.02 and 8.16\tev. $D$ mesons were also measured in fixed target mode with LHCb for \pX{He} collisions at $\sqrtsnn = 86\gev$ and \pX{Ar} collisions at $\sqrtsnn = 110\gev$.
\end{itemize}

\subsection{Fixed target experiments}
\label{sec:fixed-target}

\begin{figure}[tb!]
\centering
\includegraphics[width=\columnwidth]{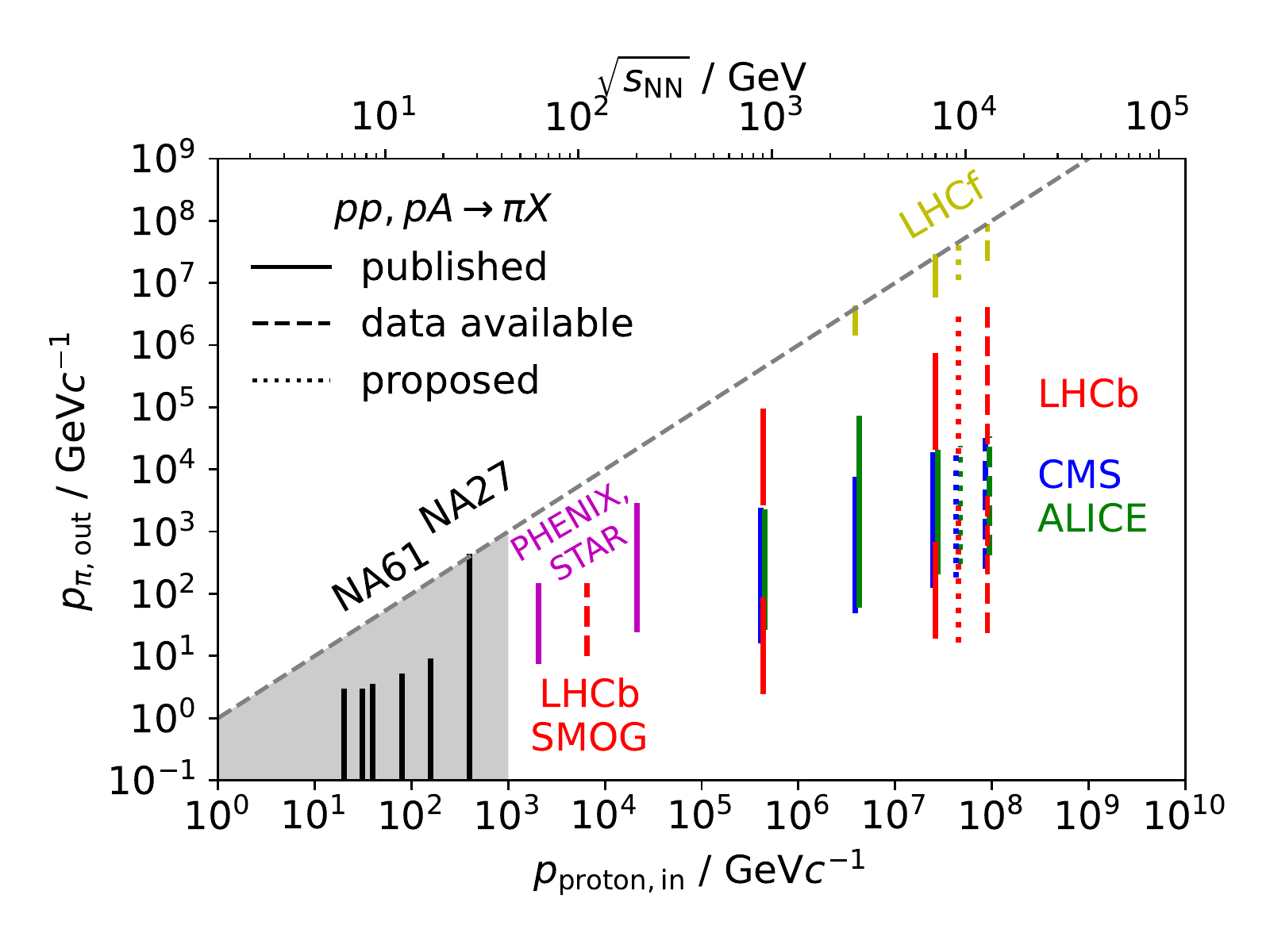}
\caption{Overview of data on pion spectra in \pp, \pPb collisions as a function of the momentum of the incoming proton \citep{Adare:2011vy,Agakishiev:2011dc,CMS:2012xvn,Abelev:2014laa,Abgrall:2013qoa,AguilarBenitez:1991yy,ALICE:2015ial,LHCb:2012lfk,LHCf:2015rcj,ALICE:2011gmo}. The proposed \pO collisions are indicated. In case of LHCb, only hadron yield ratios were measured and not full spectra.}
\label{fig:pion_spectra_data}
\end{figure}

\begin{figure*}
 \includegraphics[width=0.49\linewidth]{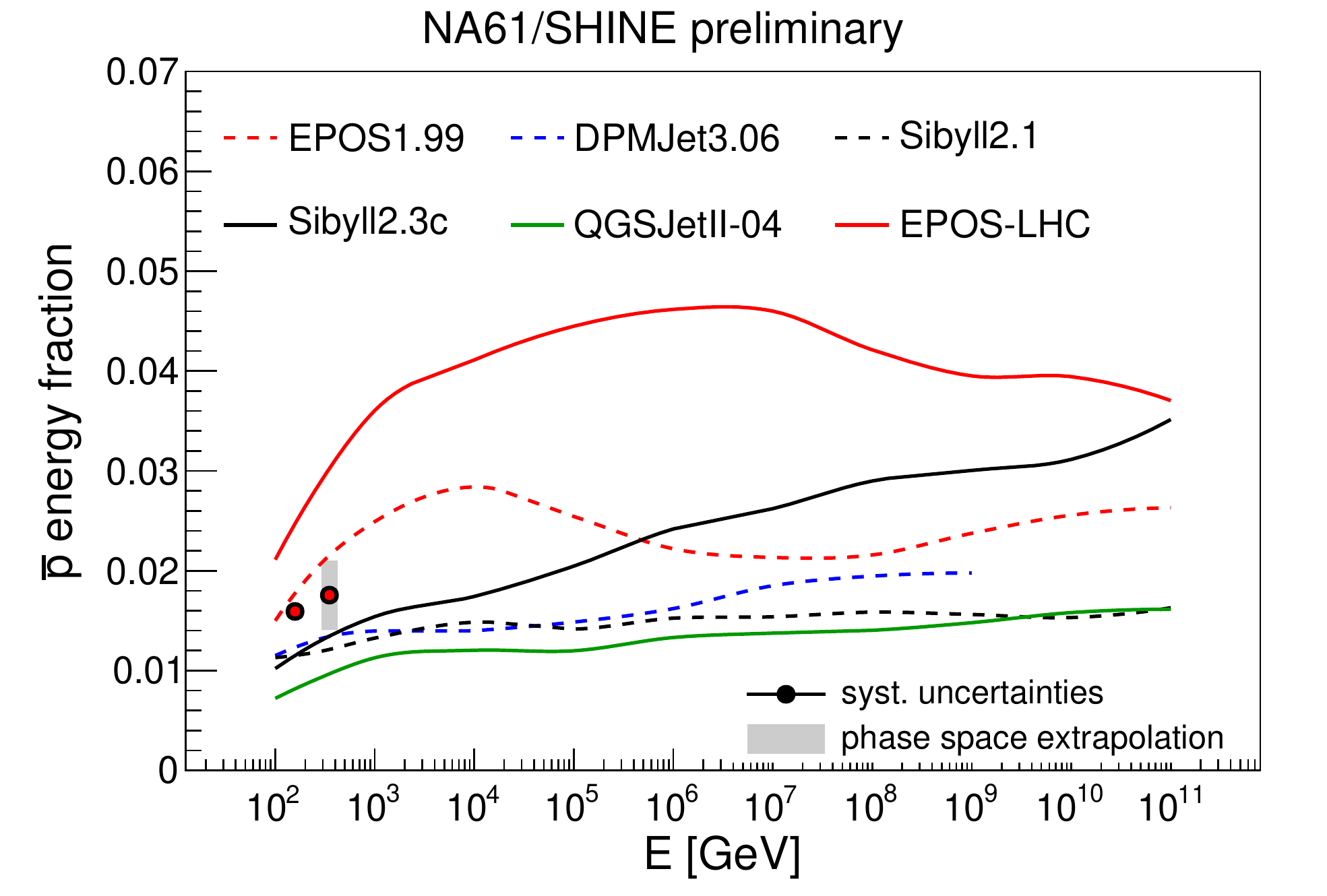}
 \includegraphics[width=0.49\linewidth]{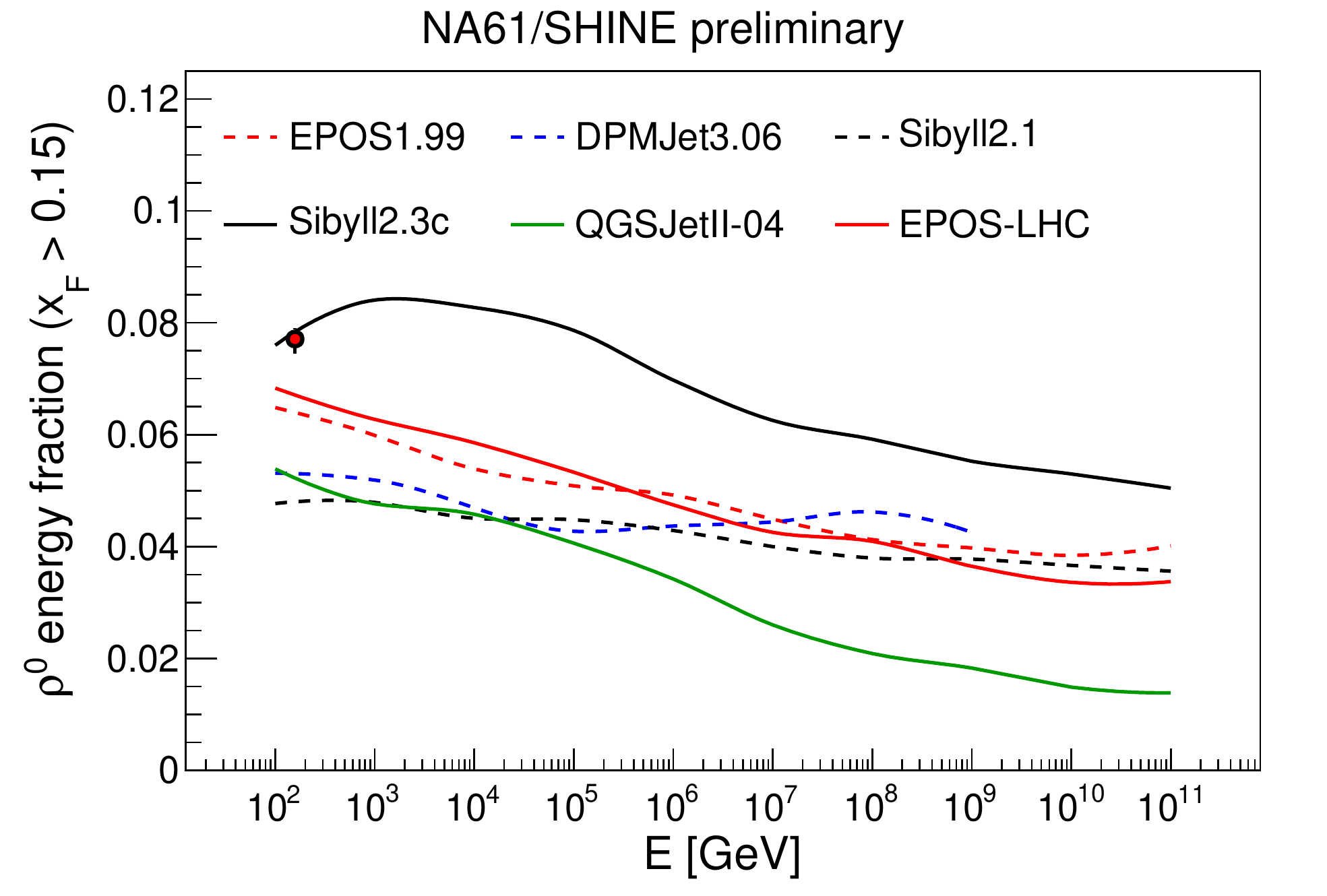}
 \caption{Energy fraction transferred to anti-protons (left) and
   $\rho^0$-mesons (right) in \piX{C} collisions as measured by \NASixtyOne (data points)
   and as predicted by hadronic interaction models over the whole
   range of beam energies relevant for air showers (images from \citet{Prado:2018wsv,Unger:2019nus}).}
  \label{fig:efrac}
\end{figure*}

This review focuses on the input from the LHC toward a solution of the Muon Puzzle in air showers, based on the results discussed in \sect{meta-analysis} that show a distinct onset of the muon discrepancy at energies which correspond to the TeV scale in the nucleon-nucleon cms-system. There are, however, also opportunities to further improve our knowledge on hadronic interactions in later stages of an air shower. A fixed-target setup can cover a large phase space for the measurement of particle productions relevant for air showers as discussed by~\cite{Meurer:2005dt}, as shown in \fg{pion_spectra_data}, and allows for great flexibility in the regard to the target nucleus, to study nuclear effects as a function of the system size. In this section, we discuss the two fixed-target experiments with high relevance for air shower physics at cms-energies that are one to two orders below the TeV scale, the \NASixtyOne experiment and the LHCb experiment in fixed-target mode.

\begin{itemize}
\item \emph{\NASixtyOne}. The main part of the \NASixtyOne experiment\footnote{\NASixtyOne is a double name. NA61 is the CERN designation and SHINE stands for SPS Heavy Ion and Neutrino Experiment.} \citep{Abgrall:2014xwa} is a set of large-acceptance time projection chambers (TPCs) with two superconducting magnets that have a combined bending power of 9\si{Tm}, resulting in a precise measurement of particle momenta and excellent particle identification capabilities via the specific energy loss in the TPC volumes. The setup is very flexible in regard to projectiles and targets, which allowed it to collect a large variety of data. Of particular interest for air shower physics are measurements of identified hadron spectra in \pX{C} interactions at 31\,\gevc \citep{Abgrall:2011ae,Abgrall:2011ts, Abgrall:2015hmv}, in \pp interactions from 20 to 158\gevc \citep{Aduszkiewicz:2020msu,Aduszkiewicz:2019ldi,Aduszkiewicz:2017sei,Aduszkiewicz:2015dmr,Abgrall:2013qoa,Abgrall:2013wda}, and in \piX{C} interactions at 158 and 350\gevc \citep{Aduszkiewicz:2017anm,Prado:2017hub,Prado:2018wsv,Unger:2019nus}. Resonances like \kshort, K$^{*0}$, $\Lambda$, $\omega$ and \rhozero were also measured. In the case of the data on \piX{C} and in the \pX{C} interactions, carbon is a very good proxy for air.

A key result for air showers is shown in \fg{efrac}, the total energy fractions transferred to anti-protons and \rhozero mesons in \piX{C} interactions. These fractions were obtained by integrating the $p \, \dd n/\dd p$ spectra including an extrapolation up to the full beam momentum \citep{Prado:2017hub}. The muon production in an air shower scales with the energy retained in the hadronic cascade. Since baryon number is conserved in subsequent interactions, this energy increases with the amount of baryon production. The forward \rhozero production is important since it is an alternative to the charge exchange reaction $\pi^- + p \to \piz + n + X$; leading \piz production incurs a large loss of hadronic energy, since there is a high probability of $\piz\rightarrow\gamma\gamma$. On the other hand, the measured anti-proton fraction constrains the production of $p$, \pbar, $n$, and \nbar. The sum of the energy fractions of these particles is about at the same level as the one going into \rhoz mesons. The comparison of the \NASixtyOne data to predictions of hadronic models reveals that none of the existing attempts to describe interactions in air showers succeeds in reproducing both energy fractions at the same time. These discrepancies get potentiated by the fact that a hadronic cascade initiated by a $10^{11}\gev$ particle will traverse all beam energies shown in \fg{efrac}. Measurements at energies beyond the SPS would be highly desirable to constrain the full air shower development. While baryon production can in principle also be measured in \pp collisions if universality in regard to the projectile can be assumed, a study of \rhozero production in the charge exchange reaction requires a $\pi^-$ beam and accelerating pion beams at the LHC is not foreseen. \cite{Petrov:2009wr} has investigated the prospect to measure the $\pi^+$-$p$ cross-section indirectly by tagging pion-exchange interactions between colliding protons, which is technically feasible. It would require tagging events with a neutron in a zero-degree calorimeter and a rapidity gap next to the neutron. One could use such events to infer the inelastic cross-section and to study inclusive hadron production.

\item \emph{LHCb in fixed-target mode}. The SMOG (System for Measuring the Overlap with Gas) is a device to inject small amounts of noble gases directly into the LHC beam pipe around the LHCb collision point \citep{Aaij:2014ida}. It was designed to precisely measure the radial profile of the LHC beams to accurately compute the luminosity of colliding proton bunches, as an alternative to van-der-Meer scans \citep{Aaij:2011er,Aaij:2014ida}. A combination of both techniques yielded the most accurate luminosity measurements so far at a bunched-beam collider \citep{Aaij:2014ida}.

This system has been used to study fixed-target interactions with a variety of noble gases. So far, interactions of proton and lead beams with helium, neon, and argon have been recorded with centre-of-mass energies $\sqrtsnn$ of 68, 87, and 110\gev. LHCb has access to the highest energies ever obtained in a fixed-target experiment and closes the energy gap between previous fixed target experiments and the TeV scale. Measurements of the charm production in \pX{He} and \pX{Ar} \citep{LHCb:2018jry} and anti-proton production in \pX{He} have been published \citep{LHCb:2018ygc}. The forward acceptance of LHCb in collider mode corresponds to an acceptance at mid-rapidity $-2.5 < \eta_\text{cms} < 0.5$ in the nucleon-nucleon cms frame when running in fixed-target mode. The fixed-target measurements of \pX{A} collisions in the mid-rapidity region offer important opportunities to study the hadron composition and its modification as a function of the nuclear target.

The SMOG device is currently upgraded \citep{Barschel:2020drr} by installing a gas storage cell with open windows upstream of the vertex locator. This upgrade allows for injecting non-noble gasses, in particular hydrogen, nitrogen, and oxygen, and at higher densities to increase the interaction rates by two orders of magnitude. The new device was been designed explicitly to allow fixed-target experiments with LHC beams and is also motivated by air shower physics. The increased luminosity will improve the accuracy of studies of charm and beauty production, while air shower physics will profit from the study of the hadron composition in different projectile-target system combinations.
A particularly interesting opportunity with SMOG2 for air shower physics will arise when the LHC is run with oxygen beams.
\end{itemize}

\subsection{Prospects with oxygen beams in the LHC}
\label{sec:oxygen}

\begin{figure}[tb]
\centering
\includegraphics[width=0.9\columnwidth]{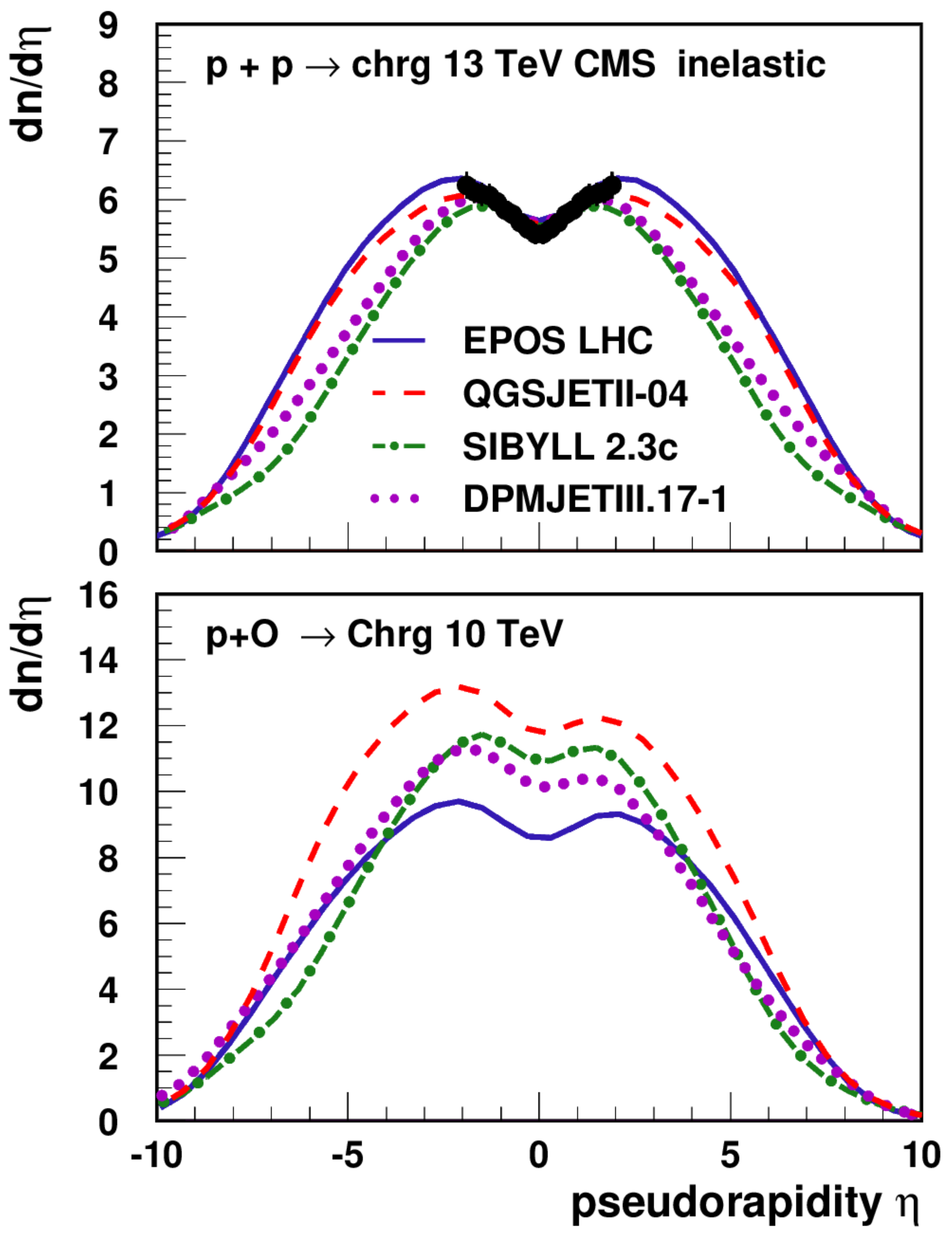}
\caption{Charged particle spectra from different hadronic generators in \pp at 13\tev and in \pO at 10\tev. The data points are from CMS \citep{CMS:2015zrm}.}
\label{fig:eta_spectra_pp_pO}
\end{figure}

\cite{Citron:2018lsq} discuss the science case for and technical feasibility of a pilot run with oxygen beams at the LHC at the end of Run~3 to study \pO and \OO collisions at 10 and 7\tev, respectively, in response to a long-standing request by the astroparticle community. The pilot run has been approved and is expected in 2023 or 2024. Details about the planned run are given by \cite{Bruce:2021hii}.

The primary motivation to request oxygen beams is the fact that measurements in \pp and \pPb are both not representative of the first interaction in an air shower, while \pO and \OO are both suitable references \citep[section 11.3]{Citron:2018lsq}. To identify simple laws and universal features of proton-ion collisions, an intermediate system between \pp and \pPb collisions is needed  as indicated by \fg{collision_systems}, since an interpolation based on two points is ambiguous. Caution is particularly warranted in light of the recent surprising findings of collective effects in high multiplicity \pp and \pPb collisions that were previously discussed in section 3.

There is a considerable theoretical uncertainty in the extrapolation of hadron production parameters from \pp to \pO, as demonstrated in \fg{eta_spectra_pp_pO}. Most hadronic interaction models (with the exception of EPOS) are only tuned to \pp collisions and then extrapolated to \pX{air} collisions. Models agree to better than 5\,\% at 13\tev in \pp collisions at mid-rapidity, but the spread in \pO collisions at 10\tev is 50\,\%. Similar divergence is found for other relevant aspects of hadronic collisions, for example, the \pX{air} cross-section previously shown in \fg{basic_model_features}. All measurements discussed in \tb{lhc-measurements} should be performed in \pO and compared to an interpolation from \pp to \pPb, especially measurements at forward rapidities.

Some measurements also become more precise in \pO. The LHCf experiment has measured the forward photon and $\piz$ flux in \pPb, but the accuracy of the measurement suffers from a large background from ultra-peripheral collisions (UPCs), in which the proton interacts with a virtual photon emitted by the lead nucleus, to form e.g.\ a $\Delta(1232)$ resonance, which in turn decays into a proton and a \piz or a neutron and a charged pion. This process occurs with a similar or even larger rate in \pPb collisions compared to strong interactions. Thus, LHCf uses a model-dependent correction which introduces a leading systematic uncertainty \citep{LHCf:2014gqm}. Since the UPC rate scales with $Z^2$, it is completely negligible in \pO. \cite{CMS:2015nfb} faced a related issue in their measurement of the \pPb inelastic cross-section, but the measurement was not as severely affected.

Interesting opportunities with oxygen beams also arise for the LHCb experiment running in fixed-target mode. If oxygen gas is injected, it would allow LHCb to study \OO collisions at $\sqrtsnn = 81\gev$ at mid-rapidity, $-2.5 < \eta_\text{cms} < 0.5$. If hydrogen gas is injected, the O-$p$ system at $\sqrtsnn = 115\gev$ probes the mid to near forward region $-0.5 < \eta_\text{cms} < 2.5$ in the nucleon-nucleon system. The acceptance further forward is an advantage compared to the mirrored system in which the proton beams collide with oxygen gas.

\section{Summary}

The muon deficit in air shower simulations has been experimentally established with $8\si{\sigma}$ evidence \cite{Dembinski:2019uta,Cazon:2020zhx,Soldin:2021wyv}. Measurements with small model-dependence were important to establish the muon deficit and were provided by the Pierre Auger Observatory and the IceCube Neutrino Observatory. The muon deficit has an onset at $\sqrtsnn \approx 8\tev$ in the nucleon-nucleon cms-system, followed by a linear increase with logarithm of the energy. The source of the deficit therefore should be observable in high-energy collisions at the LHC. The most likely explanation consistent with all available data is a small modification to the hadron production that reduces the energy fraction carried by photons, which originate mostly from \piz decays, in soft hadronic collisions. Such a modification has a compounded effect on the hadronic cascade, so that only a comparably small modification is required. A small modification would not destroy the consistency of current air shower simulations with other air shower data; the first two moments of the depth of shower maximum and the intrinsic fluctuations of the muon number in air showers that were recently measured by the Pierre Auger Observatory for the first time.

An enhancement of strangeness production that qualitatively matches this desired behaviour was observed at the LHC by ALICE in the mid-rapidity region \citep{ALICE:2016fzo}. There is no consensus in regard to the theoretical explanation of the phenomenon, but the most important point for air showers is its apparent universality \citep{Anchordoqui:2019laz}. The modification does not depend on the collision energy and the collision system, only on the multiplicity of the event. A preliminary study \citep{Baur:2019cpv} suggests that the effect could resolve the muon discrepancy. Since hadron production at mid-rapidity does not have a direct impact on air showers, the ALICE result by itself does not yet solve the Muon Puzzle. Required are measurements of the forward hadron production at a pseudo-rapidities $\eta > 2$ in the region relevant for air showers. Further needed are studies of light hadron production in the forward region and of the hadron composition, and in particular strangeness production, as a function of the charged particle multiplicity in the collision systems from \pp and \pPb, and possibly \PbPb. The LHCb experiment is in a unique position to perform these so far missing measurements. Also important are direct measurements of the ratio of the electromagnetic and hadronic energy flow with the CASTOR experiments in proton-lead collisions in addition to those in proton-proton collisions.

In light of these findings and in regard to the current divergence of generators of hadronic interactions when applied to \pO collisions, studies of \pO collisions at the LHC as proposed in \cite{Citron:2018lsq} are essential to understand the evolution of hadron production between \pp and \pPb, which should be performed with all LHC experiments to characterise hadron production over a wide acceptance in rapidity. To improve the accuracy of air shower predictions for the depth of shower maximum, also a precise measurement of the \pO cross-section is highly desired which could be provided by TOTEM or ATLAS-ALFA and the very forward photon production, which could be measured with LHCf, if the runs with oxygen beams are not postponed to Run 4 of the LHC.

A solution to the muon deficit from the LHC would have a large impact on the astroparticle community. Research on high-energy cosmic rays would directly benefit, but also research with gamma-ray and neutrino observatories, since cosmic rays generate the background for these observatories. In the field of cosmic-ray experiments, future upgrades will provide more muon data. The Pierre Auger Observatory is upgrading its array (AugerPrime) to give each surface detector muon separation capabilities \citep{Aab:2016vlz}. The radio observation of highly-inclined air showers will combine a calorimetric measurement of the shower energy using radio data with muon measurements in the surface detectors. The IceCube Neutrino Observatory is working on a study of the GeV and TeV muon component in air showers, which can be measured simultaneously by combining data from the surface array and the deep in-ice detector \citep{DeRidder:2017alk}. An extension of IceTop, the surface array of IceCube, with scintillator and radio detectors \citep{Haungs:2019ylq,Schroder:2019suq} will provide a calorimetric measurement of the shower energy and will further enhance the sky coverage, energy range, and accuracy of hybrid muon measurements. The future measurements with AugerPrime and IceCube will significantly contribute to the solution of the Muon Puzzle in EAS.

\section{Acknowledgements}

We thank Roger Clay, Felix Riehn, and Lorenzo Sestini for their valuable comments on this review.

\section{Declarations}

\subsection{Funding}

A.F.\ contributed as JSPS International Research Fellow (JSPS KAKENHI Grant Number 19F19750). L.C.\ wants to thank for the financial support by
OE - Portugal, FCT. J.A.\ acknowledges support from the Heisenberg programme of the Deutsche Forschungsgemeinschaft (DFG), GZ: AL 1639/5-1. H.D., B.S., and KH.K. acknowledge support from the Verbundforschung of the German national agency BMBF. D.S.\ acknowledges support from the U.S. National Science Foundation (award number PHY-1913607). W.R.\ acknowledges the funding by the Deutsche Forschungsgemeinschaft under grant number DFG RH 35/9-1. R.U.\ and T.P.\
acknowledge support by the High Performance and Cloud Computing Group at the Zentrum für Datenverarbeitung of the University of Tübingen, the state of Baden-Württemberg through bwHPC and the German Research Foundation (DFG) through grant no INST 37/935-1 FUGG.

\subsection{Conflicts of interest/Competing interests}

Not applicable.

\subsection{Availability of data and material}

This review uses only previously published data.

\subsection{Code availability}

Not applicable.

\bibliographystyle{spr-mp-nameyear-cnd}
\bibliography{main,gsf,lhc_measurements_ALICE,lhc_measurements_ATLAS,lhc_measurements_CMS,lhc_measurements_LHCb,lhc_measurements_LHCf,lhc_measurements_TOTEM}

\end{document}